\documentclass[prd,aps,twocolumn,a4paper,floatfix,amsmath,amssymb,superscriptaddress]{revtex4-1}
\usepackage{graphicx,color,array,float}

\newcommand{\rb}{{\bar r}}
\newcommand{\tb}{{\bar t}}

\newcommand{\p}{\partial}
\newcommand{\br}{\nonumber \\ &&}
\newcommand{\tul}{\left(-{\tilde u\over\ell}\right)}
\newcommand{\tvl}{\left(-{\tilde v\over\ell}\right)}
\newcommand{\hul}{\left(-{\hat u\over\ell}\right)}
\newcommand{\hvl}{\left(-{\hat v\over\ell}\right)}
\newcommand{\cA}{{\cal A}}
\newcommand{\tcA}{\tilde{\cal A}}
\newcommand{\bcA}{\bar{\cal A}}
\newcommand{\hcA}{\hat{\cal A}}
\newcommand{\ccA}{\check{\cal A}}
\newcommand{\cR}{{\check R}}
\newcommand{\cf}{{\check f}}
\newcommand{\cx}{{\check x}}
\newcommand{\cn}{{\check n}}
\newcommand{\crho}{{\check\rho}}
\newcommand{\cF}{{\check F}}

\begin{document}


\title{Scalar field critical collapse in 2+1 dimensions}

\author{Joanna Ja\l mu\.zna}
\affiliation{Institute  of  Physics,  Jagiellonian  University,  Krak\'ow,  Poland}
\author{Carsten Gundlach}
\affiliation{Mathematical Sciences, University of Southampton,
  Southampton SO17 1BJ, United Kingdom}
\author{Tadeusz Chmaj} 
\affiliation{Cracow University of Technology, Krak\'ow, Poland}
\affiliation{Institute of Nuclear Physics, Polish  Academy of Science, Krak\'ow, Poland }


\begin{abstract}

We carry out numerical experiments in the critical collapse of a
spherically symmetric massless scalar field in 2+1 spacetime
dimensions in the presence of a negative cosmological constant and
compare them against a new theoretical model. We approximate the true
critical solution as the $n=4$ Garfinkle solution, matched at the
lightcone to a Vaidya-like solution, and corrected to leading order
for the effect of $\Lambda<0$. This approximation is only $C^3$ at the
lightcone and has three growing modes. We {\em conjecture} that
pointwise it is a good approximation to a yet unknown true critical
solution that is analytic with only one growing mode (itself
approximated by the top mode of our amended Garfinkle solution). With
this conjecture, we predict a Ricci-scaling exponent of $\gamma=8/7$
and a mass-scaling exponent of $\delta=16/23$, compatible with our
numerical experiments.

\end{abstract}

\maketitle

\tableofcontents


\section{Introduction}



\subsection{Critical collapse}


Starting with Choptuik's investigation of scalar field collapse
\cite{Choptuik1993}, and since then generalised to many other systems
\cite{GundlachLRR}, critical collapse is concerned with the threshold
of black hole formation in the space of initial data. A practical way
of investigating this threshold is to pick any one-parameter family of
asymptotically flat initial data, with parameter $p$, such that for
$p>p_*$ the data form a black hole, and for $p<p_*$ they do not.

More specifically, ``type II'' critical collapse is concerned with the
case where the black hole mass can be made arbitrarily small at
the threshold. A necessary condition for this to happen is that the
system of Einstein equations and matter evolution equations is
scale-invariant, or effectively scale-invariant on sufficiently small
length scales. As far as we know, exact scale-invariance is also
sufficient for the existence of type II critical collapse.

In type II critical collapse in $d+1$ spacetime dimensions, for
$p<p_*$ (``subcritical'' data), the maximum value of curvature (say
the Ricci scalar) achieved on the spacetime scales as
\begin{equation}
\label{Ricciscaling}
|R|_{\rm max}\sim (p_*-p)^{-2\gamma}
\end{equation}
and for $p>p_*$ (``supercritical'' data), the black hole mass scales
as
\begin{equation}
\label{massscaling}
M_{BH}\sim (p-p_*)^\delta
\end{equation}
where in $d\ge 3$
\begin{equation}
\label{deltafromgamma}
\delta=\gamma(d-2).
\end{equation}
The relation (\ref{deltafromgamma}) follows essentially from
dimensional analysis, with $d-2$ the dimension (in gravitational units
$c=G=1$) of mass (or energy). The exponent $\gamma$ depends on the
type of matter and spacetime dimension, but is universal for all
1-parameter families of initial data.

In a small spacetime region just before the point of maximum
curvature, or just before the formation of an apparent horizon,
the spacetime and matter field are approximated by a ``critical
solution'' which is again universal for a given system and spacetime
dimension. The critical solution has three defining properties: it is
regular, scale-invariant (continuously self-similar, CSS) or
scale-periodic (discretely self-similar, DSS), and it has precisely
one unstable mode. Continuous self-similarity means that there is a
conformal Killing vector field $K$ such that ${\cal
  L}_Kg_{ab}=-2g_{ab}$. In coordinates $(x,T)$ adapted to CSS and
spherical symmetry (but otherwise general), such that $K=\p/\p T$,
this means that the metric takes the form
\begin{eqnarray}
\label{generalCSS}
ds^2&=&\ell^2e^{-2T}[A(x)\,dT^2+2B(x)\,dT\,dx
+C(x)\,dx^2 \br +R^2(x)\,d\Omega_{d-1}^2], 
\end{eqnarray}
where $\ell$ is an arbitrary length scale. This functional form of the
metric is invariant under gauge transformations of the form
\begin{equation}
x\to F(x), \qquad T\to T+G(x).
\end{equation}
[In DSS, in adapted coordinates, the metric takes the same form, with
$A$, $B$, $C$, $R$ (and $F$, $G$) now depending periodically on $T$
with some scale-echoing period $\Delta$.] 

The most general ansatz for a massless scalar field that is
compatible, via the Einstein equations with $\Lambda=0$, with
continuous self-similarity of the metric is the Christodoulou ansatz
\cite{Christodoulou}
\begin{equation}
\label{Christodoulouansatz}
\phi(x,T)=cT+f(x)
\end{equation}
for some constant $c$. [For DSS, $f=f(x,T)$ depends also on $T$ with
  period $\Delta$.] The constant $c$ does not depend on the choice of
similarity coordinates. The spherical scalar field critical solution
in higher dimensions is DSS with $c=0$ but, as we shall see later, in
2+1 dimensions it seems to be CSS with $c\ne 0$.

In a spherically symmetric critical solution, the regular centre
corresponds to one value of $x$. $T=\infty$ (for all $x$) represents a
single spacetime point at the centre, the accumulation point, where
the curvature blows up. Another value of $x$ corresponds to the past
lightcone (or soundcone, for fluid matter) of the accumulation
point, where the critical solution must also be regular. The critical
solution can
be continued in $x$ to the future lightcone of the accumulation
point. Beyond the future lightcone, there is no unique continuation,
but that part of the critical solution is not relevant for critical
collapse.

If we choose $T$ to be timelike or null, we can interpret it both
as a time coordinate on spacetime and as the logarithm of scale in
renormalisation group theory. From self-similarity and the existence
of precisely one unstable mode, using a little dynamical systems
theory and dimensional analysis, one can then derive both
universality and the above scaling relations. $\gamma$ turns out to be
the inverse Lyapunov exponent of the one unstable mode.

This scaling argument \cite{KoikeHaraAdachi1995,GundlachLRR} goes
roughly as follows: the closer $p$ to $p_*$, the smaller the initial
value of the one growing mode, the longer (larger $T$) the
spacetime stays close to the critical solution. But larger $T$
also means scalar field variation on smaller length scales,
and hence larger curvature, before the solution either starts
dispersing or forms an apparent horizon.

For a spherically symmetric massless scalar field in the presence of a
negative cosmological constant, critical collapse has been
investigated in 3+1 dimensions \cite{BizonRostworowski2011}. In higher
dimensions, critical collapse has been investigated in
\cite{Birukouetal} for $\Lambda=0$, and in
\cite{JalmuznaRostworowskiBizon2011} for $\Lambda<0$. A cosmological
constant (of either sign) obviously breaks scale-invariance, but one
would expect it to become negligible in regions of sufficiently large
curvature, and hence in the regime where type II critical phenomena
are seen. Indeed this seems to be the case in 3+1 and higher
dimensions. A further effect of a negative cosmological constant is to
replace asymptotic flatness with asymptotically anti-deSitter (adS)
boundary conditions. The only boundary conditions for a massless
scalar field compatible with the Einstein equations are totally
reflecting. As a consequence, it appears that arbitrarily weak generic
initial data collapse after sufficiently many reflections off the
boundary. (But see \cite{MaliborskiRostworowski2013} for exceptions to
this). However, at the thresholds $p_{*0}$, $p_{*1}$, $p_{*2}$ for
black hole formation after zero, one, two, and so on, reflections the
same type II critical phenomena are seen as in asymptotically flat
spacetime. Because of the reflecting boundaries, all the mass must
fall into the black hole eventually, but the mass of the apparent
horizon when it first forms does scale, with the same $\gamma$ as the
black hole mass in asymptotically flat spacetime.


\subsection{2+1 dimensions}


The situation is quite different in 2+1 dimensions. First, this is the
critical dimension for the wave equation, meaning that the scalar
field energy ($|\nabla\phi|^2$ integrated over $d$ space dimensions)
is dimensionless. Similarly, for gravity the black hole mass and the
2+1 dimensional equivalent of the Hawking mass are
dimensionless. This already indicates that any mass scaling cannot be
derived using the standard dimensional analysis argument. Secondly, in
the absence of a cosmological constant there are no black hole
solutions, and finite mass regular initial data cannot form an
apparent horizon dynamically. 

Standard gauge choices in spherical symmetry in $2+1$ spacetime
dimensions are polar-radial coordinates $(\bar r,\bar t)$,
\begin{equation}
\label{metrictr}
ds^2=e^{2\alpha}\left(-e^{2\beta}d\tb^2+d\rb^2\right)+\rb^2\,d\theta^2,
\end{equation}
where the area radius $\rb$ is a coordinate, 
and double null coordinates $(u,v)$,
\begin{equation}
\label{metricuvr}
ds^2=-e^{2\cA}du\,dv+\rb^2\,d\theta^2,
\end{equation}
where $\rb$ is a metric coefficient.
With $u=:t-r$ and $v=:t+r$, this can also be written as
\begin{equation}
\label{ds2Psi}
ds^2=e^{2\cA}(-dt^2+dr^2)+\rb^2\,d\theta^2.
\end{equation}
[In $d+1$ dimensions, the same coordinate choices exist, with
$d\theta^2$ replaced by the line element on the unit $(d-1)$-sphere.]

In 2+1 dimensions, the field equations 
\begin{equation}
\label{fieldequations}
  G_{ab}+\Lambda g_{ab}=\kappa 
[\nabla_a\phi\nabla_b\phi-{1\over 2}g_{ab}(\nabla\phi)^2],
\quad \nabla^2\phi=0
\end{equation}
for the metric (\ref{metricuvr}) are 
\begin{eqnarray}
\label{phiuveqn}
2\rb\phi_{,uv}+\rb_{,u}\phi_{,v}+\rb_{,v}\phi_{,u}&=&
0,\\
\label{Auvequation}
-4\cA_{,uv}-2\kappa\phi_{,u}\phi_{,v}+\Lambda e^\cA&=&0, \\
\label{rbuveqn}
-2\rb_{,uv}+\Lambda e^\cA\rb&=&0, \\
\label{rbuuequation}
\rb_{,uu}-2\cA_{,u}\rb_{,u}+\kappa \rb \phi_{,u}^2&=&0, \\
\rb_{,vv}-2\cA_{,v}\rb_{,v}+\kappa \rb \phi_{,v}^2&=&0.
\end{eqnarray}
These are the field equations that we will use in the theory
Section~\ref{section:theory} below. 

In 2+1 dimensions, if $\Lambda=0$, then from (\ref{rbuveqn})
$\rb_{,uv}=0$. In a region containing a regular centre, one can then
make the same gauge choice $\rb=(v-u)/2$ as in flat spacetime. But,
always in 2+1 dimensions, the coefficients of the spherical wave
equation (\ref{phiuveqn}) depend only on $\rb$, not on $\cA$, and so
the matter evolution equation is not modified by curvature. This is
one intuitive way of seeing why gravitational collapse cannot occur in
2+1 with $\Lambda=0$.

However, in the presence of a negative cosmological constant
$\Lambda=:-1/\ell^2$ black holes do exist in 2+1 spacetime dimension,
and can be formed from regular data. These black holes are the BTZ
solutions, which in polar-radial coordinates are given by
\begin{equation}
ds^2=-\left({\rb ^2\over\ell^2}-M\right)d\tb^2
+\left({\rb ^2\over\ell^2}-M\right)^{-1}d\rb ^2+\rb ^2\,d\theta^2.
\end{equation}
Although this looks similar to the Schwarzschild-adS solution in
higher dimensions, it is locally flat. This is because in 2+1
dimensions, the Ricci tensor determines the Weyl tensor, and so a
vacuum region is not only Ricci-flat but flat. The BTZ solution with
$M=-1$ is the $2+1$-dimensionsonal adS spacetime. All other BTZ
solutions with $M<0$ have a naked conical singularity, while the BTZ
solutions with $M>0$ are black hole solutions. This mass gap between
the ground state and the smallest black hole is another feature of 2+1
dimensions. Regular initial data with $-1<M<0$ cannot form a black
hole (although they can develop arbitrarily large curvature
\cite{BizonJalmuzna2013}.)

There seems to be a dilemma for type II critical collapse: in order to
form a black hole at all, a cosmological constant is needed, but for
curvature and mass scaling to occur, it must be dynamically
negligible. 

It is convenient to introduce the local mass function $M(u,v)$ defined
by
\begin{equation}
\label{BTZmass}
M=:{\rb ^2\over\ell^2}-(\nabla \rb)^2.
\end{equation}
This is the 2+1 dimensional equivalent of the Hawking mass for
spherical symmetry in 3+1 dimensions, and has similar properties: it
is constant in vacuum, while in the presence of matter it increases
with $\bar r$ on any spacelike surface in regions where $(\nabla
\rb)^2>0$. A spherically symmetric marginally outer-trapped surface
(MOTS) is given by $\rb_{,v}=0$, and so its mass is given by
$\rb^2/\ell^2$, as is the mass of the BTZ horizon.


\subsection{Previous work}


The first numerical simulations of critical collapse of a spherically
symmmetric scalar field in 2+1 dimensions with a negative cosmological
constant were carried out by Pretorius and Choptuik
\cite{PretoriusChoptuik} and Husain and Olivier
\cite{HusainOlivier}.

In order to avoid the complications associated with the reflecting
boundary conditions, Pretorius and Choptuik, like others in 3+1 and
higher dimensions after them, focused on the scaling of maximum
curvature and the mass of the apparent horizon when it first appears.
They found that for each of several one-parameter families of initial
data they examined, there was a $p_*$ such that the maximum of the
Ricci curvature scaled as (\ref{Ricciscaling}) where $\gamma \simeq
1.2\pm0.05$. They also gave evidence for a universal CSS critical
solution. They claimed also that the apparent horizon mass at first
appearance scales as
\begin{equation}
\label{MAHscaling}
M_{\rm FMOTS}\sim (p-p_*)^\delta  
\end{equation}
with $\delta=2\gamma$, although their Figs.~4 and 5 correctly suggest a mass
scaling exponent somewhere between 0 and 1. [We use the terminology
  FMOTS for for ``first marginally outer trapped surface'', as the
  terminology ``apparent horizon mass'' is ambiguous in this context; see
  Sec.~\ref{section:massscaling} below.] Their theoretical argument
for $\delta=2\gamma$ is that the dimensionless mass $M$ and area
radius $\bar r$ of an apparent horizon are related by $M_{\rm AH}=\bar
r^2_{\rm AH}/\ell^2$, and $r_{\rm AH}$ should scale as suggested by
its dimension. We shall correct this argument in
Sec.~\ref{section:deltaderivation}. Husain and Olivier found apparent
horizon mass scaling with $\delta\simeq 0.81$, consistent with our
results, but their data are fairly far from criticality.

On the grounds that $\Lambda$ should be dynamically negligible in
critical collapse, Garfinkle \cite{Garfinkle} looked for exactly CSS
solutions for $\Lambda=0$ that are analytic between the two values of
$x$ corresponding to the centre and to the past lightcone of the
accumulation point (the standard procedure in higher dimensions). As
we shall review in Sec.~\ref{section:garfinkle}, he found a family of
these parameterised by $n=1,2,3,\dots$. The $n=1$ solution is the
Friedmann-Robertson-Walker solution. In hindsight it is surprising
that these solutions exist, as we have seen that with $\Lambda=0$
gravity does not affect the scalar field and so cannot regularise it,
something that is essential for the existence of regular CSS solutions
in higher dimensions. The Garfinkle solution is also in closed form,
whereas critical solutions for spherical massless scalar field
collapse in higher dimensions can only be constructed numerically (but
see \cite{ReitererTrubowitz} for an existence proof of the Choptuik
critical solution in 3+1 dimensions).

Garfinkle \cite{Garfinkle} noted that the $n=4$ solution showed good
agreement with the numerical data of Pretorius and Choptuik inside the
lightcone. However, the lightcone is also an apparent horizon,
whereas the critical solution in higher dimensions has no trapped
surfaces. Furthermore, the analytic continuation of the Garfinkle
solution through the lightcone has a spacelike central curvature
singularity, for all $n$. This means that it is the CSS equivalent of
a black hole, rather than a critical solution. (We will fix these
problems in Secs.~\ref {section:continuations},
\ref{section:LambdaGarfinkle}, \ref{section:Lambdanull} and
\ref{section:outeransatz} below.)

Ignoring these obvious problems of the Garfinkle solution, Garfinkle
and Gundlach \cite{GarfinkleGundlach} computed its perturbation
spectrum, by making the standard requirement that perturbations be
analytic at both the centre and lightcone. As we shall review in
Sec.~\ref{section:garfinklepert}, they found that the Garfinkle
solution with parameter $n$ has $n-1$ unstable modes. This then raised
the problem that the $n=2$ Garfinkle solution does not fit the
numerical data, while the $n=4$ Garfinkle solution, which does, has
three growing modes. We have no theoretical solution for this problem,
but we will show numerically in
Sec.~\ref{section:evolveamendedGarfinkle} that our modified $n=4$
Garfinkle solution appears to have only one growing mode when evolved
with $\Lambda<0$.


\section{Numerical results}


\subsection{Numerical method}


We experimented with a time evolution code using polar-radial
coordinates, the standard coordinate choice for critical collapse in
higher dimensions. However, as we want to continue the evolution after
the time slicing crosses the apparent horizon, we have changed over to
the numerical method of Pretorius and Chopuik \cite{PretoriusChoptuik}.

The metric ansatz is essentially (\ref{ds2Psi}), but reparameterised as
\begin{equation}
\label{PCmetric}
ds^2=\cos^{-2}\left({r\over \ell}\right)e^{2A}(-dt^2+dr^2)
+\ell^2\tan^2\left({r\over \ell}\right)e^{2B}\,d\theta^2,
\end{equation}
so that
\begin{equation}
\cA=A-\ln[\cos(r/\ell)], \qquad \rb=\ell\tan(r/\ell)\exp B.
\end{equation}
This brings the timelike infinity of asymptotically anti-de Sitter
spacetimes to $r=\ell\pi/2$ and the centre $\rb=0$ to $r=0$. Note
that the adS spacetime is given by $A=B=0$. We refer the reader to
\cite{PretoriusChoptuik} for the field equations in these coordinates.

The metric effectively represents the metric in double-null
coordinates $u:=t-r$ and $v:=t+r$ (which go through apparent or event
horizons), but the numerical algorithm evolves it on a grid in $t$ and
$r$, time-stepping in $t$.

Both $A$ and $B$ obey wave equations and are evolved from initial data
at $t=0$. The residual gauge freedom is $u\to u'(u)$ and $v\to
v'(v)$. We fix this in part by setting the initial data $B=B_{,t}=0$
at $t=0$. With $\phi$ and $\phi_{,t}$ also set freely, the initial
data for $A$ and $A_{,t}$ are then determined by the Hamiltonian and
momentum constraints. During the evolution, we impose the gauge fixing
boundary conditions $A=B_{,r}=0$ at the adS timelike infinity $r=1$,
and the regularity boundary conditions $A=B$ and $A_{,r}=B_{,r}=0$ at
the centre $r=0$.

The Hamiltonian and momentum constraints become singular on any time
slice that contains a trapped surface, but we solve them only on the
initial slice. The evolution equations remain regular at a trapped
surface.

We choose units such that $G=2$ and $\ell=\pi/2$, so that $r=0$
represents the regular centre and $r=1$ the adS boundary. We choose
$\Delta t/\Delta r=1/64$, and typically $4096$ equally spaced grid
points in $r$.

For a geometric analysis of the results, and in particular for looking
for self-similarity near the centre, more geometric coordinates fixed
at the centre are helpful. This will be discussed in
Sec.~\ref{section:CSStest} below. 


\subsection{Evolution of fine-tuned generic initial data}
\label{section:genericdata}


In scalar field critical collapse in 3+1 and higher dimensions there
is a clear distinction between two outcomes. Either the scalar field
forms a black hole, and the remaining scalar field escapes to
infinity, or the scalar field disperses, leaving behind flat
spacetime. With a negative cosmological constant, there are the twin
complications that a scalar wave that disperses initially can collapse
after one or more reflections at the outer boundary, and that more
scalar field can fall into an initially small black hole after
reflection.  However, locally in space and time there are still two
distinct outcomes, at least as long as the initial data are on scales
much smaller than the scale $\ell$ set by $\Lambda=-1/\ell^2$.  From
now on, the previously arbitray length scale $\ell$ in
(\ref{generalCSS}) is set by the cosmological constant for
definiteness. Hence $T>0$, from (\ref{generalCSS}), indicates
spacetime scales smaller than $\ell$.

In 2+1 dimensions, the situation appears initially more confusing. The
Ricci scalar at the centre either blows up while increasing
monotonically, or it goes through one or more extrema before blowing
up a short time later. Similarly, the mass of the first
MOTS appearing anywhere on a time slice (what \cite{PretoriusChoptuik}
call the apparent horizon mass) behaves in a non-monotonic way with $p$.

We adopt the working definition of $p_*$ that for $p>p_*$, $|R(0,t)|$
monotically increases and blows up at finite $t$, while for $p<p_*$ it
goes through at least one maximum and minimum before blowup. We shall
see that with this definition, $|p-p_*|$ controls all scaling
phenomena. This is in itself an important observation, as it strongly
indicates that the scaling is controlled by a single growing mode of a
self-similar critical solution.

For the scalar field initial data we choose approximately ingoing
(that is $\phi_{,t}=\phi_{,r}$) Gaussian or kink profiles located at
$r_0=0.2$ with width $\sigma=0.05$.  Their amplitude $p$ is a free
parameter used for fine-tuning the initial data to the black hole
threshold.  Note that both chosen families of initial data are the
same as considered in \cite{PretoriusChoptuik}, which allows us to
compare results. We find that $M_{\rm tot}(p_*)\simeq 0.003$ for both
these two families. All plots and numbers presented in the current
Subsection~\ref{section:genericdata} use the Gaussian family, but we
have checked that we obtain the same results for the kink data.

The absolute value of $p_*$ for any given one-paramter family is
irrelevant and depends on the parameterisation. However, with $p_*$ of
order one, $p-p_*$ is a meaningful measure of the amount of
fine-tuning. For simplicity, we use the terminology ``sub10'' for
initial data with $p\simeq p_*-\exp (-10)$ and ``super10'' for
$p\simeq p_*+\exp (-10)$. The best fine-tuning we have achieved is of
the order of $\exp(-26)$.


\subsubsection{Ricci scaling at the centre}

\begin{figure}[!htb]
\includegraphics[scale=0.7, angle=0]{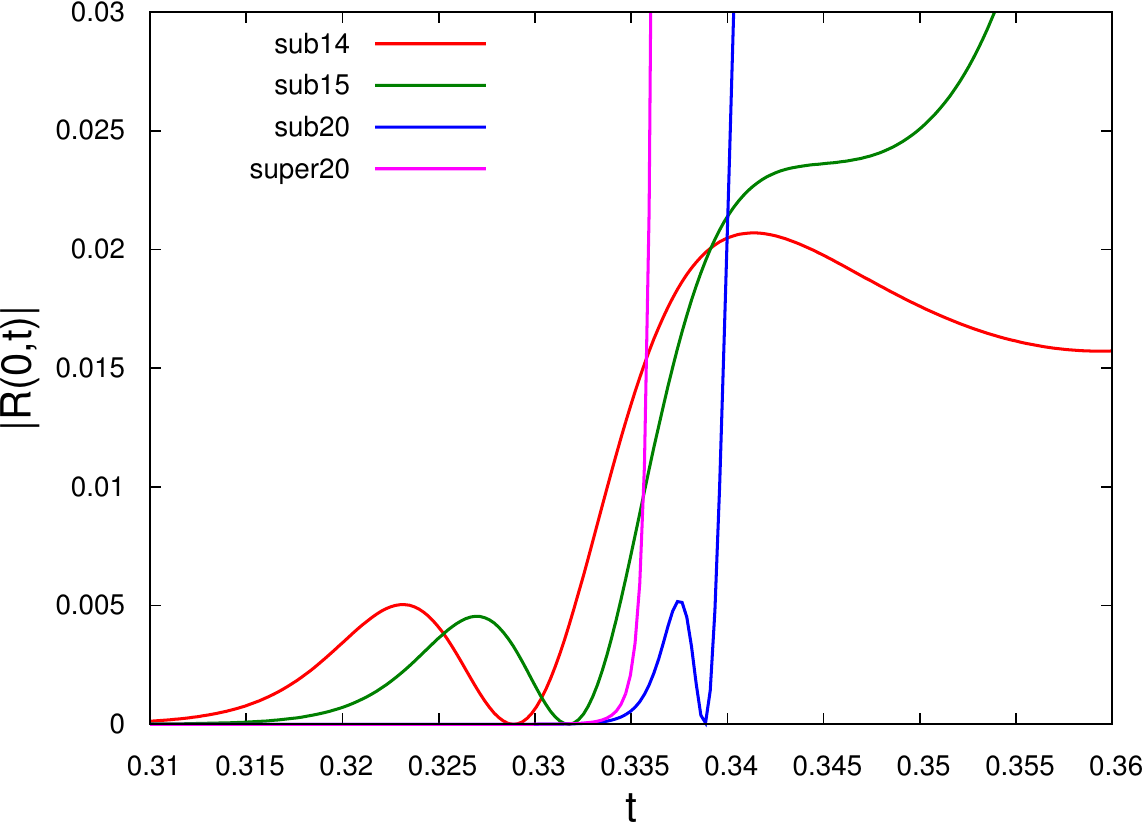}
\caption{ $|R(0,t)|$ against $t$ for representative values of $p$. The
  vertical axis has been rescaled to give the first maxima
    approximately the same value. The magenta curve super20 is for
  supercritical data, with immediate blowup of Ricci, while the other
  lines show representative subcritical data: sub20 has a maximum and
  minimum of Ricci followed by blowup (for $t>0.36$), sub14 has two
  maxima and minima followed by blowup, and sub15 represents the
  transition between these last two cases, where the second maximum
  and minimum merge.}
\label{figure:Rvst}
\end{figure}

As stated above, we define $p_*$ so that for $p>p_*$, $|R(0,t)|$
monotically increases until blowup, while for $p<p_*$ there is at
least one maximum and minimum before blowup. For subcritical data
further away from criticality than approximately sub15, the Ricci
scalar at the centre goes through a second maximum and minimum before
blowup. Fig.~\ref{figure:Rvst} illustrates this for representative
values of $p$.  Going further away from criticality, the second
minimum and eventual blowup moves to larger values of $t$. For about
sub8, the blowup moves to a time $t\simeq 2.3$ that indicates one
reflection from the outer boundary; see
Sec.~\ref{section:secondcriticality} below. Decreasing the
amplitude further, below about sub7 we obtain initial data with mass
below the threshold $M_{\rm tot}=0$ for black hole formation and these
data cannot form a black hole. (While we therefore cannot
  observe mass scaling for these data, we still observe Ricci scaling.)

\begin{figure}[!htb]
\centering
\includegraphics[scale=0.7, angle=0]{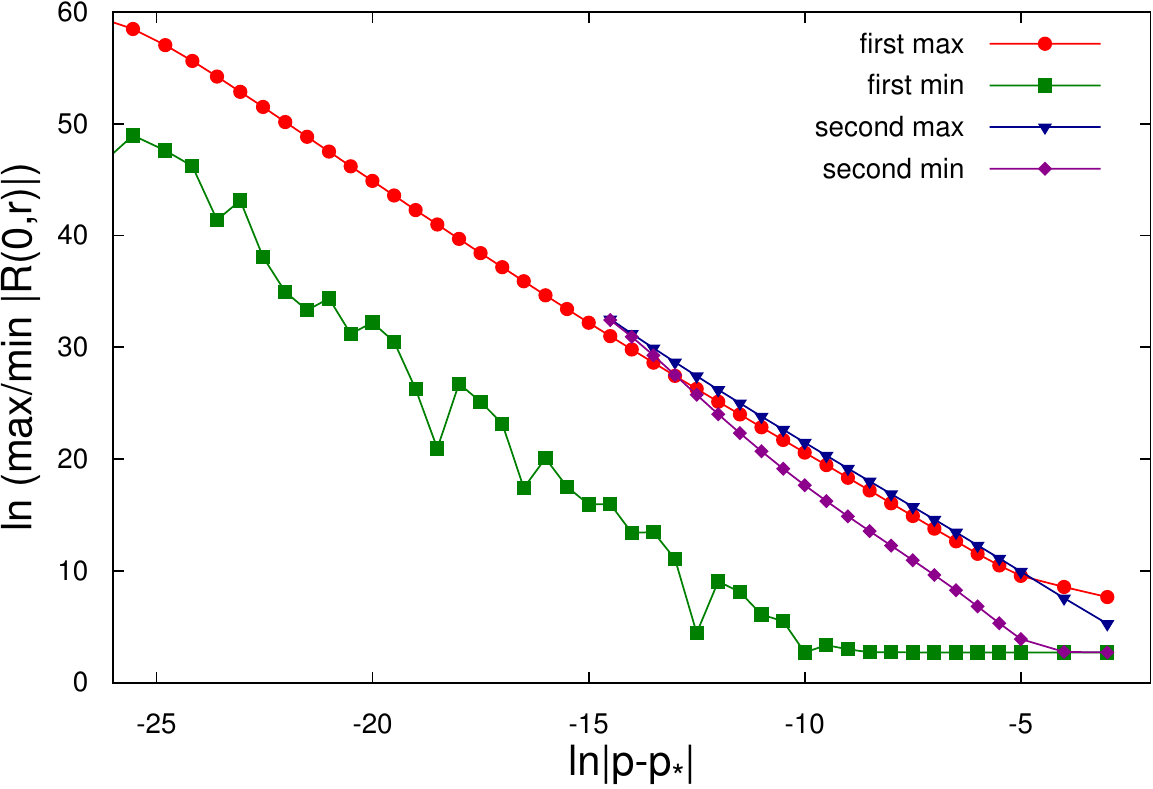}
\caption{Scaling of the {\em values} of the maxima and minima of
  $|R(0,t)|$. 
  At approximately sub15, the
  second maximum and minimum merge and disappear, see also the sub15
  curve in Fig.~\ref{figure:Rvst}. The slope for the minima is
  slightly but significantly different from that for the maxima. The
  horizontal segments of the value of the first minimum and second
  minimum are dominated by the $\Lambda$ term in $R=\kappa
  G(\nabla\phi)^2+6\Lambda$. The slope for the first maximum obtained
  by a least-square fit on the fitting interval $[-5,-26]$ is
  {$\gamma=1.2345 \pm 0.0076$}. The standard deviation $\sigma$ cited
  here and for similar slopes in the following is based on the
  hypothesis that the deviations from a straight line are
  independently normally distributed, and so does not take into
  account systematic error, which is clearly larger. However, we note
  that the deviation from the theoretical value of $8/7$ is $12\sigma$
  or $8\%$.}
\label{figure:R_scaling}
\end{figure}

The scaling of the maxima and minima of the {\em value} of the Ricci
scalar at the centre is shown in Fig.~\ref{figure:R_scaling}.  For
subcritical data, the first local maximum of $|R(0,t)|$ scales as in
(\ref{Ricciscaling}) with $\gamma \simeq 1.23(4)$, the same value, to
within our numerical precision, as found by
\cite{PretoriusChoptuik}. We determine $p_*$ to high precision by
fitting to the Ricci scaling law (\ref{Ricciscaling}). The critical
value $p_*$ defined in this way is consistent with the
definition we have given before, but can be determined more accurately
in practice.

The first minimum also scales, with $\gamma \simeq 1.4(7)$. Further
away from criticality than approximately sub10, the first minimum
reaches a floor set by the cosmological constant, $R\simeq 6\Lambda$.
Extrapolating beyond the limit of our fine-tuning, the scaling of the
first maximum and first minimum would suggest that they merge at
sub38. However, this extrapolation is probably incorrect, as by
definition we would expect them to merge precisely at $p=p_*$.

The value of the second maximum scales with $\gamma \simeq 1.17(8)$,
similar to the first maximum, and the second minimum with $\gamma
\simeq 1.48(9)$. At approximately sub15 its value agrees with the
second maximum, and at this point the second maximum and minimum merge
and disappear.  The second miminum reaches the same floor as the first
maximum, but only at sub2 and then at large $r$, which is out of the
range of critical phenomena at first implosion.

\begin{figure}[!htb]
\includegraphics[scale=0.7, angle=0]{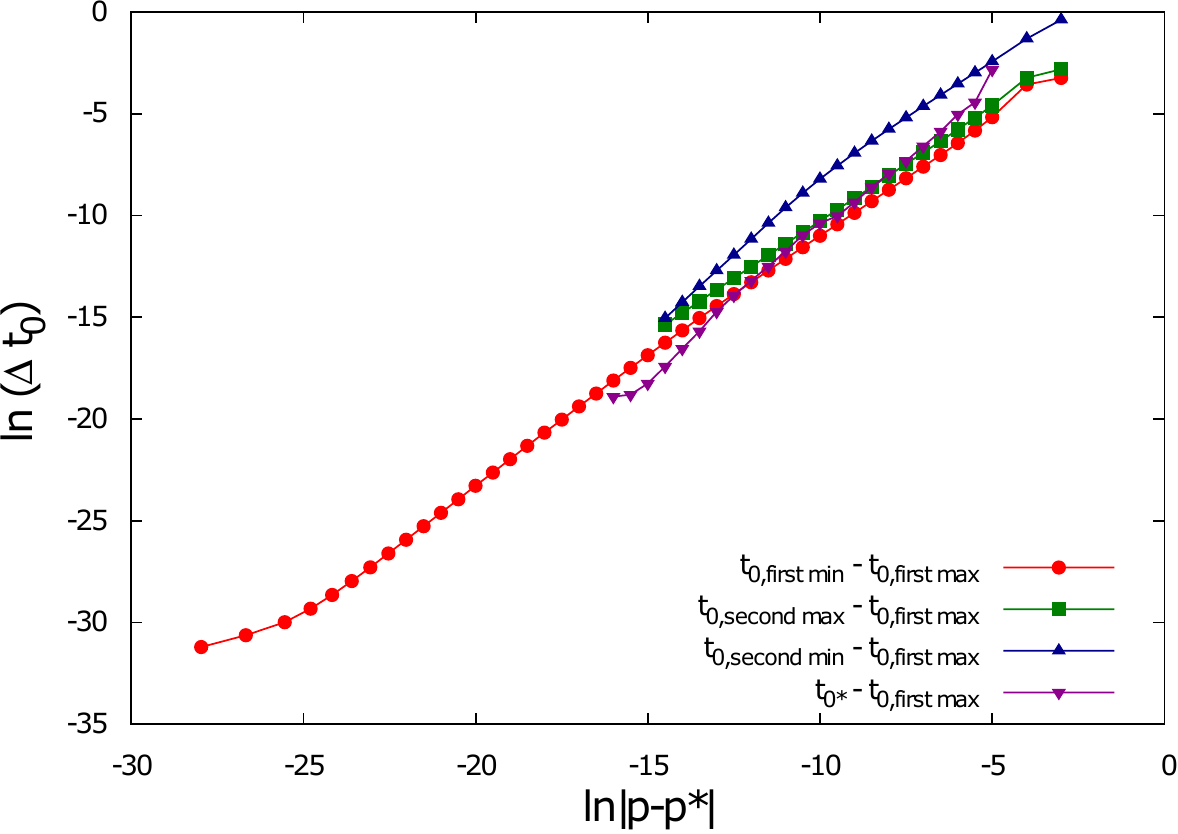} 
\caption{Scaling of the {\em location} of the minima and maxima of
  Ricci in proper time at the centre $t_0$.  A least-squares fit
   of the proper time between the first maximum and first
    minimum to a straight line on the fitting interval $[-5,-26]$
  gives {$\gamma=1.2410 \pm 0.0067$}. The fitted value differs from
  our theoretical value $\gamma=8/7\simeq 1.1429$ by $15\sigma$, or
  $9\%$.}
\label{figure:time_scaling}
\end{figure}

Fig.~\ref{figure:time_scaling} shows the scaling of the {\em
  locations}, in proper time at the centre $t_0$, of the first
minimum, second maximum, and second minimum, all with respect to the
first maximum, as well as the location of the first maximum with
respect to the accumulation point $t_0=t_{0*}$. The scaling exponents are 1.2(2),
1.12(7), 1.2(8) and 1.4(3) respectively, see
Fig.~\ref{figure:time_scaling}. (The reason that we do not use the
accumulation point as our primary reference point is that its location
$t_{0*}$ is obtained by curve-fitting, and is therefore less accurate
than the location of the extrema with respect to each other.)

\begin{figure}[!htb]
\includegraphics[scale=0.7, angle=0]{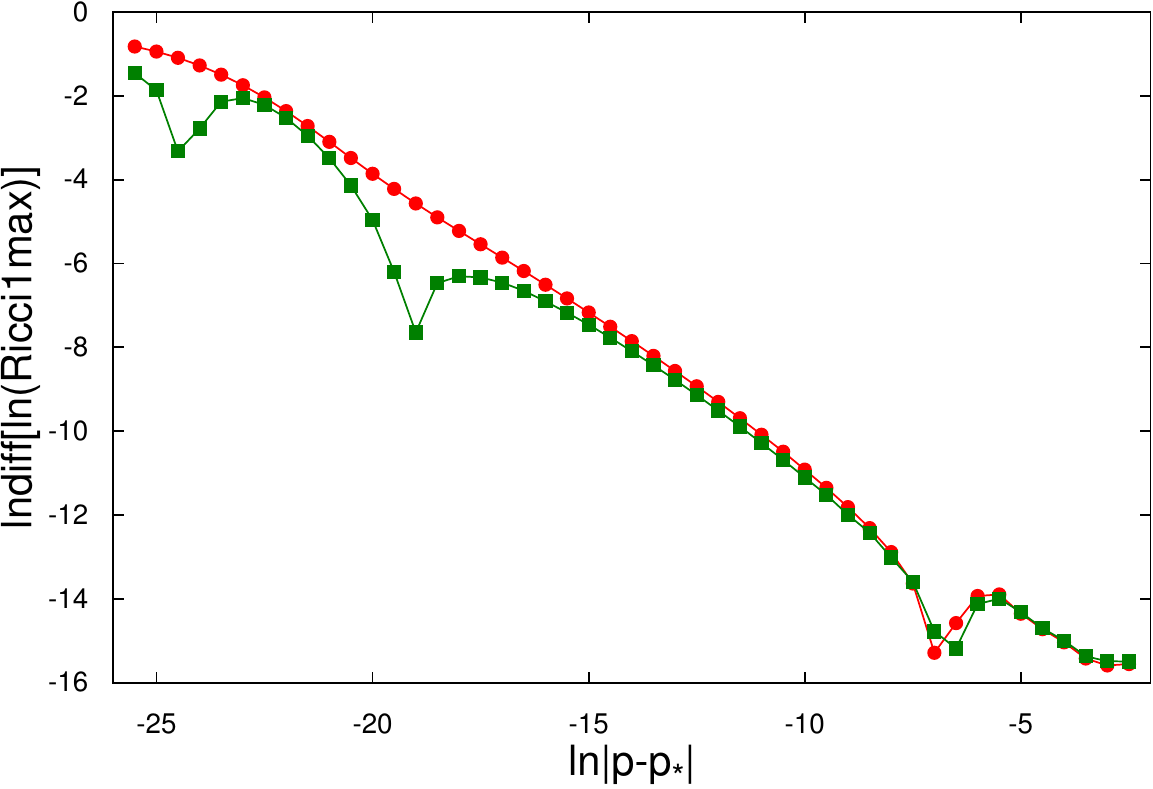}
\caption{Plots of $\ln(\ln R_{1k}-\ln R_{4k})-C$ (red dots) and
  $\ln(\ln R_{2k}-\ln R_{4k})$ (green squares) against
  $\ln(p-p_*)$. Here $R_{1k}$, $R_{2k}$ and $R_{4k}$ stand for the
  value of the first maximum of the Ricci scalar at the centre in
  evolutions with 1024, 2048 and 4096 grid points, and $p_*$ is
  shorthand for the relevant value for each of these resolutions,
  obtained by bisection. The fact that the lower resolution curve lies
  on or above the higher resolution curve when shifted down by
  $C=\ln[(4^4-1)/(2^4-1)]$ demonstrates 4-th order convergence with
  resolution. The value of the curves gives an estimate of $\ln$ of
  the numerical error in $\ln R$. The linear dependence on
  $\ln(p-p_*)$ is not related to the underlying scaling law, but shows
  that the relative error in Ricci increases with fine-tuning. }
\label{figure:riccifirstmax_gauss_convergence}
\end{figure}

Checking pointwise convergence in $(r,t)$ of our time evolutions is
difficult in the critical regime because of the sensitive dependence
on initial data. At best we can compare scalar quantities such as
$M(x,T)$ at fine-tuning ``sub$n$'' for the same $n$ at different
numerical resolutions. (Note that $p_*$ itself is
resolution-dependent). This works for $M$ and $\bar r$, but not for
$f$ and $R$. However, physical results such as Ricci and mass scaling
should converge with resolution. In
Fig.~\ref{figure:riccifirstmax_gauss_convergence} we demonstrate that
the first maximum of the Ricci scalar as a function of $\ln(p-p_*)$
converges with resolution to better than fourth order from sub3 to
sub22.

\subsubsection{Apparent horizon mass scaling}
\label{section:massscaling}


The scaling argument \cite{GundlachLRR} only determines the size and
hence mass of the black hole when it first forms, in a regime where
the transition from the critical solution to black hole formation is
still universal up to an overall scale. However, in asymptotically
flat spacetimes and for massless scalar field matter, little
additional mass falls inlater (when the scaling argument no longer
holds), so one effectively has a scaling law for the asymptotic black
hole mass. (In a cosmological context, there may be significant
infall \cite{Hawke}.) In 2+1 dimensions, the cosmological constant can
never be neglected where collapse takes place, and so the local
scaling argument breaks down already.  Furthermore, for $\Lambda<0$ in
any dimension all the mass eventually falls into the black hole
because of reflecting boundary conditions. Therefore Pretorius and
Choptuik focused on the mass at the first appearance (with respect to
a given time slicing) of a marginally outer trapped surface (MOTS),
which they call the apparent horizon mass. To explain the phenomenology we
observe, we need to use a more explicit terminology, as follows.

We assume spherical symmetry. We shall use the term MOTS to denote any
point $(r,t)$ where $\rb_{,v}=0$. We shall call the union of all MOTS
the apparent horizon (AH), parameterised in coordinates as a curve $t=t_{\rm
  AH}(r)$. It bounds the region of outer-trapped spherically symmetric
$(d-1)$-surfaces (circles in 2+1). It is easy to see using the field
equations that the AH $\rb_{,v}=0$ is spacelike for $\phi_{,v}\ne 0$
(meaning that energy crosses the horizon) and outgoing null for
$\phi_{,v}=0$. What Pretorius and Choptuik denoted by apparent horizon
mass $M_{\rm AH}$ is the mass $M=\rb^2/\ell^2$ of the first appearance
of a MOTS for a given time slicing, that is the {\em absolute}
minimum of the AH curve $t=t_{\rm AH}(r)$ with respect to the time
coordinate $t$. For clarity, we shall call this the first MOTS (FMOTS).

\begin{figure}
\centering
\includegraphics[scale=0.5, angle=270]{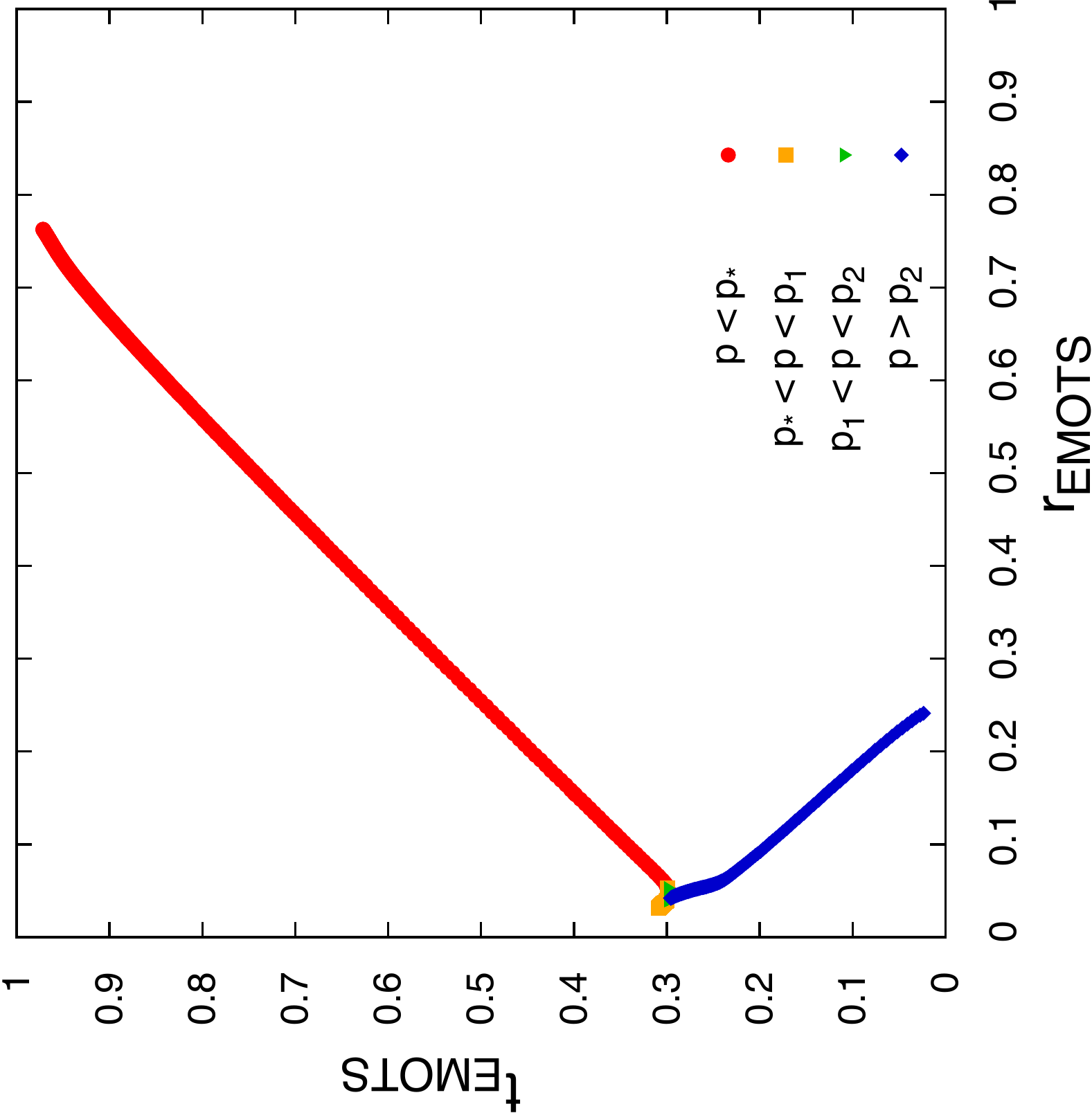}
\hfill
\includegraphics[scale=0.6, angle=270]{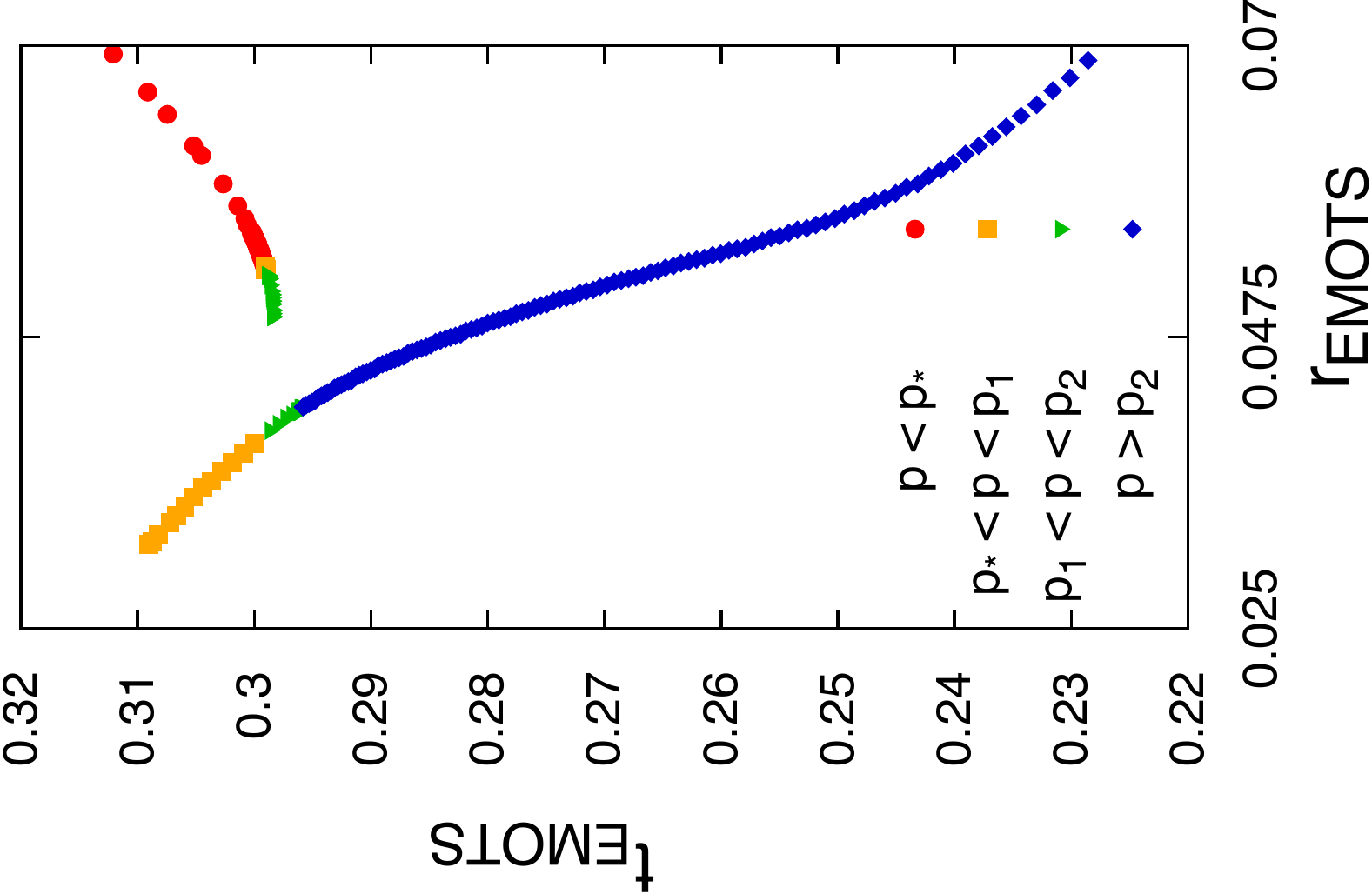}
\caption{Both plots show the track $[r_{\rm EMOTS}(p),t_{\rm
    EMOTS}(p)]$ of the inner and outer EMOTS. The lower plot is a
  closeup. Blue diamonds ($p>p_2$) is the regime where the AH curve
  $t_{\rm AH}(r)$ has only one local minimum (EMOTS), green
  upside-down triangles
  ($p_1<p<p_2$) the regime where there are two EMOTS but the inner
  one, which scales, appears first (i.e. at smaller $t$) and so may be
  considered as the FMOTS, and orange squares ($p_*<p<p_1$) the regime
  where the outer EMOTS appears first. The transition between the last
  two regimes causes the jump in the FMOTS mass at $p=p_1$ in
  Fig.~\ref{figure:mass_jump}. Note that $r$ and $t$ are drawn to the
  same scale, and so radial null rays are at 45 degrees. The top right
  end of the subcritical (red circles) track is approximately at sub10
  (corresponding also to the left edge of the upper plot in
  Fig.~\ref{figure:mass_jump}), but the curve does not end there.}
\label{figure:tr_curve}
\end{figure}

\begin{figure}[!htb]
\centering
\includegraphics[scale=0.37, angle=270]{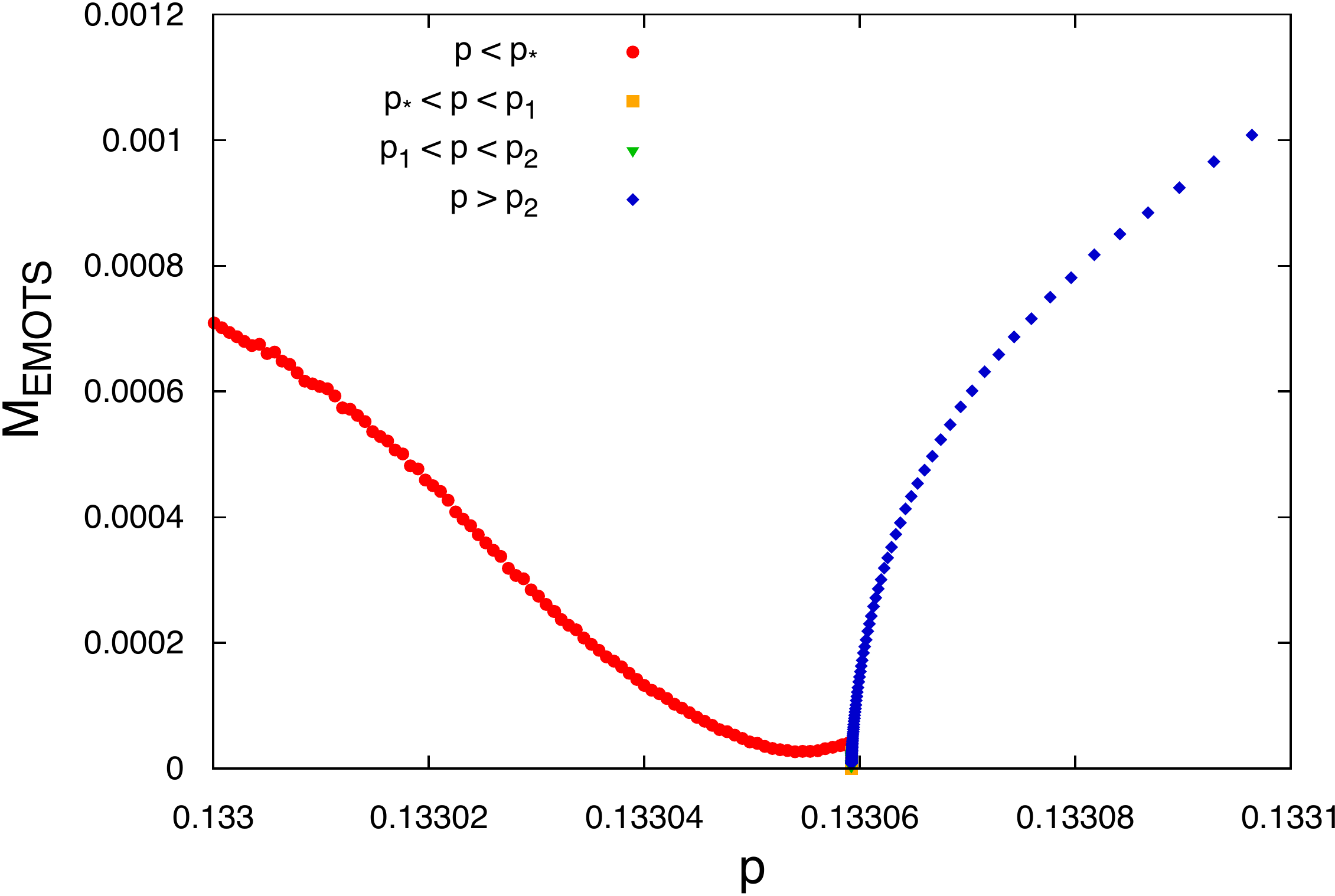}
\hfill
\includegraphics[scale=0.4, angle=270]{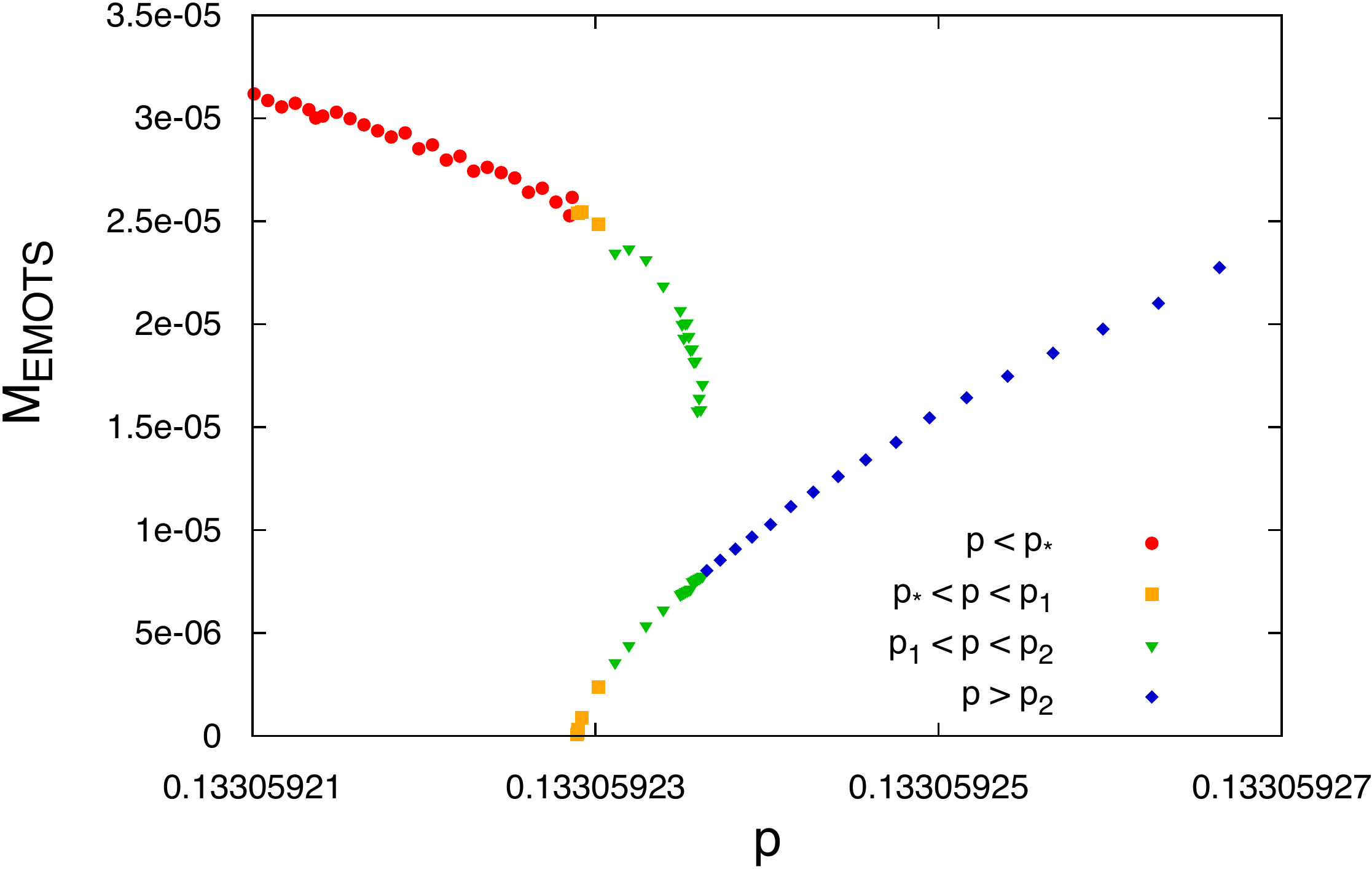}
\caption{The top plot shows the behavior of $M_{\rm FMOTS}(p)$. At
  $p=p_1$, $M_{\rm FMOTS}(p)$ appears to jump to a larger value. The
  bottom plot is a closeup focusing on the amplitudes close to
  $p_1$. The colors and symbols for the different ranges of the parameter $p$
  are the same as in Fig.~\ref{figure:tr_curve}. In the supercritical case
  $p>p_*$, the only EMOTS left is the outer one, which does not
  scale.  We expect the two curves to join at $p=p_2$, but the we
  cannot resolve this numerically.}
\label{figure:mass_jump}
\end{figure}

For the ingoing Gaussian data, the plot of $M_{\rm FMOTS}(p)$ shows
power-law scaling down to a very small value of $M$ at $p=p_1>p_*$,
but then $M$ jumps to a larger value and varies only slowly with
$p$. This is shown in the upper plot of
Fig.~\ref{figure:mass_jump}. This apparent jump is explained simply by
the AH curve having two {\em local} minima for the range $p_*<p<p_2$,
which includes $p_1$, see the lower plot. We shall refer to such a
local minimum of $t_{\rm AH}(r)$ as an earliest MOTS (EMOTS). It is
helpful to consider the tracks of both EMOTS in the $(r,t)$ plane
(Fig.~\ref{figure:tr_curve}), together with a plot of their masses
against $p$ (Fig.~\ref{figure:mass_jump}).

At some very large value $p_0 \simeq 200$ of $p$ (compare this to $p_*
\simeq 0.133059$) there is only one EMOTS, and it is located on the
initial slice $t=0$ at some large $r$ and $M$. As $p$ is decreased
from $p_0$, the EMOTS moves to smaller $r$ (on track that is
approximately null) and smaller $\rb$ and hence $M$. At $p=p_2$
(approximately sub19) the single EMOTS splits into two. To the limit
of our fine-tuning of the initial data, the inner EMOTS approaches
zero $r$ and $M$ as $p \to p_*$. 

For $p < p_*$, there is no inner EMOTS, and the outer EMOTS, whose
mass does not scale, moves to larger $r$ and $t$ with decreasing $p$
on an approximately null track, until at sub10 it approaches the outer
boundary. Presumably it will
then move back in, but we have not followed this further. For
$p_*<p<p_1(<p_2)$ the outer EMOTS appears first, so if one looks only
for the first appearance of a MOTS, for any $r$, its mass appears to
jump at $p=p_1$ from the mass of the inner EMOTS to that of the outer
EMOTS.

\begin{figure}[!htb]
\centering
\includegraphics[scale=0.65, angle=0]{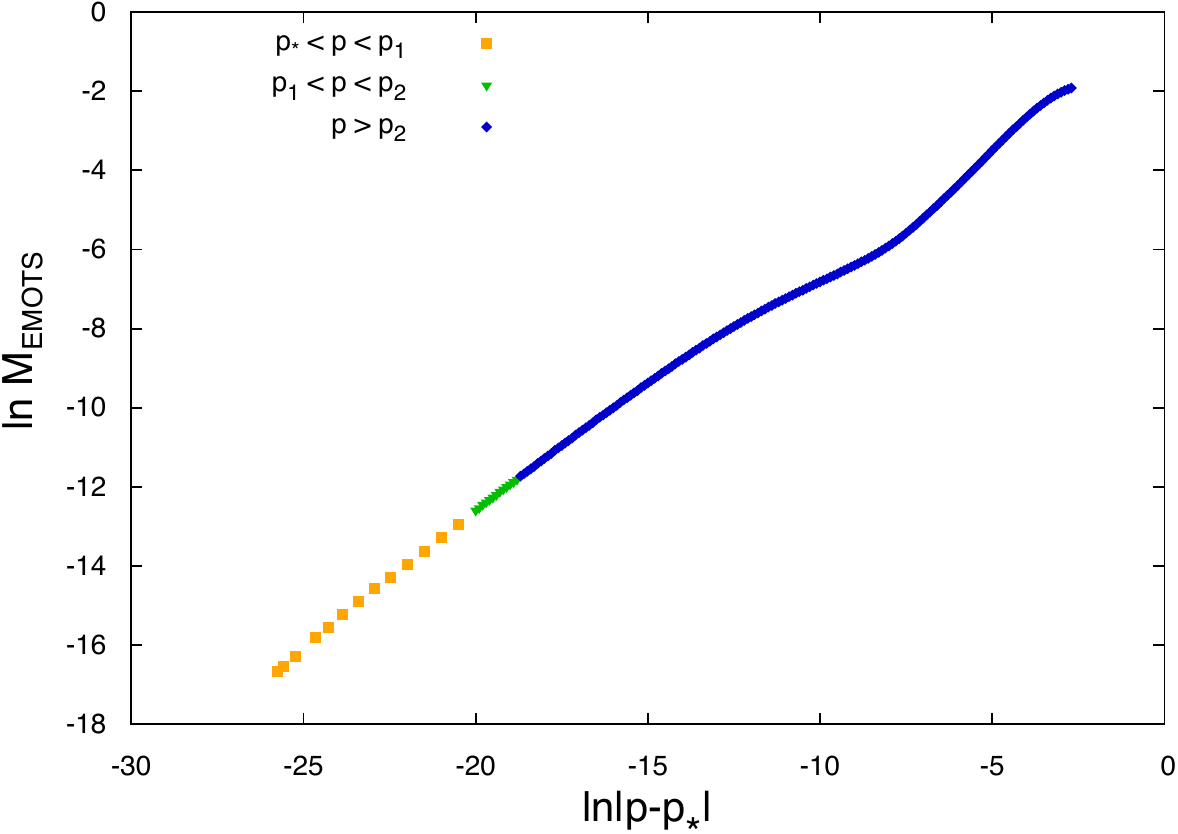}
\caption{Scaling of the {\em inner} EMOTS mass (\ref{MAHscaling}) with
  an exponent {$\delta\simeq 0.6839 \pm 0.0023$}, for
  supercritical data $p>p_*$, a deviation from our theoretical value
  of $16/23$ is $5\sigma$ or $2\%$. The colors are the same as in
  Fig.~\ref{figure:tr_curve}. The fitting interval is $[-26,-17]$.  }
\label{figure:Mscaling_fig}
\end{figure}

As far as our fine-tuning reaches, the mass of the {\em inner} EMOTS
scales as (\ref{MAHscaling}) with $\delta \simeq 0.68(4)$, see
Fig.~\ref{figure:Mscaling_fig}.  This value is roughly similar to the
value $\delta \simeq 0.81$ of \cite{HusainOlivier} (but 
different from the $\delta=2\gamma\simeq 2.50$ of
\cite{PretoriusChoptuik}).

\begin{figure}[!htb]
\centering
\includegraphics[scale=0.8]{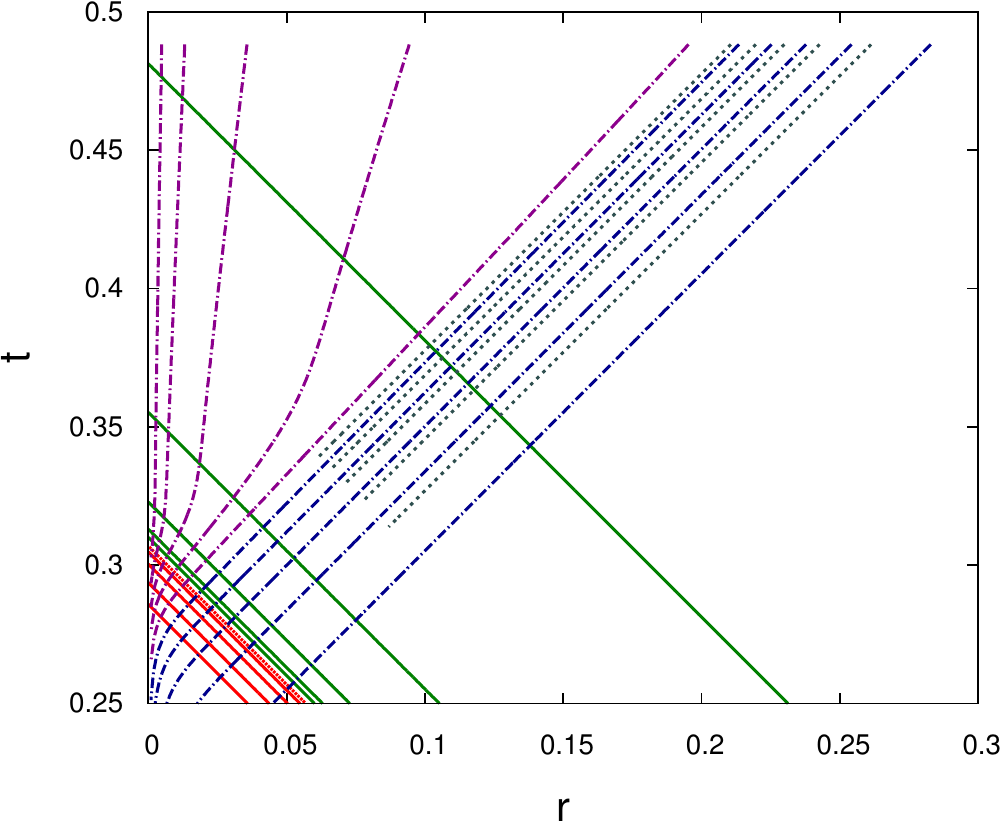}
\caption{Spacetime diagram of the sub10 evolution: contours of
  $-\ln(-\tilde v)$ for $\tilde v<0$ (red, solid lines) and of
  $-\ln\tilde v$ for $\tilde v>0$ (green, solid lines), both from $9$
to $13$ in steps of 1, contours of $-\ln\rb$ from $4$ to $8$ (blue, dash-dotted lines) and
  from $9$ to $13$ (magenta, dash-dotted lines) in steps of $1$,
  and contours of $T=-\ln(-\tilde u)$ for $\tilde u<$0 from $4$ to $8$
  in steps of $1$ (gray dotted lines, shown only for $v>0.4$ for
  clarity). This figure illustrates several things: the automatic zoom
  we get in our numerical coordinates, $r$ and $t$ corresponding to
  much smaller physical scales everywhere to the future of the
  accumulation point, and the transition of $\rb$ from spacelike to
  ``almost null'' and back to spacelike.}
\label{figure:vtilderbar_rt_contours_sub10}
\end{figure}

As $p_*$ is the same for both Ricci and mass scaling, to within our
accuracy of fine-tuning (sub26 and super26), the exponents $\gamma$
and $\delta$ must also be related. As we do not have the exact
critical solution, we cannot give a complete derivation of this
relation, but a tentative derivation of $\delta$ and $\gamma$ based on
an amended Garfinkle solution and approximate single growing mode is given
below in Secs.~\ref{section:gammaderivation} and
\ref{section:deltaderivation}.


\subsubsection{Second criticality}
\label{section:secondcriticality}

\begin{figure}[!htb]
\centering
\includegraphics[scale=0.35, angle=270]{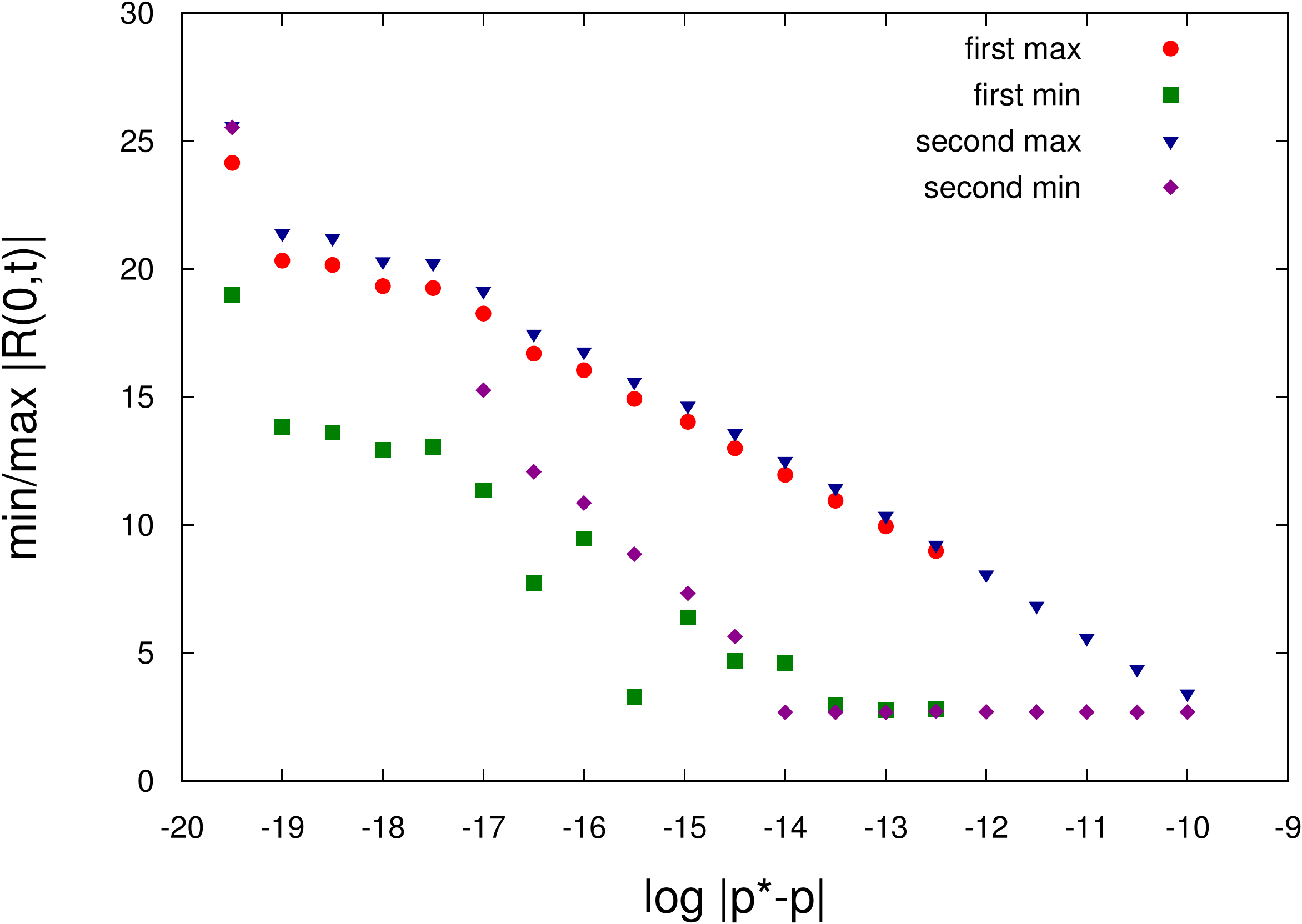}
\caption{Scaling of minima and maxima of $|R(0,t)|$ for subcritical
  evolutions with respect to the {\em second} critical amplitude
  $p_{*1}$. From a linear fit to the log-log plot we obtain the
  critical exponent $\gamma\simeq 1.065 \pm 0.44$
   which is slightly different from $\gamma\simeq 1.23(4)$ obtained for
  $p_{*0}$. The fitting interval is
  $[-19,-12]$.}
\label{figure:R_scaling2}
\end{figure}

Decreasing $p$ further, we again find critical phenomena after reflection at
the outer boundary. This means that we fine-tune the amplitude such
that the initially ingoing Gaussian reflects off the center, moves
towards the outer boundary, reflects off it and collapses while
approaching the center for the second time. The bisection is again
based on the behavior of the Ricci scalar at the centre. We find that
there is a second critical amplitude $p_{*1}\simeq p_{*0}-\exp(-8)$
such that the maxima and minima of the Ricci scalar for subcritical
evolutions scale according to \eqref{Ricciscaling}. This is
demonstrated in Fig.~\ref{figure:R_scaling2}. (We use $p_{*n}$ to
denote the critical amplitude after $n$ reflections, with our original
$p_*=:p_{*0}$.)

Note that $p_{*1}$ is itself only sub8 with respect to $p_{*0}$,
so that scaling maxima and minima are covered up by the
initially ingoing part of the initial data. We were also
unable to fine-tune as accurately as for the first criticality. At
about sub20 relative to $p_{*1}$, the Ricci scalar still has maxima
and minima, but their values fail to scale. We believe this is due to
loss of numerical accuracy.

The accumulation point for immediate critical collapse, for the
Gaussian initial data, was located at $t_{*}\simeq 0.34$ in coordinate
time and $t_{0*}\simeq 0.2374$ in proper time at the centre. For
critical phenomena after one reflection, for the same family of
initial data, the corresponding values are $t_{*1}\simeq 2.24$ and
$t_{0*1}\simeq 0.293$. Note that the two accumulation points are
separated by $\Delta t\simeq 2$, consistent with the intuitive picture
of reflection at the outer boundary, but that they are separated in
proper time only by $\Delta t_0\simeq 0.05$. This is due to the fact
that $A$ and $B$ jump down across the future lightcone of the first
accumulation point to $A\sim B\sim -6$ and then remain small. Hence
after first near-criticality, $r$ and $t$ correspond to much smaller
physical scales than before, but by definition $r=1$ is still the
outer boundary and the light-crossing time is therefore still $\Delta
t=2$. See also Fig.~\ref{figure:vtilderbar_rt_contours_sub10} for an
illustration of this memory effect in the sub10 evolution.

As a consequence of this separation of scales, the wave going back out
in (first) near-subcritical evolutions comes back in what is a very
short time at the centre and interacts with the aftermath of first
criticality. First and second criticality therefore overlap in time,
and this may explain why they are also close in $p$, in the sense that
the scaling regimes overlap.

Furthermore, if we we compare the constant factors in front of the two
Ricci scaling laws $|R|_{\rm max} \simeq C_0(p-p_{*0})^{-2\gamma}$ and
$|R|_{\rm max} \simeq C_1(p-p_{*1})^{-2\gamma}$, we find that
$C_1\simeq 10^{-6}C_0$. This may also be a consequence of the jump
down in $A$ and $B$. 

\begin{figure}[!htb]
\centering
\includegraphics[scale=0.35, angle=270]{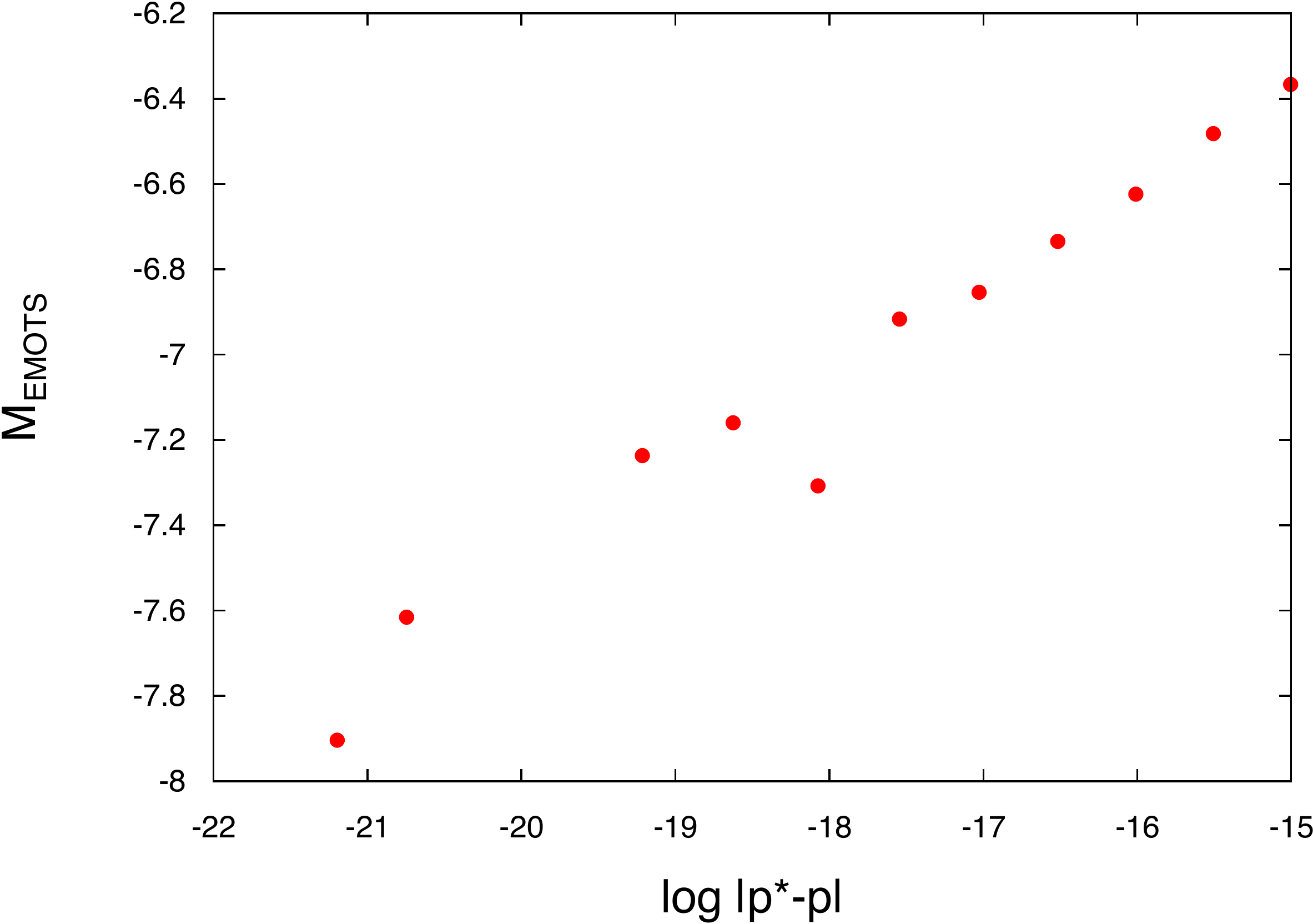}
\caption{The mass of the EMOTS for supercritical data after one
  reflection. The value of the critical exponent $0.23$ found from a
  linear fit to the log-log plot differs from the value found for the
  data before any reflection. Note that $p_{*1}$ is obtained from the
  second Ricci scaling, not this plot.}
\label{figure:mass_scaling2}
\end{figure}

For second-supercritical data we also looked for evidence of mass
scaling. The supercritical data with respect to $p_{*1}$ can be also
supercritical with respect to $p_{*0}$, and therefore to see second
mass scaling one has to look at the proper range of amplitudes. We
find some evidence that for second supercritical data the mass of an
apparent horizon roughly behaves according to \eqref{massscaling}, but
with a critical exponent {$\delta\simeq 0.23$}, significally different
from the $\delta\simeq 0.68(4)$ found in first criticality. The
evidence is presented in Fig.~\ref{figure:mass_scaling2}. We have no
theoretical explanation of the discrepancy in the mass scaling
exponent, but as the scaling appears to be very noisy anyway, the
discrepancy may be just numerical error due to loss of resolution.

\subsubsection{Self-similarity inside the lightcone}
\label{section:CSStest}


We now examine the claim that a CSS critical solution is observed
\cite{PretoriusChoptuik}, and that inside the lightcone it agrees with
the $n=4$ Garfinkle solution \cite{Garfinkle}.

Recall that we denote by $t_0$ the proper time at the centre, starting
with $t_0=0$ at $t=0$. In a half-diamond bounded on the left by $r=0$,
we can rescale $u$ and $v$ to new double null coordinates $(\tilde
u,\tilde v)$, so that both correspond to $t_0-t_{0*}$ on the central
worldline $\rb=0$, which by ansatz is at $u=v$ and so is also at
$\tilde u=\tilde v$, and where $t_{0*}$ denotes the accumulation point
in central proper time. A plot of the contour lines of $\ln \tilde
u$ and $\ln \tilde v$ in a near-critical evolution,
see Fig.~\ref{figure:vtilderbar_rt_contours_sub10} for sub10, shows
that our numerical algorithm provides some automatic zooming in, which
means we can resolve self-similarity over many $e$-foldings in
scale without mesh refinement -- to optimally resolve self-similarity,
these lines should be equally spaced.

The first task is to find the accumulation point. With the scalar
field at the centre in the Garfinkle solution given by $\phi(0,T)=
c\ln(t_{0*}-t_0)+{\rm const}$, we make a linear fit
\begin{equation}
\left({d\phi\over dt_0}\right)^{-1}={t_0-t_{0*}\over c}
\end{equation}
for $c$ and $t_{0*}$. 
We can then compute 
\begin{equation}
\tilde u(u)=t_0(u)-t_{0*}, \qquad \tilde v(v)=t_0(v)-t_{0*}
\end{equation}
from $t_0(t)$ and $t_{0*}$. To see CSS, this needs to be done
separately for each $p$, but $t_{0*}$ and $c$ depend only weakly on
$p$ and have a limit as $p\to p_*$. We have fitted $c(p)$ by the
quadratic function $c(p)=c_*+c_1(p_*-p)+c_2 (p_*-p)^2$. For subcritical
evolutions of our Gaussian initial data, a least squares fit gives
$c_*=0.26381 \pm 0.00018$, $c_1=127.531 $, $c_2=-159484$ for the
fitting interval $[\text{sub}8, p_*]$. This range of $c$ is equivalent
to $n=3.986\pm0.038$. Hence we can strongly rule out any $n$ other
than 4.

In the following, we denote by $\tcA$ the value of $\cA$ in the
preferred double-null coordinates $(\tilde u,\tilde v)$. It is given
in terms of the numerically evolved metric coefficient $A$ as
\begin{equation}
\label{tcAdef}
\tcA=A(r,t)-{1\over 2}\left[A(0,t-r)+A(0,t+r)\right]-\ln[\cos(r/\ell)].
\end{equation}

Following \cite{GarfinkleGundlach}, we then define similarity
coordinates $(x,T)$ by
\begin{equation}
\label{xTdef}
x:=\left(\tilde v\over \tilde u\right)^{1\over 2n}, \qquad T:=-\ln\tul,
\end{equation}
for $n$ a positive integer. Hence the regular centre is given by $x=1$
and the lightcone by $x=0$. We also define
\begin{eqnarray}
\label{Rdef}
R(x,T)&:=&\ell^{-1}e^{T}\rb, \\
f(x,T)&:=&c^{-1}\phi - T - d,
\end{eqnarray}
where $d$ is a family-dependent, dynamically irrelevant constant. The
solution is then CSS if and only if $\tcA$, $M$, $R$, $f$ and
$\lambda$ are functions of $x$ only. These functions for the
countable family of Garfinkle solution are reviewed in
Sec.~\ref{section:garfinkle}. 

Finally, we define
\begin{equation}
\lambda:=-{s\over \tilde u},
\end{equation}
where $s$ is the affine parameter along outgoing null geodesics,
measured away from the centre, and normalised so that the inner
product of $\p/\p s$ with the 4-velocity of the central observer is
$-1$. With the centre at $v=u$, this gives
\begin{equation}
s_{,v}(u,v)={1\over 2}e^{2\cA(u,v)-\cA(u,u)},
\end{equation}
which we integrate along each line of constant $u$. In particular, 
\begin{equation}
\label{dsdvtilde}
s_{,\tilde v}(\tilde u,\tilde v)={1\over 2}e^{2\tcA(\tilde u,\tilde v)}.
\end{equation}
The rescaled 
\begin{equation}
\bar\lambda(x):=\lambda(x)/\lambda(0)
\end{equation}
is a function of $x$ only in CSS, and having tested this, we
will later use it as the similarity coordinate in place of $x$.

\begin{figure}[!htb]
\centering
\includegraphics[scale=0.65]{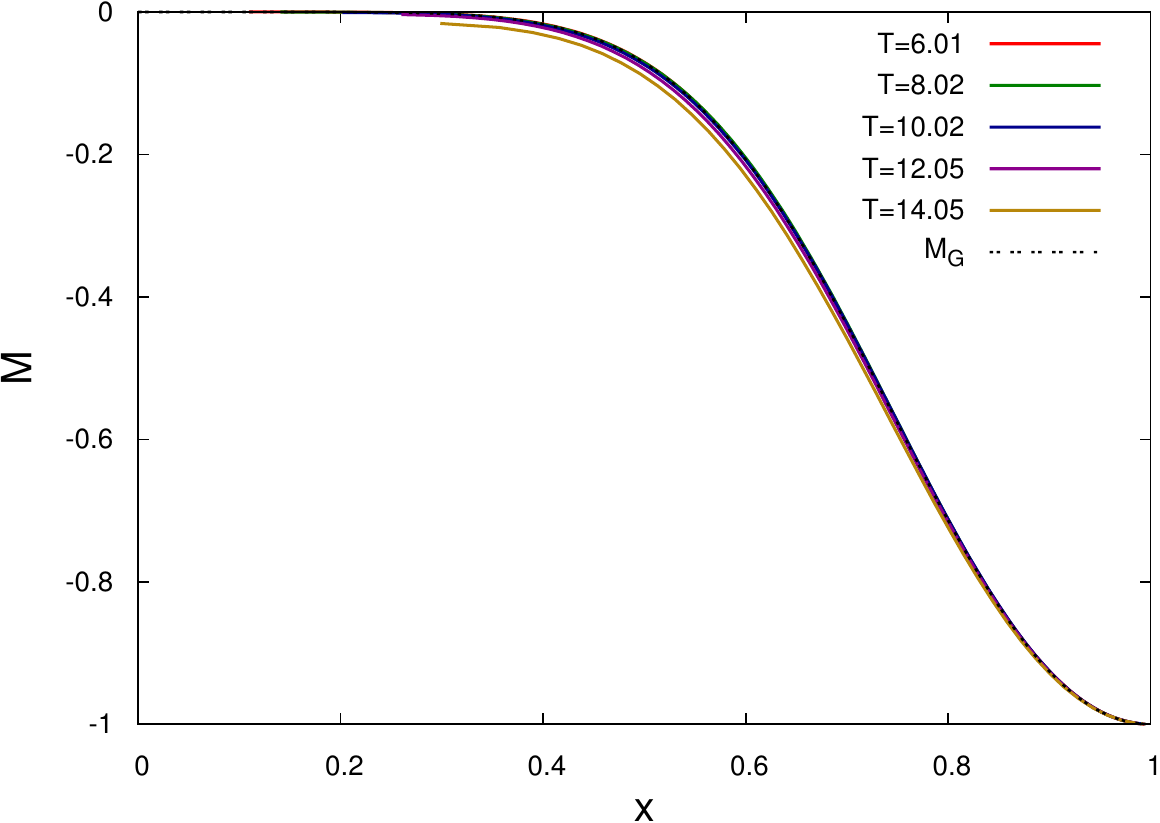}
\caption{Comparison of the mass function $M[x(r,t),T(r,t)]$ in
  the sub25 evolution at five $T=\textrm{const}$ moments
  against the $n=4$ Garfinkle solution $M_G(x)$. The numerical data
  are plotted for $T=6.01$ (red), $T=8.02$ (blue), $T=10.02$
  (magenta), $T=12.05$ (gray) and $T=14.05$ (orange). The Garfinkle
  solution is denoted by a dotted black line. The $\ell^2$ norm of the
  difference between numerical results and the Garfinkle solution for
  $n=4$ calulated for $T=10.02$ is $0.013$.  This norm is around $60$
  times larger for $n=3,5$ and over $200$ times larger for $n=2,6$.  }
\label{figure:masscomparison}
\end{figure}

\begin{figure}[!htb]
\centering
\includegraphics[scale=0.65]{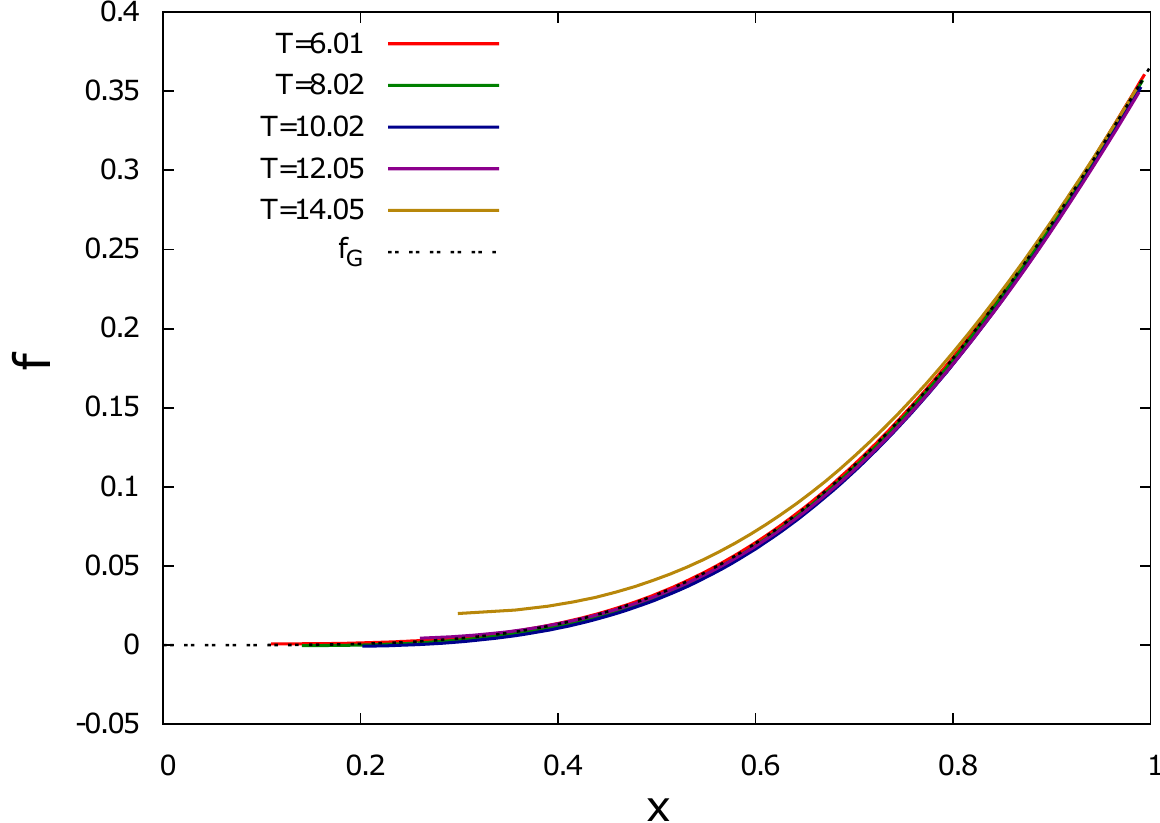}
\caption{Comparison of the $f[x(r,t),T(r,t)]$ in the sub25 evolution
  given at five $T=\textrm{const}$ moments with $f_G(x)$. Even more
  clearly than by the comparison of $f$ against $f_G$, $n=3$ and $n=5$
  are ruled out by our estimate of $n$ from the $cT$ dependence of
  $\phi(r,t)$.}
\label{figure:phicomparison}
\end{figure}

\begin{figure}[!htb]
\centering
\includegraphics[scale=0.65]{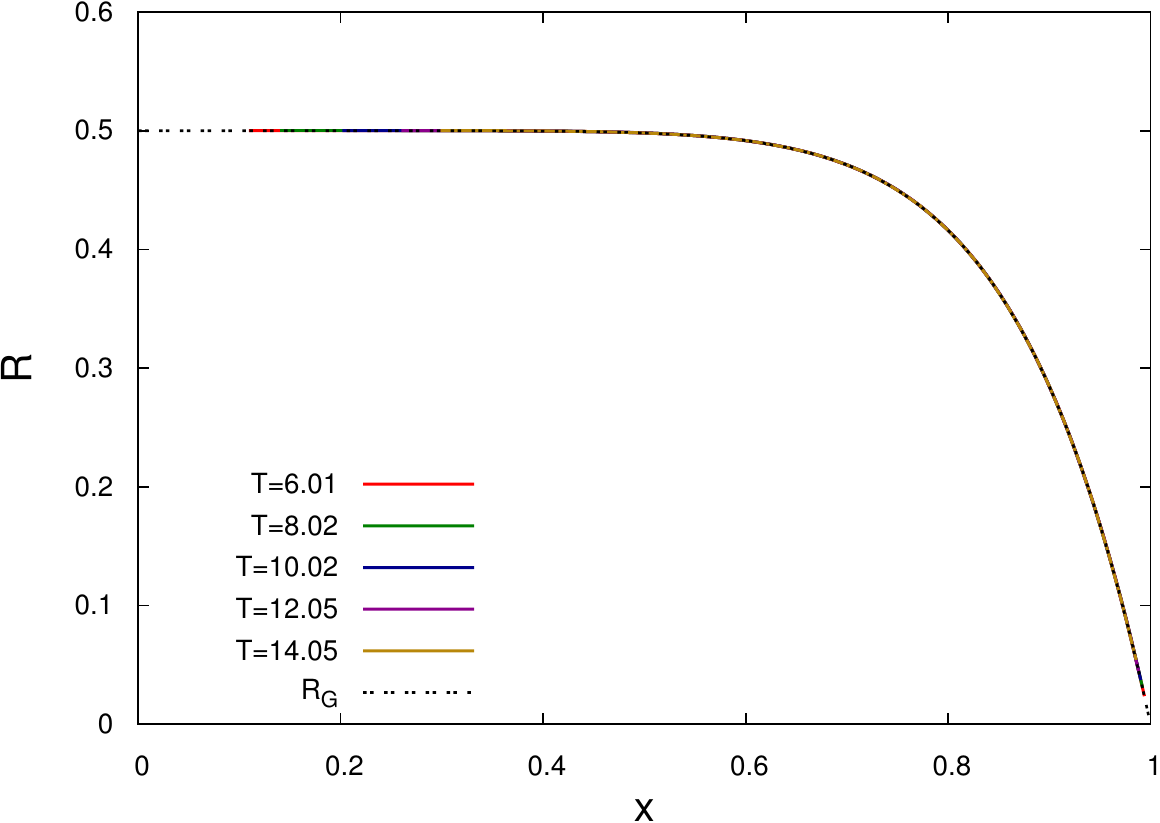}
\caption{Comparison of $R[x(r,t),T(r,t)]$ in the sub25 evolution
against $R_G(x)$. Again, the numerical data are given at five
$T=\textrm{const.}$ moments.}
\label{figure:Rcomparison}
\end{figure}

\begin{figure}[!htb]
\centering
\includegraphics[scale=0.65]{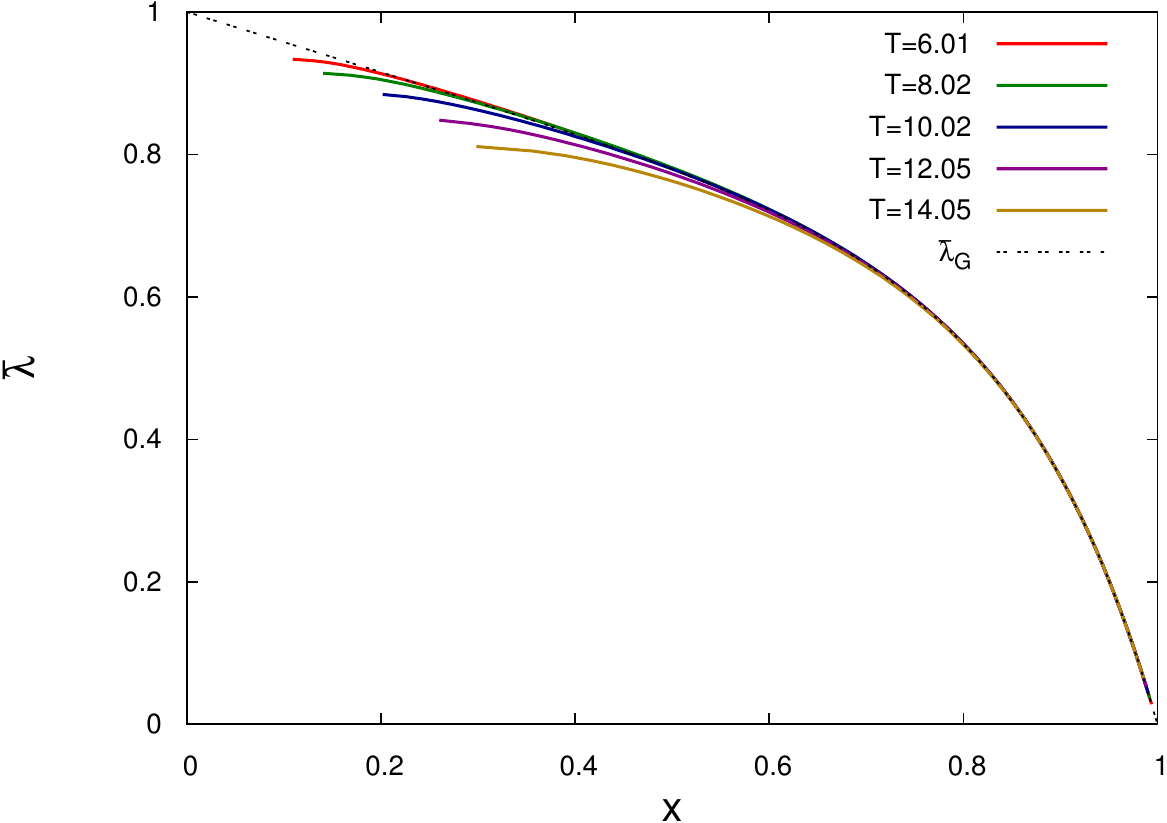}
\caption{Comparison of $\bar\lambda[x(r,t),T(r,t)]$ in the sub25
  evolution inside the lightcone for five moments of $T$ against
  $\bar\lambda_4(x)$.}
\label{figure:lambdabarxT_sub25}
\end{figure}

Fig.~\ref{figure:masscomparison} shows a comparison of the mass
function $M=M_G(x)$ of the Garfinkle solution with $n=4$ against
$M[x(r,t),T(r,t)]$ of the sub25 evolution. There is good agreement
everywhere between the regular centre and the lightcone ($0\le x\le
1$), over the range $6\le T\le 14$, which means that the solution is
CSS inside the lightcone over 8 $e$-foldings of scale,
all of which are much smaller than the scale $\ell$ set by the
cosmological constant.  Fig.~\ref{figure:phicomparison} shows a
similar comparison of $f_G(x)$, against
$f[x(r,t),T(r,t)]$, where the constant $d$ depends on the
family of initial data (but not on $p$) and has been determined by fitting.
 Figs.~\ref{figure:Rcomparison} and \ref{figure:lambdabarxT_sub25} show
the corresponding tests for $R$ and $\bar\lambda$.

Even though there is good numerical evidence that the critical
solution inside the lightcone is the $n=4$ Garfinkle solution (up to
small corrections in powers of $\Lambda$), we keep $n$ generic in the
following for clarity of presentation.


\subsubsection{Outside the lightcone}


Garfinkle \cite{Garfinkle} compared his exact solution with the
numerical evolutions of \cite{PretoriusChoptuik} only inside the
lightcone. Here we will go significantly beyond the lightcone. We will
see that the analytic continuation of the Garfinkle solution is
definitely ruled out, but that a different, $C^3$, continuation
proposed in Sec.~\ref{section:continuations} below, and which we call
the null continuation, appears to be at least a rough approximation to
the true critical solution.

The best choice of data for this comparison would appear to be an
evolution with the best available fine-tuning, as there we expect to
see the critical solution most clearly. However, in near-critical
evolutions, even subcritical ones, the evolution ends in a central
singularity very soon after the accumulation point of the CSS
regime. This is different from critical collapse in 3+1 and higher
dimensions, where subcritical evolutions go to essentially vacuum
after the CSS regime (in the case $\Lambda<0$, at least until the next
reflection at the outer boundary). Hence we also consider the sub10
evolution, which corresponds to the closest we can get to critical
initial data while still having a significant evolution in $t$ after
the accumulation point of the CSS region. In sub10, we have access to
large positive values of $v=t-r$, but because $A$ and $B$ are very
negative in this regime, this does not correspond to large values of
the proper retarded time $\tilde v$ or area radius $\rb$, and so we
are not far away in this sense from the accumulation point. See again
Fig.~\ref{figure:vtilderbar_rt_contours_sub10} in this context.

Recall that $\tilde v$ is normalised to proper time at the regular
centre, so it is not defined outside the past of blowup at the
centre. Moreover, even in subcritical evolutions, where blowup occurs
significantly after the accumulation point, spacetime at the centre
after the accumulation point is not expected to be self-similar. Hence
we cannot use the similarity coordinate $x$ based on $\tilde v$ and
$\tilde u$ outside the lightcone of the critical solution. We use
$\bar\lambda$ instead. It is given in terms of $x$ for both the
Garfinkle solution and its null continuation in
Sec.\ref{section:analyticlambdabar} below. 

 Fig.~\ref{figure:Tlambdabarrt_contours_sub25_ex} shows contour
  lines of $T$, $\bar\lambda$ and $x$ in the $(r,t)$ plane for the
  sub25 evolution with singularity excision. Near the center the
  contour lines of $x$ and $\lambda$ are approximately parallel, as
  one would expect in a CSS spacetime. Near the lightcone, they are
  not even approximately parallel, and the contour line
$\bar\lambda=0$ is not particularly close to the past lightcone of the
accumulation point. (The contour line $x=0$ is precisely the past
lightcone of the accumulation point by definition.) This disagreement
is already visible in Fig.~\ref{figure:lambdabarxT_sub25}, but appears
more clearly here because both $x$ and $\bar\lambda$ vary very slowly
with respect to $t$ and $r$ near the lightcone. We believe that the
origin of the discrepancy is that the true critical solution has a
symmetry that is approximately CSS only inside the lightcone, but
changes over to a different symmetry outside the lightcone in analytic
manner; see Sec.~\ref{section:outeransatz} below. Hence we expect some
deviation from CSS already as we approach the lightcone from the
inside.

Fig.~\ref{figure:Tlambdabarrt_contours_sub10} shows contour
  lines of $T$, $\bar\lambda$ and $x$ for the sub10 evolution. The discrepancy
  between $x$ and $\lambda$ is visible here, too. Sub25 gives us the
  larger range of $T$ (better fine-tuning), while sub10 gives us the
  larger range of $\bar\lambda$ (larger $t$ before the simulation
  stops). Overlaying the two sets of $T$ and $\bar\lambda$ contour
  lines in Fig.~\ref{figure:Tlambdabarrt_contours_sub10_sub25} shows
that the $T$ contours are essentially the same, while the $\bar\lambda$
contours differ significantly for $T\gtrsim 6$, as does the coordinate
location of the accumulation point. Yet when we plot $M$, $R$ and $f$
against $(\bar\lambda,T)$, the two evolutions agree perfectly with
 the Garfinkle solution, and therefore each other, inside the
lightcone.

 By comparing $M$, $f$ and $R$ with the null-continued Garfinkle
  solution,
Figs.~\ref{figure:MlambdabarT_sub25_ex}-\ref{figure:RlambdabarT_sub25_ex}
for sub25 and
Figs.~\ref{figure:MlambdabarT_sub10}-\ref{figure:RlambdabarT_sub10}
for sub10  also demonstrate that the analytic continuation is
clearly ruled out, while the null continuation appears more
plausible. In sub10, the strongest indication of this is that $M\simeq
0$ outside the lightcone, while the evidence from $R$ and $f$ is
somewhat less clear.

\begin{figure}[!htb]
\centering
\includegraphics[scale=0.8]{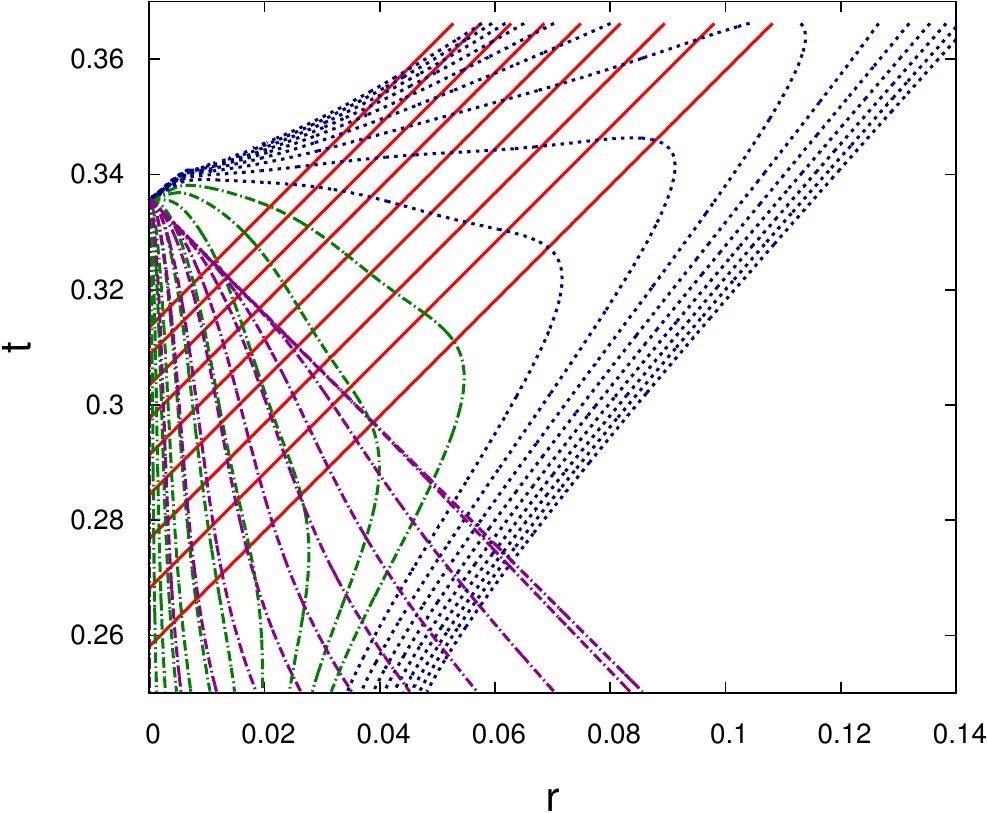}
\caption{Contour lines of $T$ from $6$ to $14$ (red, solid) in steps of
  $1$, contour lines of $\bar\lambda$ from $0$ to $1$ (green,
  dash-dotted) and from $1.1$ to $2$ (blue, dotted) in steps of $0.1$
  and contour lines of $x(r,t)$ (magenta dash-dotted for $1\ge x\ge 0$
  in steps of $0.1$) in the sub25 evolution. Excision allows us to
  access a wider range of $\bar\lambda$ for $6<T<14$. (Excision begins
  at the centre at $t=0.34$ and spreads to $r=0.05$ before the
  evolution stops at $t=0.365$.) The contour lines of $x$ and
  $\bar\lambda$ are approximately parallel near the centre, but not
  near the lightcone. This is clearer here than in
  Fig.~\ref{figure:lambdabarxT_sub25} based on the same data. Note that
  while $\bar\lambda=1$ is approximately null for $6<T<14$, it is
  significantly to the future of the past lightcone $x=0$ of the
  accumulation point.}
\label{figure:Tlambdabarrt_contours_sub25_ex}
\end{figure}

\begin{figure}[!htb]
\centering
\includegraphics[scale=0.65]{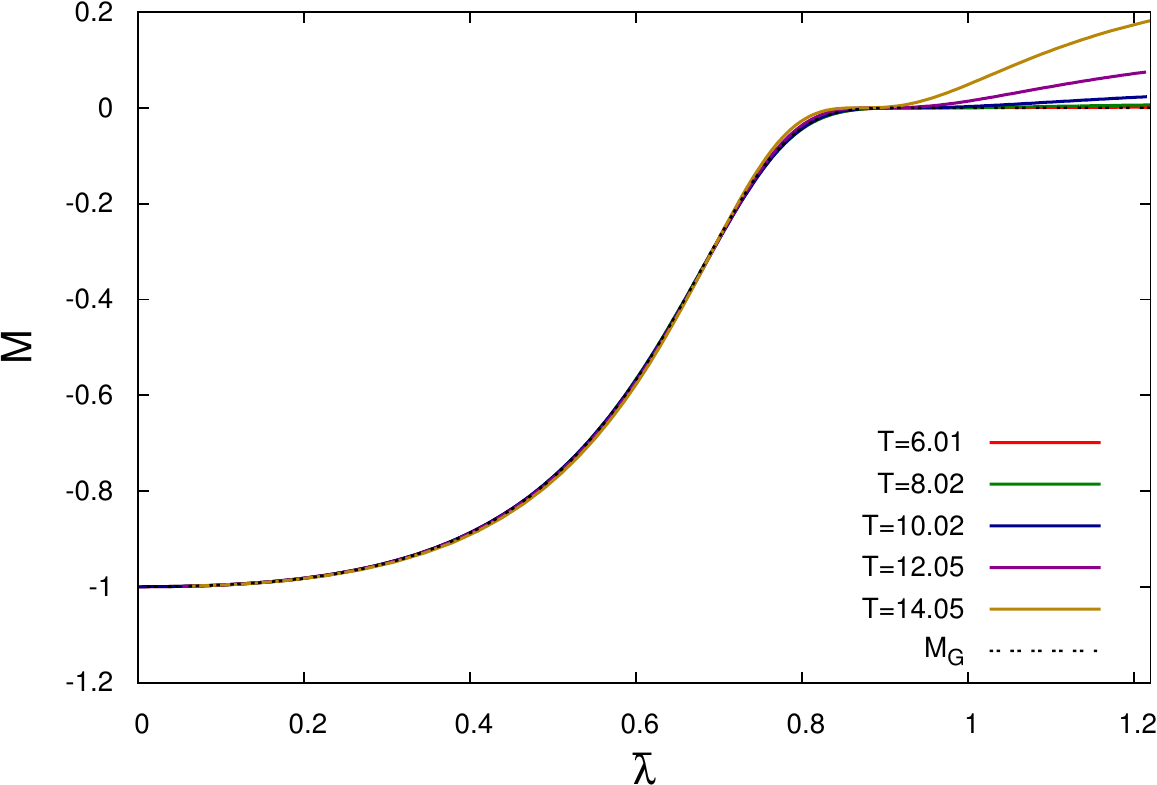}
\caption{$M(\bar\lambda,T)$ in the sub25 evolution
  with excision for five values of $T$ in the interval $[6,14]$.}
\label{figure:MlambdabarT_sub25_ex}
\end{figure}

\begin{figure}[!htb]
\centering
\includegraphics[scale=0.65]{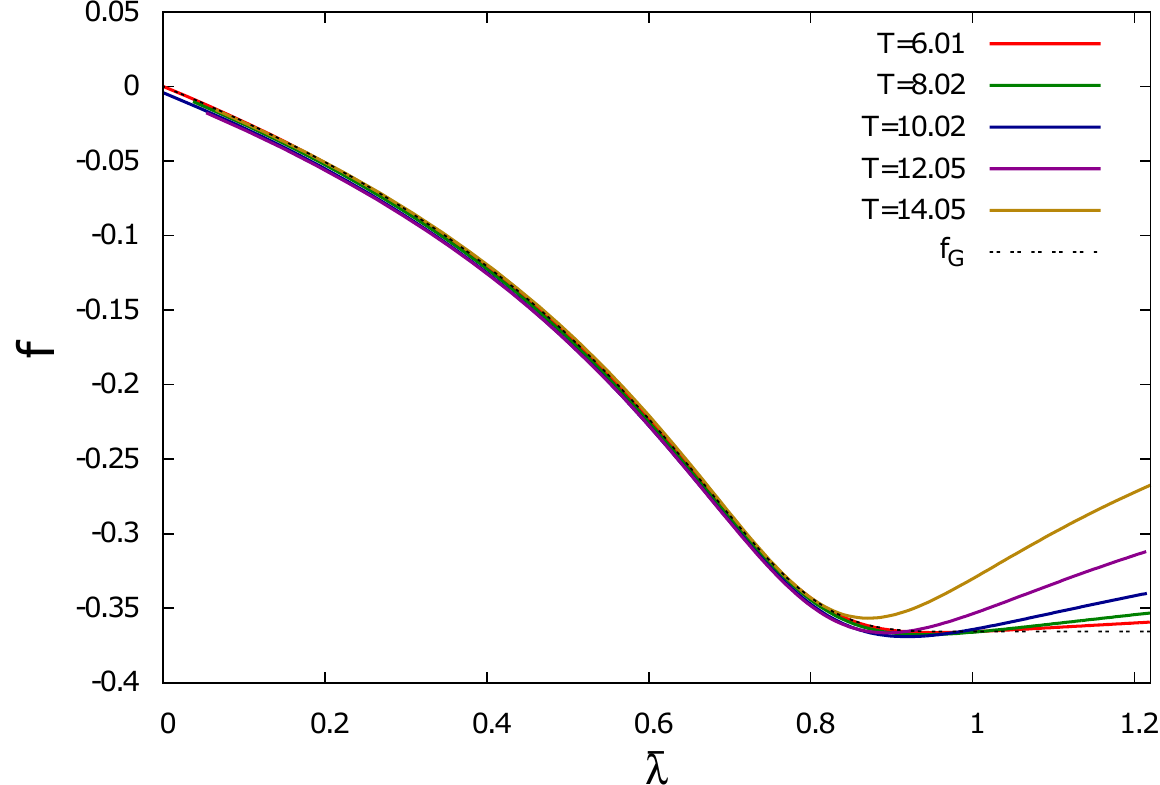}
\caption{$f(\bar\lambda,T)$ in the sub25 evolution
  with excision for five values of $T$.}
\label{figure:phi-cT_lambdabarT_sub25_ex}
\end{figure}

\begin{figure}[!htb]
\centering
\includegraphics[scale=0.65]{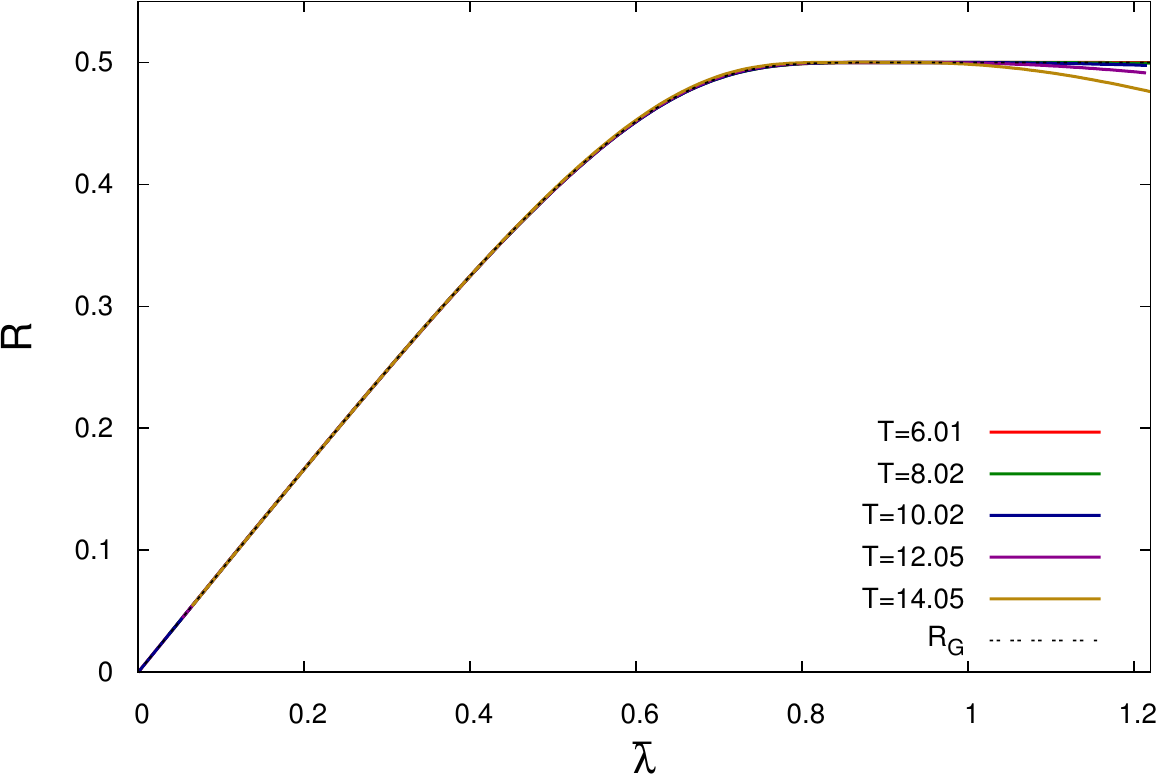}
\caption{$R(\bar\lambda,T)$ for $6<T<14$ in the sub25 evolution
  with excision for five values of $T$.}
\label{figure:RlambdabarT_sub25_ex}
\end{figure}

\begin{figure}[!htb]
\centering
\includegraphics[scale=0.8]{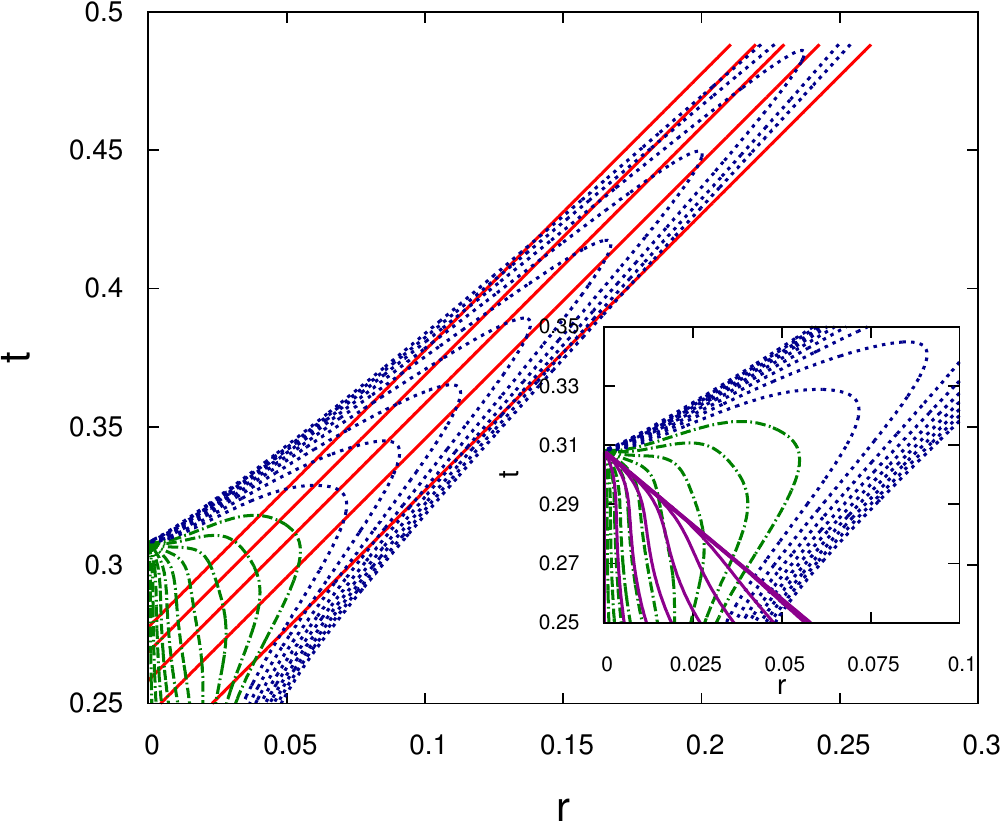}
\caption{Contour lines of $T$ from $4$ to $8$ (red, solid) in steps of
  $1$, and contour lines of $\bar\lambda$ from $0$ to $1$ (green,
  dash-dotted) and from $1.1$ to $2$ (blue, dotted) in steps of $0.1$,
  in the sub10 evolution. The code stops much later than in the sub25 evolution,
  allowing us to access a much larger range of $\bar\lambda$. 
  The inset shows again contour lines of $\bar \lambda$ (colours and lines
  as in the main plot), and contour lines of $x$ for $0\le x <1$ 
  (magenta, solid).}
\label{figure:Tlambdabarrt_contours_sub10}
\end{figure}

\begin{figure}[!htb]
\centering
\includegraphics[scale=0.8]{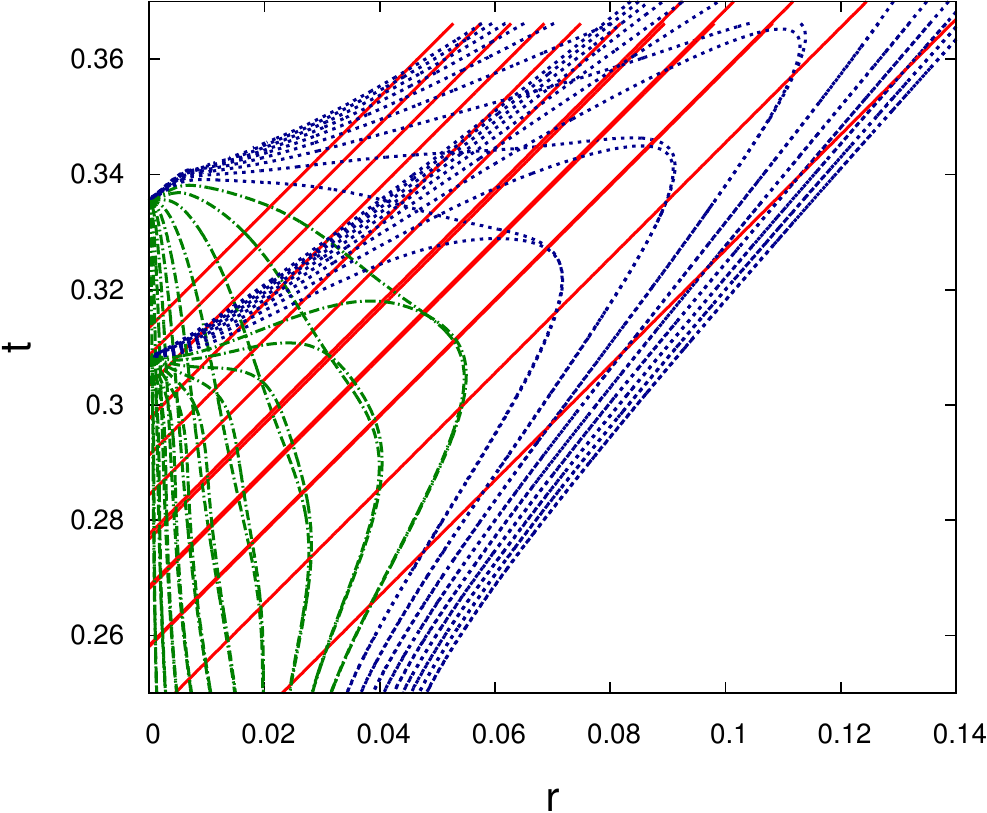}
\caption{Overlay of Figs.~\ref{figure:Tlambdabarrt_contours_sub25_ex}
  (sub25) and \ref{figure:Tlambdabarrt_contours_sub10} (sub10), but
  without the contour lines of $x$. The $T$ contour lines (from $6$ to
  $14$ for sub25 and $4$ to $8$ for sub10) essentially agree. The
  $\bar\lambda$ contour lines agree for $T\lesssim 6$. The coordinate
  time value $t_*$ of the accumulation point differs significantly
  betweeen the two evolutions ($0.31$ and $0.34$). (The proper time
  value $t_{*0}$ is essentially the same.)}
\label{figure:Tlambdabarrt_contours_sub10_sub25}
\end{figure}

\begin{figure}[!htb]
\centering
\includegraphics[scale=0.65]{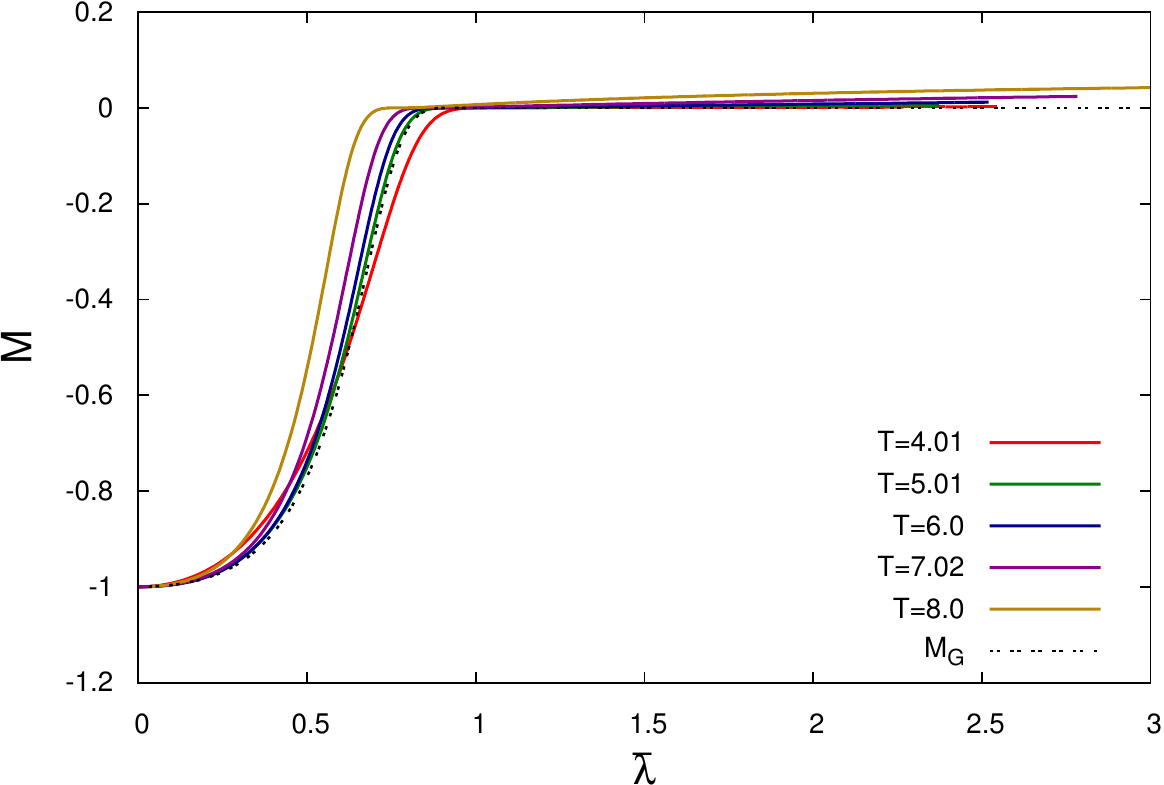}
\caption{$M(\bar\lambda,T)$ for the sub10 evolution. In the
  analytic null continuation of the Garfinkle solution,
  $M(1-\bar\lambda)=-M(\bar\lambda)$, whereas in the null continuation
  $M(\bar\lambda)=0$ for $\bar\lambda>1$. Clearly the latter fits the
  plot and the former does not.}
\label{figure:MlambdabarT_sub10}
\end{figure}

\begin{figure}[!htb]
\centering
\includegraphics[scale=0.65]{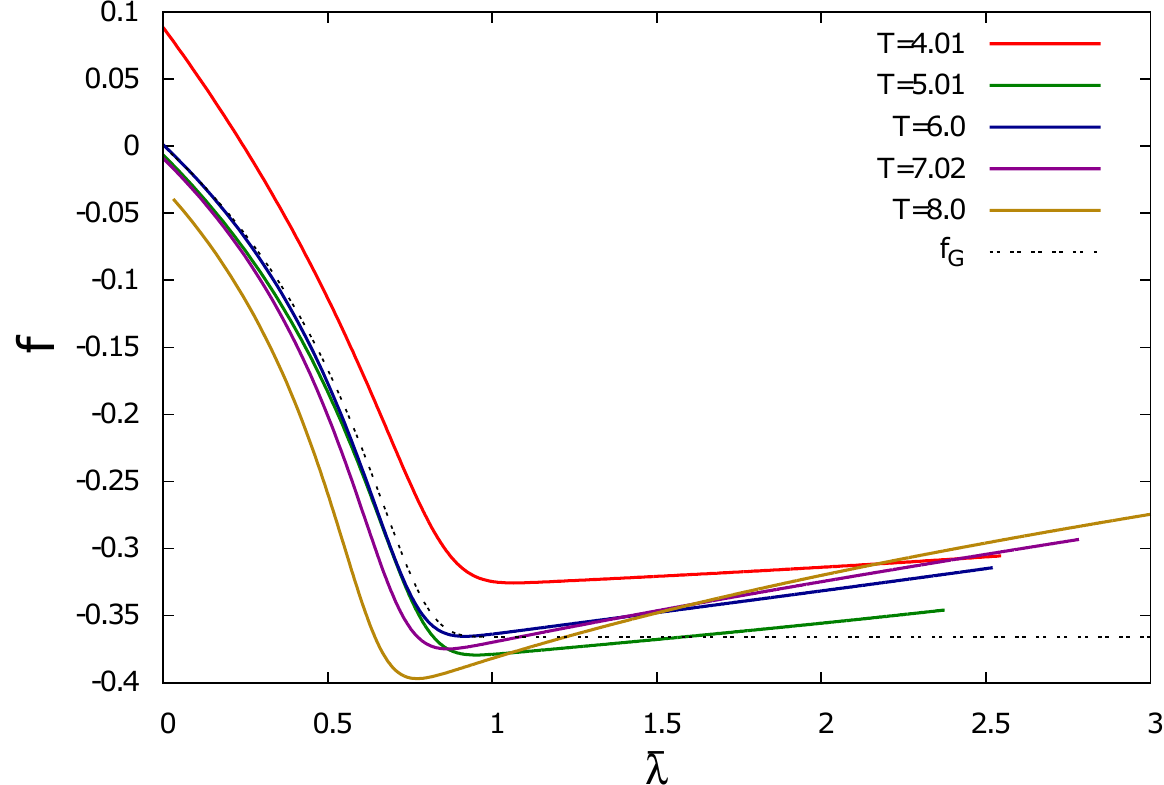}
\caption{$f(\bar\lambda,T)$ for the sub10 evolution. }
\label{figure:phi-cT_lambdabarT_sub10}
\end{figure}

\begin{figure}[!htb]
\centering
\includegraphics[scale=0.65]{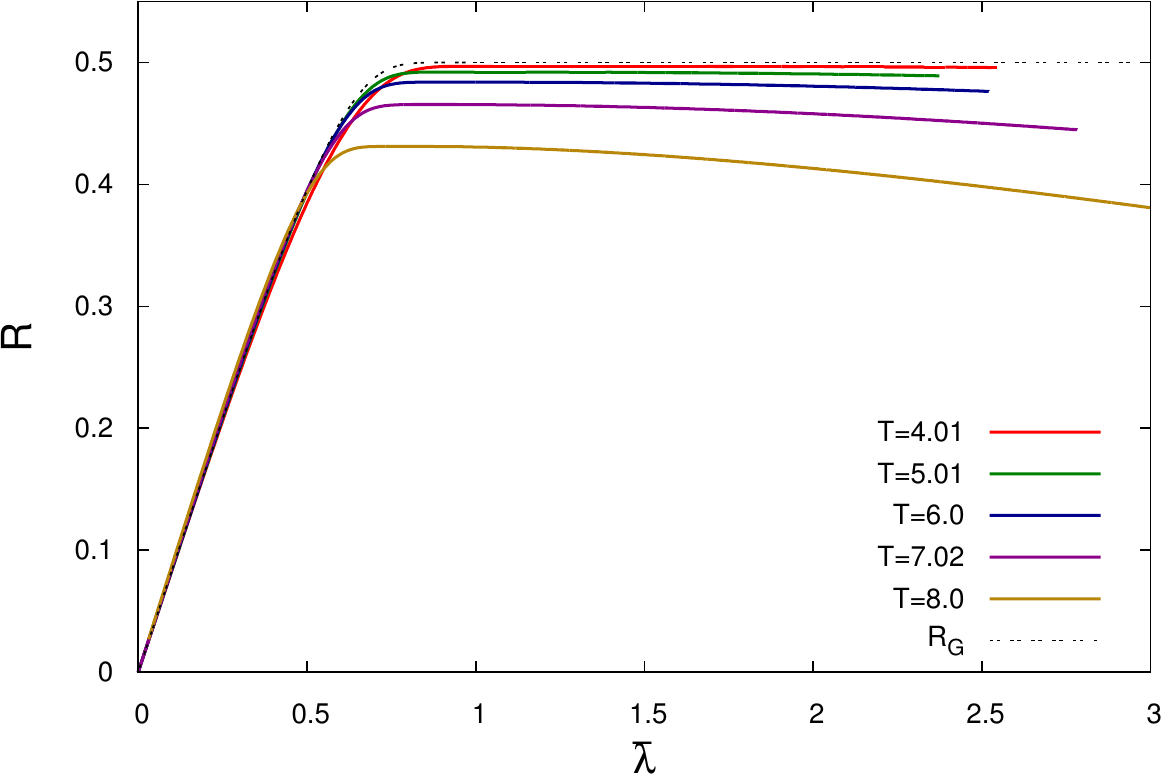}
\caption{ $R(\bar\lambda,T)$ for the sub10 evolution. In the
  analytic null continuation of the Garfinkle solution,
  $R(1-\bar\lambda)=R(\bar\lambda)$, whereas in the null continuation
  $R(\bar\lambda)=1/2$ for $\bar\lambda>1$. Clearly the latter fits the
  plot better than the former.}
\label{figure:RlambdabarT_sub10}
\end{figure}

Our plots of $M$, $R$ and $f$ against $\bar\lambda$ or $x$, at a range
of fixed values of $T$, show that inside the lightcone the deviations
from the $n=4$ Garfinkle solution are very small. Such deviations are
expected from a number of sources. In
Sec.~\ref{section:LambdaGarfinkle} below we compute perturbative
corrections to the Garfinkle solution for a nonvanishing
$\Lambda<0$. These are of order $\Lambda\tilde u^2=\exp-2T$ and
hence very small. We also expect one growing perturbation, which is
small by virtue of fine-tuning, infinitely many decaying
perturbations, small by virtue of large $T$, and numerical error. As
these deviations from the Garfinkle solution are unlikely to cancel
systematically, our plots indicate that they are all separately
small.

Outside the lightcone, the deviations from our proposed null
continuation of the $n=4$ Garfinkle solution are larger than inside
the lightcone. It is clear that they cannot be mainly $\Lambda$
corrections, as they increase with $T$, rather than depending on $T$
as $\exp-2T$. Rather, we believe that these deviations depend on the
initial data in a manner that does not vanish in the fine-tuning
limit. Mathematically, this may reflect that the discrete perturbation
modes of the critical solution are not complete, or that a sum over
those modes does not converge outside the lightcone. The latter could
happen because individual modes that decay more rapidly with $T$ grow
more rapidly as functions of $x$ outside the lightcone. (While we
formally construct the discrete mode spectrum in
Sec.~\ref{section:garfinklepert} below, we have only explicitly
calculated the growing modes as functions of $x$.) Yet another way of
looking at this is to note that while demanding CSS and analyticity at
the centre and the lightcone uniquely defines the countable family of
Garfinkle solutions, the null data on the lightcone define a unique
analytic continuation only if we demand CSS everywhere. 

In Sec.~\ref{section:outeransatz} below, we find an ODE system whose
solution is an exact solution of the full field equations for finite
$\Lambda<0$ outside the lightcone, and which can be matched at the
lightcone to the Garfinkle solution and its first $\Lambda$
corrections as smoothly as the null continuation itself, namely
$C^3$. Hence, this is better than the null continuation plus
$\Lambda$ corrections, but one may wonder how the two are related. As
discussed in Sec.~\ref{section:continuations} below, the bare null
continuation has a null translation invariance in addition to
spherical symmetry and CSS. Our exact outer solution has only one
continuous symmetry of a hitherto unknown type: it acts as an isometry
on the $(r,t)$ plane but a CSS on the orbits of spherical
symmetry. However, if we expand this solution into a series in powers
of $\exp-2T$, we obtain term by term the null continuation and its
perturbative $\Lambda$ corrections, so we can also think of it as an
approximate CSS symmetry. (Recall that with finite $\Lambda$, exact
CSS is impossible.) In the regime where we have plotted near-critical
solutions, even the first-order $\Lambda$ corrections are very small
compared to the zeroth order null continuation, so the deviations from
the null continuation that we see cannot be caused mainly by
these. For the same reason, we cannot distinguish the null
  continuation plus first $\Lambda$ correction from the exact solution
  of which it is the expansion.


\subsection{Evolving initial data for our amended Garfinkle solution}
\label{section:evolveamendedGarfinkle}


\subsubsection{Motivation and overview}
\label{section:Garfinklemotivation}


Our working hypothesis, compatible with the numerical results
presented so far, is that there is a true critical solution, which is
asymptotically CSS, and which has one growing mode with
$\lambda_0\simeq 7/8$. We have given strong numerical evidence that
this critical solution is very well approximated by the $n=4$
Garfinkle solution inside the lightcone. We have also given, somewhat
weaker, numerical evidence that outside the lightcone it is
approximated not by the analytic continuation of the Garfinkle
solution, but by what we have called its null extension.

The $\Lambda=0$ Garfinkle solution has a MOTS on its lightcone, and
the $\Lambda=0$ null extension has a MOTS at every point. Therefore,
on theoretical grounds, we need to add a $\Lambda$ correction
to both, which removes the MOTSs. (These corrections are de facto so
small, at least inside the lightcone, that we would have no reason to
add them only to improve agreement with our numerical data.) We shall call
this null-continued, $\Lambda$-corrected $n=4$ Garfinkle solution
the ``amended Garfinkle solution''.

Our amended Garfinkle solution still has two obvious shortcomings,
namely that both it and its linear perturbations are not analytic but
only $C^3$ at the lightcone, and that it has three growing
modes. Analyticity at the lightcone is a natural requirement if the
critical solution is required to arise from the evolution of generic
initial data. Hence the non-analyticity is not a mere technical
shortcoming, and may well be related to the incorrect number of
growing modes. Similarly, any universal critical solution can only
have one growing mode. 

In this Subsection we will give numerical evidence that, in some way that
we do not yet understand theoretically, these twin problems seem to
cancel each other out. We shall evolve initial data for our
amended Garfinkle solution, matched outside its lightcone to
asymptotically adS data, and add perturbations from one
of five families: two that we consider as generic, and the three
growing perturbation modes of the null-continued $n=4$ Garfinkle
solution. We shall find that these data evolve in the expected CSS
way, and that our amended Garfinkle solution with (approximately)
zero perturbation is critical in all of these five families, showing
scaling with $\gamma\simeq 8/7$ and $\delta\simeq 16/23$ in each case,
with no indication of any other growing mode. Hence we conclude that
analyticity and the presence of the cosmological constant together
somehow suppress the $\lambda=2/8$ and $3/8$ growing modes, while the
top $7/8$ mode survives.


\subsubsection{Data and results}
\label{section:Garfinkleresults}


The technical details of how we construct the initial data at $t=0$ for our
amended Garfinkle solution are given in
Sec.~\ref{section:Garfinkledata} below. Here we need to say only that
they are parameterised by $T_{\rm initial}$, the
value of $T$ at $(r=0,t=0)$, which governs the magnitude of the
$\Lambda$ corrections, the value $r_{\rm lightcone}$ of $r$ where the
lightcone of the Garfinkle solution intersects $t=0$, and the location
$r_0$ and width $\Delta r$ of the switchover from Garfinkle data to
vacuum.

We have chosen $T_{\rm initial}=10.0$ in order to make the
$\Lambda$ correction small throughout the initial data, and $r_{\rm
  lightcone}=0.3$, $r_0=0.6$ and $\Delta r=0.3$ in order to minimise
spurious mass generated by the switching. With these parameters the
total mass is $0.00635$. The $\Lambda$ correction to the initial data
is small enough not to be visible in plots, and is of course expected
to decay further as $\exp(-2T)$. Hence we can expect to compare the
time evolution of these data against the null-continued $\Lambda=0$ Garfinkle
solution within our plotting accuracy.

In the first one-parameter family of deformations of these data, we
multiply $\phi(r,0)$ and $\phi_{,t}(r,0)$ by a
factor of $1+p$. We find that the critical value is $p_*\simeq
6.68\cdot 10^{-7}$. As expected, this is small. Moreover, we find good
agreement with the $n=4$ Garfinkle solution inside the lightcone from
$t=0$ onwards, as there is no transition from generic initial data to
the Garfinkle solution.

\begin{figure}[!htb]
\includegraphics[scale=0.65, angle=0]{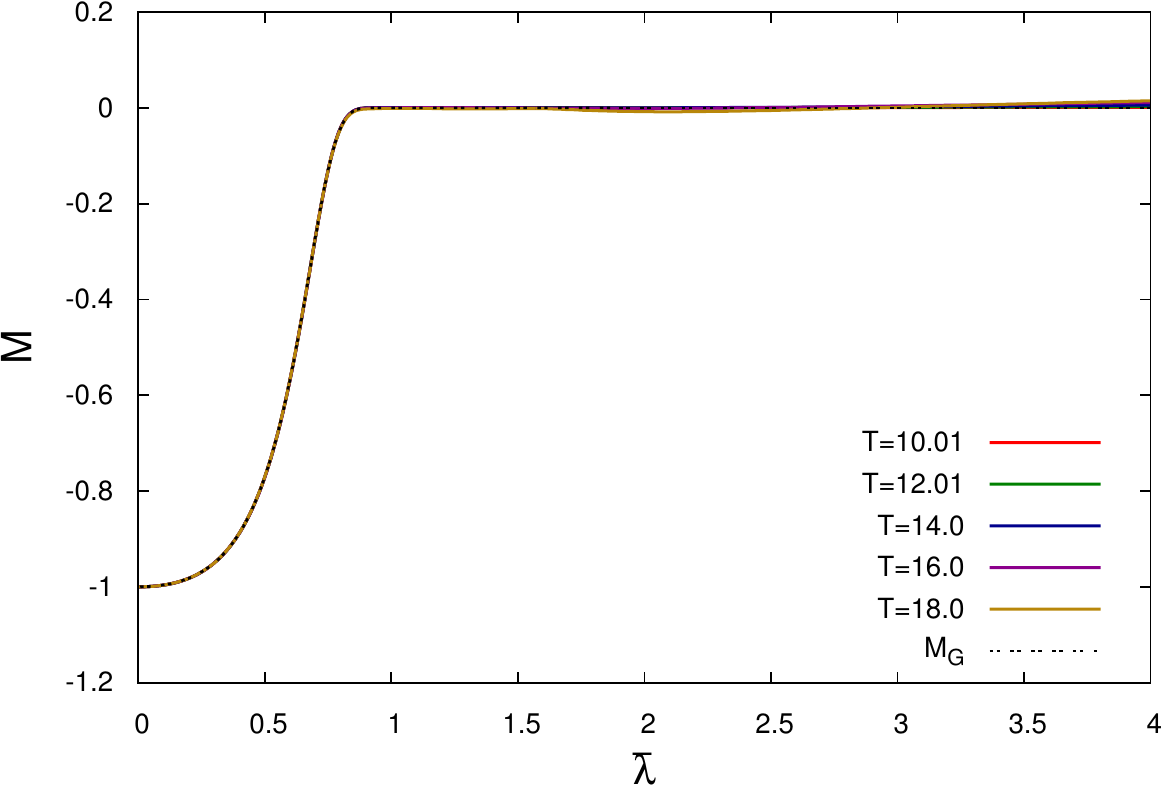}
\caption{$M(\lambda,T)$ for evolution of sub8 Garfinkle initial data for
five chosen moments of $T=\textrm{const.}$, where the scalar field is multiplied
by the factor $1+p$.
}
\label{figure:/MlambdabarT_gar_sub8}
\end{figure}

\begin{figure}[!htb]
\includegraphics[scale=0.65, angle=0]{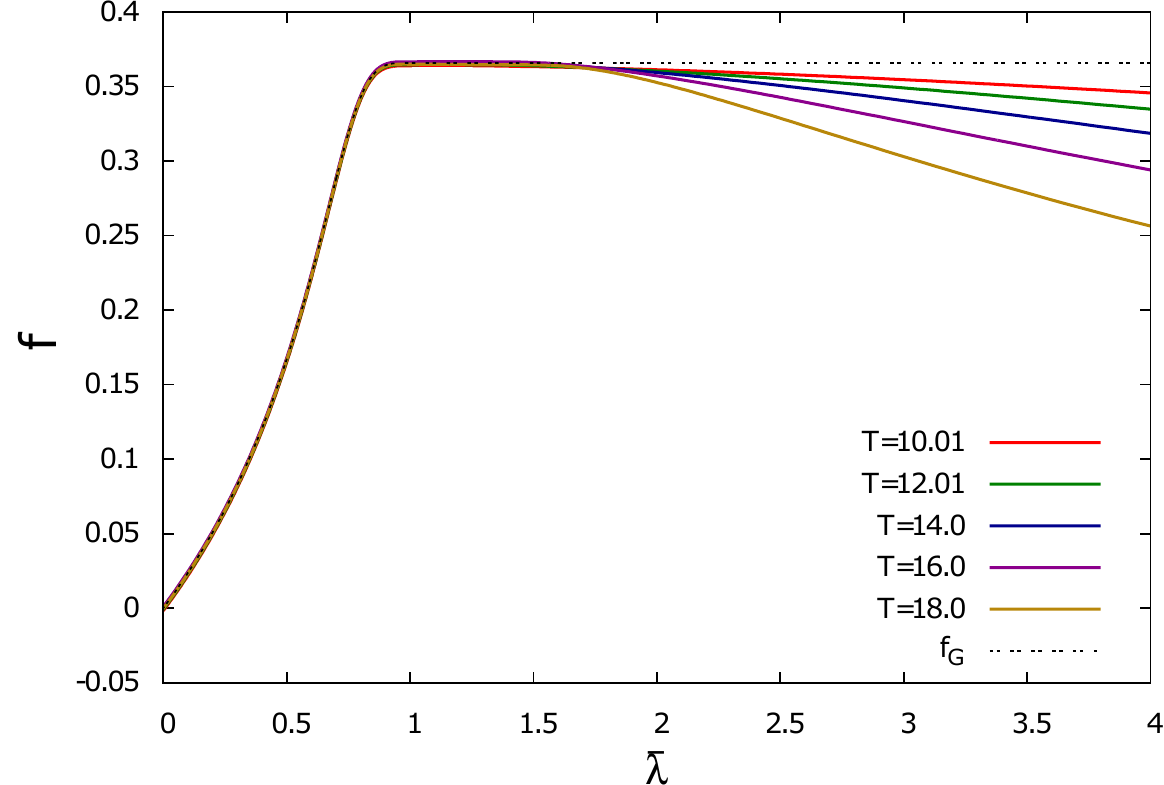}
\caption{The same for $f(\lambda,T)$.}
\label{figure:/phi-cT_lambdabarT_gar_sub8}
\end{figure}

\begin{figure}[!htb]
\includegraphics[scale=0.65, angle=0]{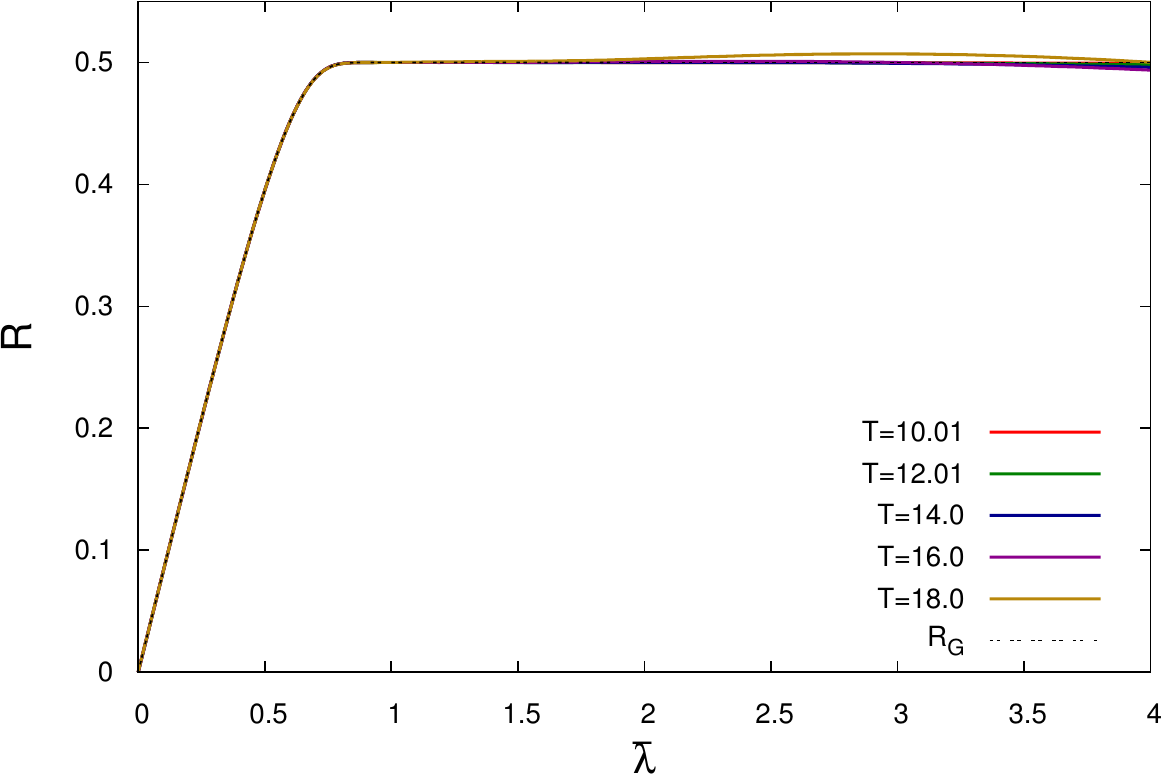}
\caption{The same for $R(\lambda,T)$.}
\label{figure:/RlambdabarT_gar_sub8}
\end{figure}

We find that $M\simeq 0$,  $f\simeq 2\ln 2$ and
$R\simeq 1/2$ outside the lightcone, as they would be in the
null continuation. This is demonstrated for the sub8 evolution in
Figs.~\ref{figure:/MlambdabarT_gar_sub8}-\ref{figure:/RlambdabarT_gar_sub8}.
 (We do not know why the deviation in $f$ is relatively much larger).
We have chosen sub8 because it is in the middle of the range of
  $\ln(p_*-p)$ where we see convergence of the Ricci scaling, and
  hence trust our evolution.

 With the same fitting procedure we used above for Gaussian initial
 data, a fit of $c(p)$ for subcritical $(1+p)$-times-Garfinkle data gives
 $c_*=-0.263871 \pm 1.7\cdot 10^{-10}$, $c_1=-0.263861$ and $c_2=5.502
 \cdot 10^{-5}$ for the fitting interval $[\text{sub}3, p_*]$. This is
 equivalent to $n=3.999036 \pm 3.7 \cdot 10^{-8}$. The values of
 constants $c_1$ and $c_2$ are much smaller than for Gaussian initial
 data, meaning that $c$ depends only very weakly on $p$. Clearly, the
 formal fitting error is an overoptimistic estimate of the error in
 $n$, but the evidence strongly suggests $n=4$ again.

In a second 1-parameter family of initial data, we take our best
approximation to the critical point of the first family and add a
Gaussian (centre $0.2$, width $0.05$) in $\phi$ and
$\phi_{,t}$ with overall amplitude $p$. The critical value for this
family is $p_*\simeq 5.5\cdot 10^{-10}$. We would expect this to be
very small, as the $p=0$ element of this family is already our best
approximation to the critical point of the first family.

We have created three other 1-parameter families of initial data by
adding one of the $m=7$, $m=3$, and $m=2$ growing perturbations of the
null-extended Garfinkle solutions to the best fine-tuned data of the
first family. We add the perturbations for both $\phi$ and $B$ and
their derivatives, with $c_2=p$, and then solve the (nonlinear)
constraints for $A$ and $A_{,t}$. The critical values are
$p_*=5.5\cdot 10^{-9}$, $5.5\cdot 10^{-12}$ and $5.5\cdot
10^{-14}$. 

\begin{figure}[!htb]
\includegraphics[scale=0.7, angle=0]{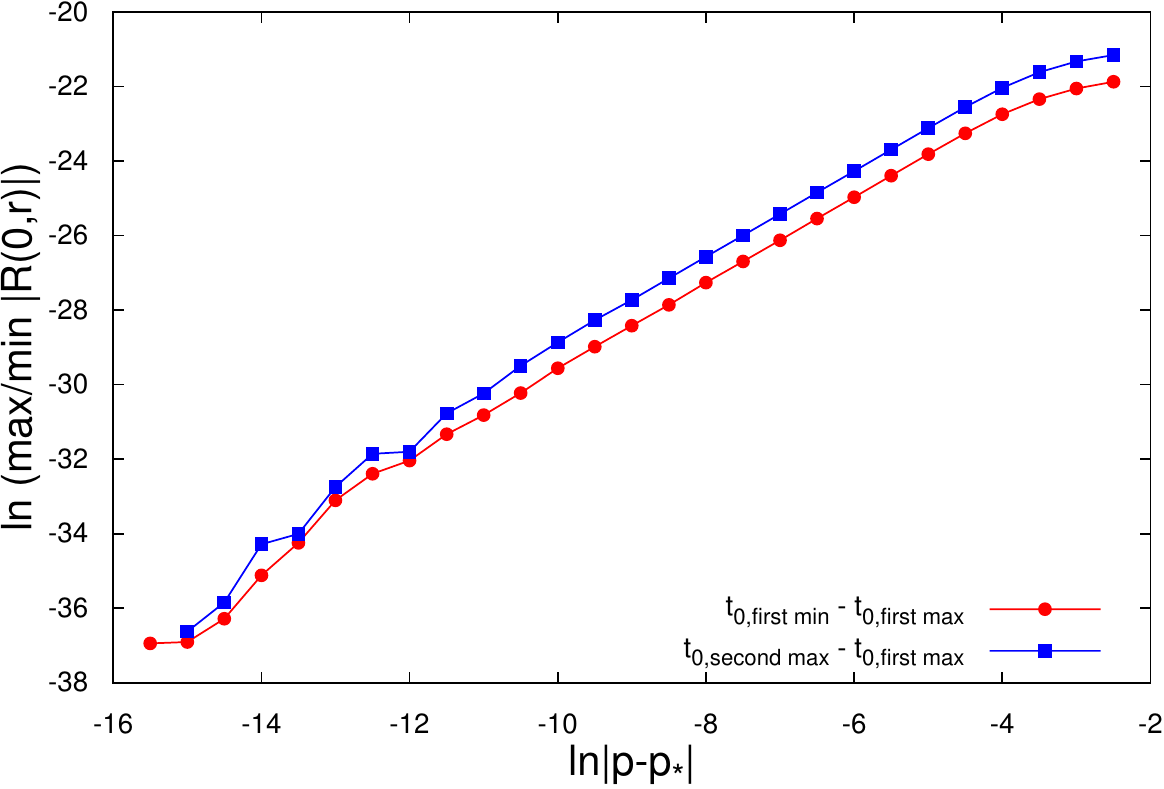} 
\caption{ Log-log plot of the proper time of the first minimum and
  second maximum in Ricci at the centre, relative to the first
  maximum, against $p-p_*$, for our ``amended Garfinkle solution''
  initial data, with the scalar field initial data scaled by
  $1+p$. From a fit on the interval $[-11,-4]$ to the first maximum we
  find {$\gamma = 1.1441 \pm 0.0022$}, compatible with $\gamma=8/7$.}
\label{figure:time_scaling_gar}
\end{figure}

\begin{figure}[!htb]
\centering
\includegraphics[scale=0.7, angle=0]{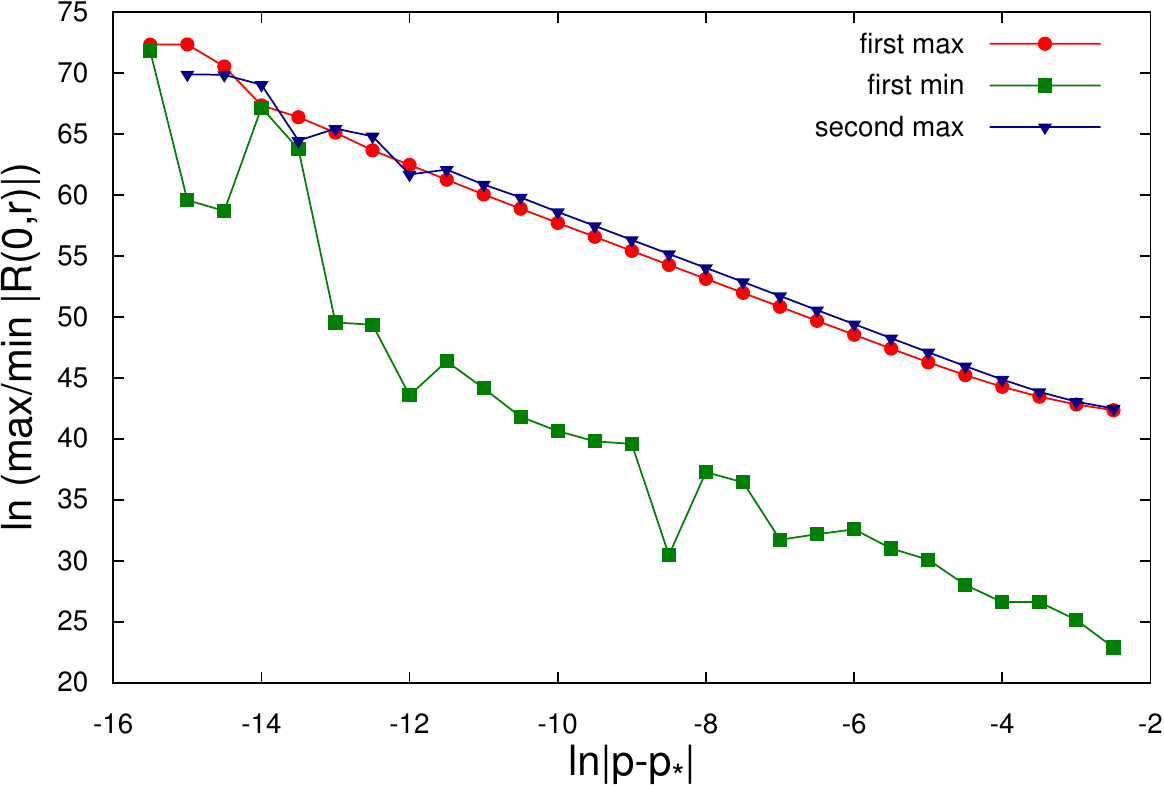}
\caption{Scaling of the values first and second maxima and first
  minimum of $ |R(0,t)|$, for the same initial data. From fit in the
  same interval as for the figure above, we have {$\gamma=1.1432 \pm
    0.0073$}, which is again compatible with theoretical value
  $\gamma=8/7$.}
\label{figure:R_scaling_gar}
\end{figure}

\begin{figure}[!htb]
\includegraphics[scale=0.7, angle=0]{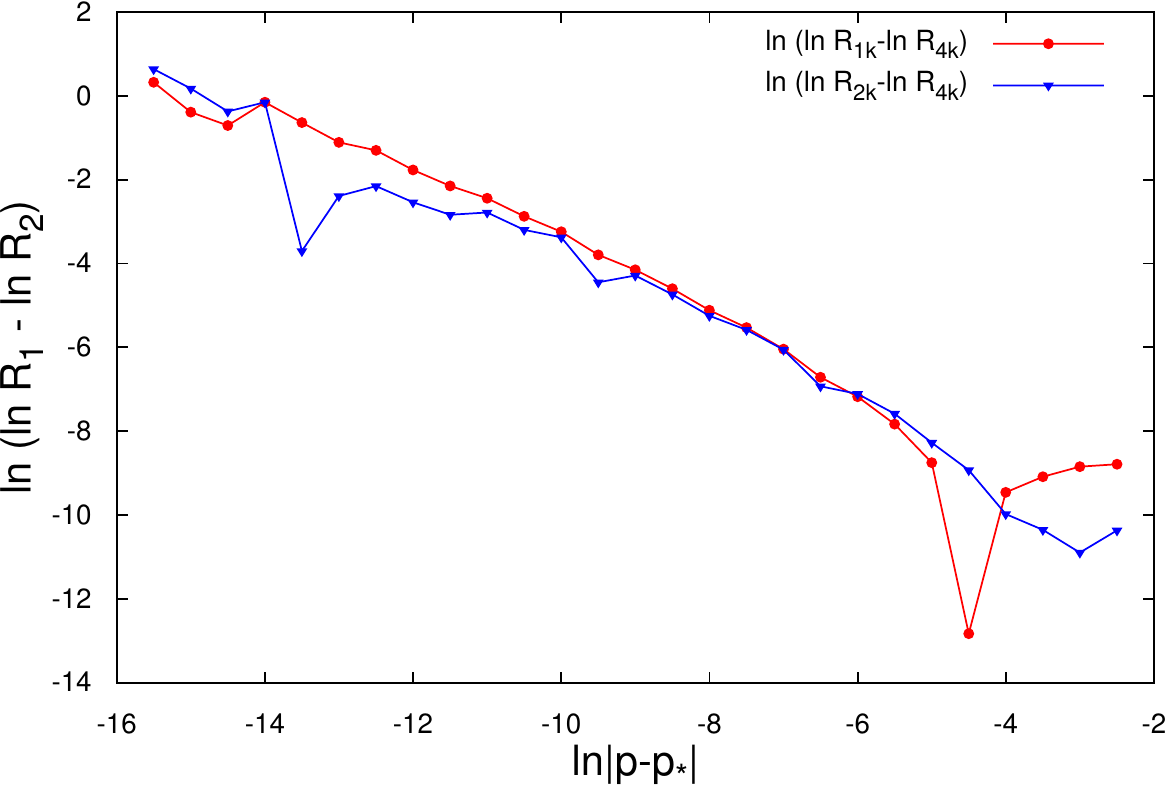}
\caption{Convergence of the first maximum of Ricci as a function of
  $p-p_*$, similar to
  Fig.~\ref{figure:riccifirstmax_gauss_convergence} but now for the
  (null-continued, $\Lambda$-corrected) Garfinkle data, fine-tuned by
  rescaling the scalar field initial data by an overall factor of
  $(1+p)$. In contrast the fine-tuned Gaussian data in
  Fig.~\ref{figure:riccifirstmax_gauss_convergence}, here
  $C=\ln[(4^1-1)/(2^1-1)]$, demonstrating 1st-order order convergence
  with resolution.}
\label{figure:riccifirstmax_gar_convergence} 
\end{figure}

All five families show similar subcritical power-law scaling of the
values and proper time locations of the extrema of the Ricci scalar at
the centre. This is demonstrated in
Figs.~\ref{figure:time_scaling_gar}-\ref{figure:riccifirstmax_gar_convergence}
for the first family. The value of $\gamma$ obatained for this family
is compatible, within our numerical accuracy, with the theoretical
value $\gamma=8/7$ and slightly different from the result obtained for
Gaussian initial data. However, the discrepancy is not
  significant if we take into account that the actual deviation of our
  data from a straight line is not random but smooth
  (i.e. systematic). By eye, a straight line with slope $8/7$ seems to
  be as good a fit as the least-squares straight line. That the
  deviation from $8/7$ is smaller than for the Gaussian initial data
might be explained by the fact that for our approximate critical
solution we start much closer to the true critical solution, and less
fine-tuning is needed to observe the scaling.

\begin{figure}[!htb]
\includegraphics[scale=0.7, angle=0]{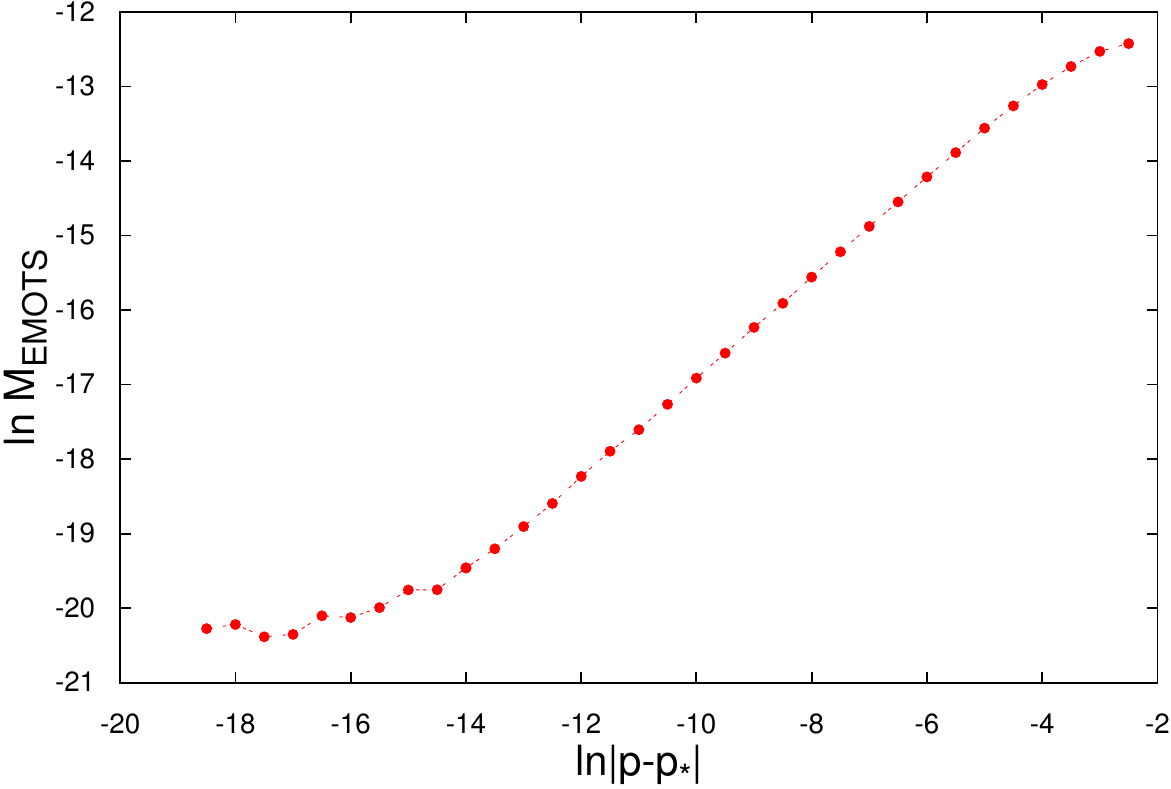}
\caption{Scaling of the EMOTS mass for supercritical evolutions of
  $(1+p)$-times-Garfinkle initial data. The fit in the interval [-14,-4] gives
  {$\delta = 0.6684 \pm 0.0017$},  which is $16\sigma$ or $4\%$ below the
  theoretical value $16/23$.}
\label{figure:mass_scaling_gar} 
\end{figure}

We have also looked at mass scaling for the supercritical
evolutions. At low fine-tuning, for example from super5 to super15 for
the $m=7$ family of initial data, we find a MOTS present already in
the initial data. (To be precise, our initial data constraint solver
fudges the MOTS, but it then appears on the first time step.)  At
larger fine-tuning, say for super17 to super27 for this family, the
EMOTS occurs at some $t>0$. We find that the mass of the MOTS in the
initial data, or the EMOTS forming later, lie on a single curve with
$\delta=16/23$. In Sec.~\ref{section:deltaderivation}, where we derive
$\delta$, we explain why it also applies for the MOTS in the
initial data in this case. The mass scaling for Garfinkle family of
initial data is presented in Fig.~\ref{figure:mass_scaling_gar}.

The $(1+p)$-times-Garfinkle family of initial data shows first-order
convergence in the intervals [sub4, sub14]  (see
    Fig.~\ref{figure:riccifirstmax_gar_convergence}) and [super3,
    super14]. For fine-tuned Garfinkle initial data plus a Gaussian
  perturbation we also see first-order convergence in the intervals
  [sub5, sub22] and [super5, super19]. For subcritical evolutions of
  $m=2$ perturbations of the Garfinkle data we observe second-order
  convergence in the interval [sub5, sub22] and first
  order-convergence for supercritical data in the interval [super3,
    super19]. For $m=3$ perturbations we again have first-order
  convergence in the intervals [sub6, sub21] and [super2,
    super20]. $m=7$ perturbations show second-order convergence for
  subcritical evolutions in the interval [sub7, sub15] followed by
  first-order convergence in [sub15, sub25], while for supercritical
  evolutions the convergence is first order in the interval
  [super3,super25]. The results obtained in fitting $\gamma$ and $\delta$
  to the power laws for all families of initial data studied are given
  in Table \ref{table:fits}.

\begin{widetext}
\begin{table*}
\begin{tabular}{| >{\centering\arraybackslash}m{0.25 \linewidth} | > {\centering\arraybackslash}m{0.18 \linewidth}| > {\centering\arraybackslash}m{0.15 \linewidth} | > {\centering\arraybackslash}m{0.18 \linewidth} | > {\centering\arraybackslash}m{0.15 \linewidth} | @{}m{0pt}@{}}
\hline
 Family of initial data & $\gamma$ & fitting interval &$\delta$ & fitting interval &\\[1.5pt]
\hline
Gaussian & $1.2345 \pm 0.0076$ & [-26,-5] & $0.684 \pm 0.023$ & [-26,-17] \\[1.5pt]
\hline
$(1+p)\text{Garfinkle}$ & $1.1441 \pm 0.0022$ & [-11,-4] & $ 0.6684 \pm 0.0017$ & [-14,-3]\\[1.5pt]
\hline
$(1+p_*)\text{Garfinkle} + \text{Gauss. pert.}$ & $1.1402 \pm 0.0056 $ & [-18,-10] & $0.6765 \pm 0.0018$ & [-19,-9]\\[1.5pt]
\hline
$(1+p_*)\text{Garfinkle} + m=2 \text{ pert.}$ & $1.1234 \pm 0.0031$ & [-18,-10] & $0.6894 \pm 0.0076 $ & [-19,-8] \\[1.5pt]
\hline
$(1+p_*)\text{Garfinkle} + m=3 \text{ pert.}$ & $1.1442 \pm 0.0084 $ & [-20,-10] & $0.6730 \pm 0.0021$ & [-20,-11] \\[1.5pt]
\hline
$(1+p_*)\text{Garfinkle} + m=7 \text{ pert.}$ & $1.14290 \pm 0.00044$
& [-25,-13] & $0.6991 \pm 0.0033$ & [-25,-10] \\[1.5pt]
\hline
 Theoretical values & $8/7\simeq 1.1429$ &  & $16/23\simeq 0.6957$ &  \\
\hline
\end{tabular}
\caption{Values of $\gamma$ and $\delta$ obtained in fitting power
  laws to the maximum of Ricci and mass of the (inner) EMOTS together
  with error bars and fitting intervals (in $\ln|p-p_*|$). Some
  fitting intervals were chosen not according to the convergence
  results, but were modified due to the fact that scaling started for
  more critical data.}
\label{table:fits}
\end{table*}
\end{widetext}


\section{Theory}
\label{section:theory}


\subsection{The Garfinkle solution}
\label{section:garfinkle}


This Subsection is based on \cite{Garfinkle} and is included here for
completeness. In the metric coefficients $\tcA$ and $R$ and
coordinates $(x,T)$ defined in
(\ref{tcAdef}-\ref{Rdef}), the general metric becomes
\begin{equation}
\label{xTmetric}
ds^2=\ell^2 e^{-2T}\left[
e^{2\bcA}\left(dx-{x\over 2n}\,dT\right)dT+R^2\,d\theta^2\right],
\end{equation}
 where we have defined the shorthand $\bcA$ by
\begin{equation}
\label{Atildedef}
e^{2\bcA}:=2nx^{2n-1}e^{2\tcA}.
\end{equation}

In order to eliminate $\kappa$ from the field equations
(\ref{fieldequations}), we define the positive dimensionless parameter
$\tilde c$ from the dimensionful parameter $c$ by
\begin{equation}
\label{ctildef}
\tilde c:=\sqrt{8\pi Gc^2}. 
\end{equation}
The Garfinkle solution \cite{Garfinkle} of the field equations
(\ref{fieldequations}) with $\Lambda=0$, denoted by the subscript
$0$, is then given by
\begin{eqnarray}
\label{tcA0val}
e^{2\tcA_0}&=&\left({1+x^n\over 2}\right)^{4\tilde
  c^2}x^{-2n\tilde c^2}, \\
\label{R0val}
R_0 &=& {1-x^{2n}\over 2},\\
\label{phi0val}
\phi_0 &=& c\left[T - 2\ln\left({1+x^n\over2}\right) \right].
\end{eqnarray}
The mass function is given by
\begin{equation}
\label{M0val}
M_0=-\left({1+x^n\over 2}\right)^{-4\tilde c^2}x^{2n-1},
\end{equation}
which takes value $M=-1$ at the centre, as necessary for regularity,
and $M=0$ at the lightcone. 

So far, the Garfinkle solution is analytic at the centre $x=1$ for any
$\tilde c$ and any positive integer $n$, but it is generically
singular at the lightcone $x=0$, because $\exp 2\bcA_0$ is
either zero or infinite there. However, with 
\begin{equation}
\label{ct2nrelation}
\tilde c^2=1-{1\over 2n},
\end{equation}
the overall power of $x$ in $\exp 2\bcA$ cancels and the
solution is also analytic at the lightcone $x=0$, with
\begin{equation}
\label{Atilde0val}
e^{2\bcA_0}=2n\left({1+x^n\over 2}\right)^{4\tilde c^2}.
\end{equation}
Hence, the Garfinkle solution is analytic at the centre and
lightcone if and only if $n=1,2,\dots$, thus restricting the possible
values of $c$.

 In order to give an analytic form of the metric also in double null
coordinates, we rescale $\tilde v$ to something that is proportional
to $x$, namely \cite{Garfinkle}
\begin{equation}
\label{hatvdef}
\hat v:=-\ell \tvl^{1\over 2n} \quad\Rightarrow\quad  x=\hvl
\tul^{-{1\over 2n}}.
\end{equation}
The metric then becomes
\begin{equation}
\label{Abarmetric}
ds^2=-\tul^{1-{1\over 2n}}e^{2 \bcA}d\tilde u\, d\hat v+(-\tilde u)^2R^2d\theta^2.
\end{equation}

Rescaling also $\tilde u$, 
\begin{equation}
\label{xThatuv}
\hat u=-\ell \tul^{1\over 2n} \quad\Rightarrow\quad
x={\hat v\over \hat u}, \quad T=-2n\ln\hul,
\end{equation}
the metric becomes
\begin{equation}
\label{Ahatmetric}
ds^2=-e^{2 \hcA}d\hat u\, d\hat v+\ell^2\hul^{4n}R^2d\theta^2,
\end{equation}
where
\begin{equation}
e^{2\hcA}:=4n^2e^{2\tcA}
\left({\tilde u \tilde v\over \ell^2}\right)^{1-{1\over 2n}}
=2ne^{2\bcA}\tul^{2(1-{1\over 2n})}.
\end{equation}
The Garfinkle solution then takes the more symmetric form \cite{CCF}
\begin{eqnarray}
\label{hatcalA0}
e^{2\hcA_0}&=&4n^2\left({\hul^n+\hvl^n\over 2}\right)^{4\tilde
  c^2}, \\
\label{hatcalB0}
\rb_0&=&\ell{\hul^{2n}-\hvl^{2n}\over 2}, \\
\label{hatphi0}
\phi_0&=&-2c
\ln\left({\hul^n+\hvl^n\over 2}\right).
\end{eqnarray}
This is again analytic at both the centre and lightcone (and of course
everywhere in between) for $n=1,2,\dots$


\subsection{Continuation beyond the lightcone}
\label{section:continuations}


As for integer $n$ the Garfinkle metric and scalar field are analytic
in $x$, they can be analytically extended to negative $x$ simply by
considering values of $x$ in the range $-1<x\le 1$.

For both even and odd $n$, every centred ring in the the region $-1<x<0$ is an
outer-trapped surface, and $x=-1$ is a future spacelike central
curvature singularity (with $M=1$ at $\rb=0$), where the Ricci scalar
blows up. In this sense, the analytically extended Garfinkle solution
could be described as a black hole that is CSS rather than stationary.

Our numerical evidence rules out the analytic continuation, but
seems to be compatible with an alternative continuation, where everything
depends only on retarded time (assuming that $\Lambda=0$, as we did in
the Garfinkle solution). We shall call this the ``null continuation''
of the Garfinkle solution, and also denote it by the suffix $0$.

At $\hat v=0$, the Garfinkle solution (\ref{hatcalA0}-\ref{hatphi0})
can be matched to
\begin{eqnarray}
\label{hatcalA0outer}
e^{2\hcA_0}&=&4n^2\left({\hul^n\over 2}\right)^{4\tilde
  c^2}, \\
\label{hatcalB0outer}
\rb_0&=&\ell{\hul^{2n}\over 2}, \\
\label{hatphi0outer}
\phi_0&=&-2c\ln\left({\hul^n\over 2}\right).
\end{eqnarray}
As the power $\hat v^n$ is unmatched, the matching is
$C^{n-1}$. Equivalently, in terms of the similarity coordinates
$(x,T)$ and the similarity variables $R$ and $\bcA$, we can match
(\ref{R0val}-\ref{M0val},\ref{Atilde0val}) to
\begin{eqnarray}
e^{2\bcA_0}&=&2^{-4\tilde c^2}, \\
R_0&=&{1\over 2}, \\
\phi_0&=&c(T+2\ln 2), \\
M&=& 0.
\end{eqnarray}
As the power $x^n$ is unmatched, the matching is again $C^{n-1}$.

From these two forms of the metric we see that the null continuation
has a translation invariance in $\hat v$, in addition to circular
symmetry and CSS. It can be thought of as an outgoing Vaidya
metric. If we think of an event horizon in spherical symmetry as an
outgoing null surface of constant area radius $\rb$, then the null
continuation is an onion with each layer representing an event
horizon, and a null singularity at its centre.


\subsection{The similarity coordinate $\bar\lambda(x)$}
\label{section:analyticlambdabar}


We can rewrite (\ref{dsdvtilde}) in generality as
\begin{equation}
\label{dsdx}
s_{,x}(x,T)=-{1\over 2}\ell e^{-T+2\bcA(x,T)}.
\end{equation}
Hence in the Garfinkle solution, $\lambda:=s/(-\tilde u)$ is given by
\cite{Garfinkle}
\begin{eqnarray}
\lambda_n(x)&=&n \int_x^1\left({1+x^n\over 2}\right)^{4\tilde c^2} dx
\\ 
&=&4^{{1\over n}-2}n\Biggl[F^2_1\left({2\over n}-4,{1\over n},{1\over
  n}+1,-1\right) \nonumber \\
&&\qquad \qquad -xF^2_1(\dots,-x^n)\Biggr].
\end{eqnarray}
This is a well-behaved function of $x$, with $\lambda_4(0)\simeq
0.8377$ and $\lambda_4'(0)\simeq -0.3536$. (Recall the centre is at
$x=1$ and the lightcone at $x=0$.) For even $n$, $\lambda$ is an
odd function of $x$. In our plots we use
$\bar\lambda:=\lambda/\lambda_4(0)$, so that the lightcone of the
$n=4$ Garfinkle solution is at $\bar\lambda =1$, and the analytically
extended $n=4$ Garfinkle solution is covered by the range $0\le \bar\lambda<
2$, with $f$ and $R$ even about $\bar\lambda=1$ and $M$ odd.
In the null-extended Garfinkle solution $\lambda(x)$ is given by
\begin{equation}
\lambda=\begin{cases}
\lambda_n(x), & x>0 \\
\lambda_n(0)+\lambda_n'(0)x, & x<0,
\end{cases}
\end{equation}
where we have imposed continuity of $s$ and $s_{,\hat v}$ at $\hat
v=0$, and hence of $\lambda$ and $\lambda_{,x}$ at $x=0$.


\subsection{Boundary conditions and gauge conditions}


To fix the residual gauge freedom in double null coordinates, we need
two gauge conditions. One of these is always taken to be $\rb=0$ at
$u=v$ in \cite{Garfinkle,GarfinkleGundlach,CCF}, that is, the centre
is at $r=0$, which also corresponds to $x=1$. 

In the Garfinkle solution \cite{Garfinkle} and in its
$\Lambda$ corrections \cite{CCF}, the second gauge condition is chosen
to be $\tcA=0$ at $r=0$. This means that $\tilde u$ and $\tilde v$ are
proper time at the centre.

However, Garfinkle and Gundlach \cite{GarfinkleGundlach} deviate from
this last gauge condition for the growing perturbations because in the
similarity coordinates $(x,T)$ these can be made regular only for a
different gauge choice. (The gauge in which each growing perturbation
is regular is linked to the gauge in which ${\cal A}=0$ at $r=0$ by
the infinitesimal gauge transformation generated by the vector field
$\xi$ given in (\ref{GGxi}) below, for a specific value of $c_1$
chosen to cancel a singularity at $x=0$.)

With the gauge choice $\rb(0,t)=0$, the absence of a conical
singularity in the metric requires that $\rb=r\exp\cA+O(r^3)$, or
equivalently $\rb_{,r}=\exp\cA$ at $r=0$. For consistency, this
requires also $\phi_{,r}=0$ and $\cA_{,r}=0$ at $r=0$. Hence we have
three regularity conditions to impose at $x=1$.

By the definition of $x$, the lightcone is at $x=0$, which corresponds
to $\tilde v=0$. The solution is analytic there if $\bcA$, $\rb$ or
equivalently $R$, and $f$ are analytic in $(x,T)$ at $x=0$. 


\subsection{Perturbations of the Garfinkle solution}
\label{section:garfinklepert}


This subsection is based on \cite{GarfinkleGundlach} and is included
here for completeness. We slightly
rewrite the ansatz of \cite{GarfinkleGundlach} as
\begin{eqnarray}
\label{cAperturbation}
\tcA&=&\tcA_0(x)+e^{kT}a(x), \\
R&=&R_0(x)+e^{kT}b(x), \\
\label{phiperturbation}
\phi&=&c[T+f_0(x)+e^{kT}h(x)],
\end{eqnarray}
where $\tcA_0$, $R_0$, $f_0$ denotes the Garfinkle solution. (The
correspondence of notation is $y=x^n$ and $H=ch$, with $a$ and $b$
having the same meaning.)

The regularity conditions on $\tcA$, $R$ and $\phi$ are
\begin{eqnarray}
\label{a1regularitycondition}
a'(1)+nka(1)&=&0, \\
\label{b1regularitycondition}
b'(1)+n[(k-1)b(1)+a(1)]&=&0, \\
\label{h1regularitycondition}
h'(1)+nkh(1)&=&0,
\end{eqnarray}
and the gauge conditions are
\begin{eqnarray}
\label{b1gaugecondition}
b(1)&=&0, \\
\label{a1gaugecondition}
a(1)&=&0.
\end{eqnarray}
Note (\ref{a1gaugecondition}) will be modified below for growing
perturbations. The ODEs for $b$, $h$ and $a$ can be solved in this
sequence, and are second-order, second-order and first-order
respectively.

The general solution of the ODE for $b$ is
\begin{equation}
\label{bxexpr}
b(x)=c_0+c_1\left[1-x^{2n(1-k)}\right],
\end{equation}
and this is pure gauge. We set $c_0=0$ to impose the gauge
condition (\ref{b1gaugecondition}) and keep the centre at $x=1$, but
leave $c_1$ arbitrary. Note that the regularity condition
(\ref{h1regularitycondition}) is obeyed for any $c_1$.

The general infinitesimal gauge transformation that preserves the
double null form of the metric in $(u,v)$ is
$\xi=f(u)\p_u+g(v)\p_v$. Preserving the gauge condition that the
centre is at $r=0$ then requires $f=g$. If we also require the gauge
transformation to be compatible with the mode ansatz
(\ref{cAperturbation}-\ref{phiperturbation}), it becomes unique up to
an overall factor, which we can choose to be $c_1$:
\begin{eqnarray}
\label{GGxi}
\xi &=& -2c_1\ell^k\left[(-u)^{1-k}{\p\over \p u}+(-v)^{1-k}{\p\over \p v}\right] \\
&=& -2c_1e^{kT}\left[{\p\over \p T}+{1\over 2n}
\left(x-x^{1-2nk}\right){\p\over\p x}\right].
\end{eqnarray}
Hence $c_1$ is also pure gauge, and the remaining gauge freedom in the
mode ansatz is parameterised precisely by $c_1$.

The general solution $h(x)$ that is regular at the centre $x=1$ is 
\begin{equation}
h(x)=c_2F_a(1-x^{2n})-2c_1{1+x^{n(1-2k)}\over 1+x^n},
\end{equation}
where $F_{a}(z)$ stands for a hypergeometric function that is
  analytic in $z$ for $|z|<1$ with $F_a(0)=1$. Hence $c_2$
multiplies a regular solution of the homogeneous ODE for $h(x)$. The other
linearly independent homogeneous solution contains a $\ln(1-x)$ term,
which is singular at the centre, and is therefore ruled out.

Not only for $b(x)$ and $h(x)$ here, but also for $a(x)$ below, all
terms proportional to $c_1$ arise as gauge transformations generated
by the vector field $\xi$ defined in (\ref{GGxi}), and all other terms
are proportional to $c_2$. In this sense any perturbation with $c_2=0$
is pure gauge.

For the special case $k=1$,
\begin{equation}
h(x)=(c_2-2c_1){1+x^{-n}\over 1+x^n}.
\end{equation}
Hence the term parameterised by $c_2$ is also pure gauge. In this
case, $\xi=-2c_1\ell(\p/\p t)$, representing just a time translation
of the Garfinkle solution. This $k=1$ time translation mode arises in
the perturbation spectrum of any self-similar solution of the Einstein
equations.

The special case $k=0$ corresponds to an infinitesimal perturbation
that takes a CSS solution into a neighbouring one. As we have seen
that the Garfinkle solution is locally unique once the gauge has been
fixed and analyticity imposed at the centre and lightcone, the $k=0$
perturbations must be pure gauge, and we need not consider them
explicitly.

For the special case $k=1/2$,
\begin{equation}
h(x)=c_2({\rm singular})-2c_1{2\over 1+x^n}.
\end{equation}
This is singular for any $c_1$, unless $c_2=0$, and so is again pure
gauge. 

The homogeneous part of $h(x)$ can be written in the form
\begin{equation}
F_a(1-x^{2n})=C_bF_b(x^{2n})+C_cx^{n(1-2k)}F_c(x^{2n}),
\end{equation}
using formula 15.3.6 of \cite{AbramowitzStegun}. Here,
\begin{equation}
C_b:={\Gamma\left({1\over2}-k\right)\over\sqrt{\pi}\Gamma(1-k)}, \quad
C_c:={\Gamma\left(k-{1\over2}\right)\over\sqrt{\pi}\Gamma(k)}.
\end{equation}

However, this does not hold when $k=\mathbb{Z}+1/2$. Then either $C_b$
or $C_c$ is formally infinite, and in fact is replaced by a $\ln x$
term.  Both hypergeometric functions are by definition regular at
$x=0$. For $k\ne \mathbb{Z}+1/2$, a necessary condition for the
$x^{n(1-2k)}$ term, and hence $F_a$ to be regular at $x=0$ is that $k=m/2n$ for
$m$ integer. From considering the three special cases above, we
already have $m\ne 0,n,2n$.

Setting $k=m/2n$ from now on, we have
\begin{eqnarray}
h(x)&=&c_2C_bF_b(x^{2n})+c_2C_cx^{n-m}F_c(x^{2n})\br
-2c_1{1+x^{n-m}\over 1+x^n}.
\end{eqnarray}
Hence this is a series in $x^n$ (or in $x^{2n}$ for $c_1=1$), plus
$x^{n-m}$ times another such series. For $m<n$, every term is
regular separately. For $n<m<2n$, there is precisely one singular
power, $x^{n-m}$, which can be cancelled by setting
\begin{equation}
\label{c1forhregularity}
c_1=c_2{C_c\over 2}.
\end{equation}

For $m>2n$, there is at least a second singular power, which cannot be
cancelled, so this must be ruled out (and $m=2n$ was already ruled
out).

Further restrictions on $m$ arise from regularity of $a(x)$ at the
lightcone. $a(x)$ takes the form
\begin{equation}
a(x)=2c_1(1-k)+\int_1^xa'(x)\,dx,
\end{equation}
where $a'(x)$ is known in terms of $b(x)$ and $h(x)$ and $a(1)$ is
determined by (\ref{a1regularitycondition}). To check regularity at
$x=0$, it sufficient to check regularity of $xa'(x)$ as $x\to 0$ by
expanding it in powers of $x$. With $k=m/(2n)$, these powers are all
integers, so we only need to look for negative powers. We find that
$xa'(x)$ is a regular series in $x^n$, plus $x^{-m}$ times a
regular series in $x^n$.

Consider first $0<m<n$. Then the only singular power is $x^{-m}$ and
it can be cancelled by choosing
\begin{equation}
\label{c1foraregularity}
2c_1={(m-2n)(2n-1)\over m(m-1)}c_2C_c.
\end{equation}
However, in the case $m=1$ this regularity condition gives $c_2=0$ and
hence this regular perturbation is pure gauge.

For $n<m<2n$, the power $x^{n-m}$ is also singular. It is cancelled by
the same condition (\ref{c1forhregularity}) that is already required
to make $h$ regular for this range of $m$. Hence we now have two
regularity conditions on $c_1$, (\ref{c1forhregularity}) and
(\ref{c1foraregularity}). They are compatible for precisely $m=2n-1$.

Hence, from regularity of $a$ at $x=0$ we have found the additional
restrictions that either $m=2n-1$ or $m<n$ with $m\ne 0,1$.

We now summarise the union of all regularity conditions: For a given
integer $n>0$, the perturbation spectrum is given by $k=m/(2n)$ where
either $m=2n-1$, or $1<m<n$, or $m<0$ with $m\ne n(1-2\mathbb{N})$. In
particular there are $n-1$ growing perturbations given by
$m=2,3,\dots,n-1$ and $m=2n-1$.


\subsection{Perturbations of the null continuation}
\label{section:nullcontcorr}


The linear perturbations of the null continuation, with $k=m/(2n)$ are
\begin{eqnarray}
\label{bxnullcont}
b(x)&=&(d_0+d_1)+d_2x, \\
h(x)&=&-2(d_0+d_1)+d_4-{2nd_2\over 1+m-n}x+d_3 x^{n-m},\\
a(x)&=&-{(1+m-4n)(d_0+d_1)+(2n-4)d_4\over 2n} \br 
+{n(2n-1)d_2\over(1+m)(1+m-n)}x
+{(1-m)d_1+2nd_5\over 2n}x^{-m} \br
+{(1-2n)d_3\over 2n}x^{n-m},
\end{eqnarray}
subject to the condition that either $m=2n-1$ or $d_2=0$. Hence this is a
5-parameter family of solutions for generic $m$, or 6-parameter in the
particular case $m=2n-1$. The parameterisation has been chosen such
that the solution with $d_0=c_0$, $d_1=c_1$ and $d_2=d_3=d_4=d_5=0$ is
pure gauge, generated by the vector field $\xi$. (We have not yet set
$c_0=0$ here, as the reason for doing so for perturbations of the
Garfinkle solutions was only to fix the centre at $x=1$.)


\subsection{Matching of perturbations on the lightcone}


For $m< 0$, when $c_1$ and $c_2$ can be chosen independently, the
perturbations $b$ and $h$ of the Garfinkle solution take the following
values on the lightcone $x=0$:
\begin{eqnarray}
b(0)&=&c_1, \\
h(0)&=&-2c_1+c_2C_c.
\end{eqnarray}
For the allowed values of $m>0$, where $c_1$ is linked to $c_2$ for
regularity at the lightcone, they take the equivalent values with
$c_1$ given by (\ref{c1foraregularity}). The value of $a(0)$ can then
be obtained from the linearisation of the $\rb_{,uu}$ field equation
(\ref{rbuuequation}), which
at $x=0$ reduces to the algebraic equation
\begin{equation}
\label{deltauueqnx=0}
a(0)+\left(k-{1\over 2n}\right)b(0)+\left(1-{1\over 2n}\right)h(0)=0.
\end{equation}

In matching to the perturbations of the null continuation, continuity
of $b$ and $h$ at $x=0$ is required by physical regularity of the
spacetime (absence of a thin shell of matter at the matching
surface). This fixes 
\begin{eqnarray}
d_1&=&c_1-d_0, \\
d_4&=&c_2C_b.
\end{eqnarray}
Continuity of $a$ follows because (\ref{deltauueqnx=0}) holds on both sides. 

We can use the remaining free parameters $d_i$ to make the metric in
coordinats $(x,T)$ more differentiable at $x=0$ by matching the lowest
non-zero powers of $x$. We begin with the case $m<0$, which implies
$d_2=0$ and allows $c_2$ and $c_1$ to be chosen independently. We set
\begin{equation}
d_3=-2c_1+c_2C_c.
\end{equation}
in order to match the coefficients of $x^{n-m}$ in $h(x)$ on both sides. The
same choice of $d_3$ matches the coefficient of $x^{n-m}$ in $a(x)$, and we can
match the coefficient of $x^{-m}$ in $a(x)$ as well by setting
 \begin{equation}
d_5={1-2kn\over 2n}d_0+c_2C_c{(2k-1)(2n-1)\over 4kn}.
\end{equation}
Although we have not formally determined $d_0$, it cancels out of the
resulting expressions for $b(x)$, $h(x)$ and $a(x)$, so that we have
specified a unique maximally differentiable continuation, in the sense
that the coefficients of all powers of $x$ on the outer side match
those of the expansion in $x$ on the inner side, and all unmatched
powers are higher. 

The case $1<m<n$ is obtained from the case $m<0$ by imposing the
particular gauge choice (\ref{c1foraregularity}). In particular, the
coefficients of the singular power $x^{-m}$ in $a(x)$ then vanishes on
both sides.

In the special case $m=2n-1$, we have the additional parameter $d_2$
on the outer side. We set the other parameters as before, and
$d_2=-c_1$. $b$, $h$ and $a$ in the outer region are now all linear
functions of $x$. Once again, this choice is maximally differentiable in the
sense above. (For $b$, it happens to be analytic, as the expressions
for $b(x)$ on both sides coincide.)

Expressing the perturbations of the null continuation in terms of the
free parameters $c_1$ and $c_2$ only, we have, for $m<0$,
\begin{eqnarray}
b(x)&=&c_1, \\
h(x)&=&(c_2C_b-2c_1)+(c_2C_c-2c_1)x^{n-m}, \\
a(x)&=&-{c_2C_b(2n-1)+c_1(m+1-4n)\over 2n} \br
+{c_2C_c(m-n)(2n-1)-m(m-1)c_1\over 2mn} x^{-m} \br
+-{(c_2C_c-2c_1)(2n-1)\over 2n}x^{n-m}.
\end{eqnarray}
The case $1<m<n$ is obtained from the case $m<0$ by imposing the
particular gauge choice (\ref{c1foraregularity}). The case $m=2n-1$
is given by 
\begin{eqnarray}
b(x)&=&c_2{C_c\over 2}(1-x), \\
h(x)&=&c_2\left[(C_b-C_c)+C_cx\right], \\
a(x)&=&c_2\left[{C_b(1-n)+C_cn\over 2n}+{C_c(1-2n)\over 4n}x\right].
\end{eqnarray}


\subsection{$\Lambda$ corrections of the Garfinkle solution}
\label{section:LambdaGarfinkle}


This subsection is based on \cite{CCF}, and is included here for completeness.
If the field equations are not scale-invariant but scale-invariance
holds asymptotically on sufficiently small scales, the critical
solution itself may be approximated by an expansion in powers of
(typical length scale of the solution)/(length scale set by the field
equations). This was discussed in generality in
\cite{universalityclasses}, where it was also shown formally that the
leading order of this expansion represents a scale-invariant solution,
and that the perturbation spectrum is given by the perturbation
spectrum of that leading order. In our current problem, the only
length scale in the field equations is $\ell$ defined by
$\Lambda=-\ell^{-2}$, and so the required expansion is one in powers
of $\exp(-T)=-\tilde u/\ell$. 

The leading order of this expansion about the Garfinkle solution was
given in \cite{CCF}. We slightly rewrite their ansatz as
\begin{eqnarray}
\tcA&=&\tcA_0(x)+\sum_{n=1}^\infty e^{-2nT}\cA_n(x), \\
R&=&R_0(x)+\sum_{n=1}^\infty e^{-2nT}R_n(x), \\
\phi&=&c[T+f_0(x)+\sum_{n=1}^\infty e^{-2nT}f_n(x)]
\end{eqnarray}
(The correspondence of notation is $\cA_1$, $R_1=F$, $f_1=H/c$,
$y=x^n$ as before, and their $c$ is our $\tilde c$, with their $u$ and
$v$ our $\hat u$ and $\hat v$.)

The field equations give five ODEs for $R_1$, $f_1$ and $\cA_1$,
equivalent to the $k=-2$ perturbation equations but with source terms
proportional to $\Lambda$. We only need to obtain a particular
integral. The general solution is then obtained by adding the homogenous
$k=-2$ perturbations $b$, $h$ and $a$ to $R_1$, $f_1$ and $\cA_1$. The regularity conditions to first order in $\Lambda$ are
\begin{eqnarray}
\cA_1'(1)-2n\cA_1(1)&=&0, \\
R_1'(1)+n[-3R_1(1)+\cA_1(1)]&=&0, \\
f_1'(1)-2nf_1(1)&=&0,
\end{eqnarray}
and the gauge conditions are
\begin{eqnarray}
R_1(1)&=&0, \\
\cA_1(1)&=&0.
\end{eqnarray}
These are equivalent to the regularity and gauge conditions
for $k=-2$ linear perturbations. 

Like $b$, $R_1$ obeys a linear second-order ODE that can be solved on
its own. The two linearly independent homogeneous solutions are known
in closed form and are $1$ and $x^{6n}$, compare (\ref{bxexpr}). This
can be used to write a particular solution in the form of an integral
using variation of parameters. Alternatively, \cite{CCF} use the fact
that $R_1$ does not appear undifferentiated to first solve for $R_1'$
using an integrating factor.

With $R_1(x)$ determined, $f_1$ obeys a linear second-order ODE that
can be solved on its own. Again the homogeneous solution is known, and
so a particular integral can be given as an integral using variation
of parameters \cite{CCF}. 

The remaining three ODEs, which are of course consistent with each
either because of the Bianchi identities, can be reduced to obtain an
algebraic expression for $\cA_1=\cA_1(R_1',R_1,f_1',f_1,x)$. As for
the linear perturbations, the first $\Lambda$ correction of the
$\rb_{,uu}$ field equation becomes an algebraic constraint on the
lightcone, namely
\begin{equation}
2n\cA_1(0)+(2n-1)f_1(0)-(4n+1)R_1(0)=0. 
\end{equation}

Either of the integral expressions for $R_1$ gives rise to a messy
expression in terms of hypergeometric functions, and we have not been
able to evaluate any integral expression for $\cA_1$ in closed
form. Hence these formal solutions are not very useful for plotting or
numerical time evolution. Instead, we obtain a numerical solution by
solving the second order ODEs for $R_1$ and $f_1$ with the boundary
conditions $R_1'(1)=R_1(1)=0$ and $f_1'(1)=f_1(1)=0$, respectively. We
then solve the $\cA_{,uv}$ field equation as a second-order ODE for
$\cA_1$ with the boundary conditions $\cA_1'(1)=\cA_1(1)=0$. (This is
numerically more robust than trying to use the algebraic expression or
first order ODE for $\cA_1$). We add a $k=-2$ linear perturbation
with $c_1$ and $c_2$ chosen to set $R_1(0)=f_1(0)=\cA_1(0)=0$.

\begin{figure}[!htb]
\includegraphics[scale=0.5]{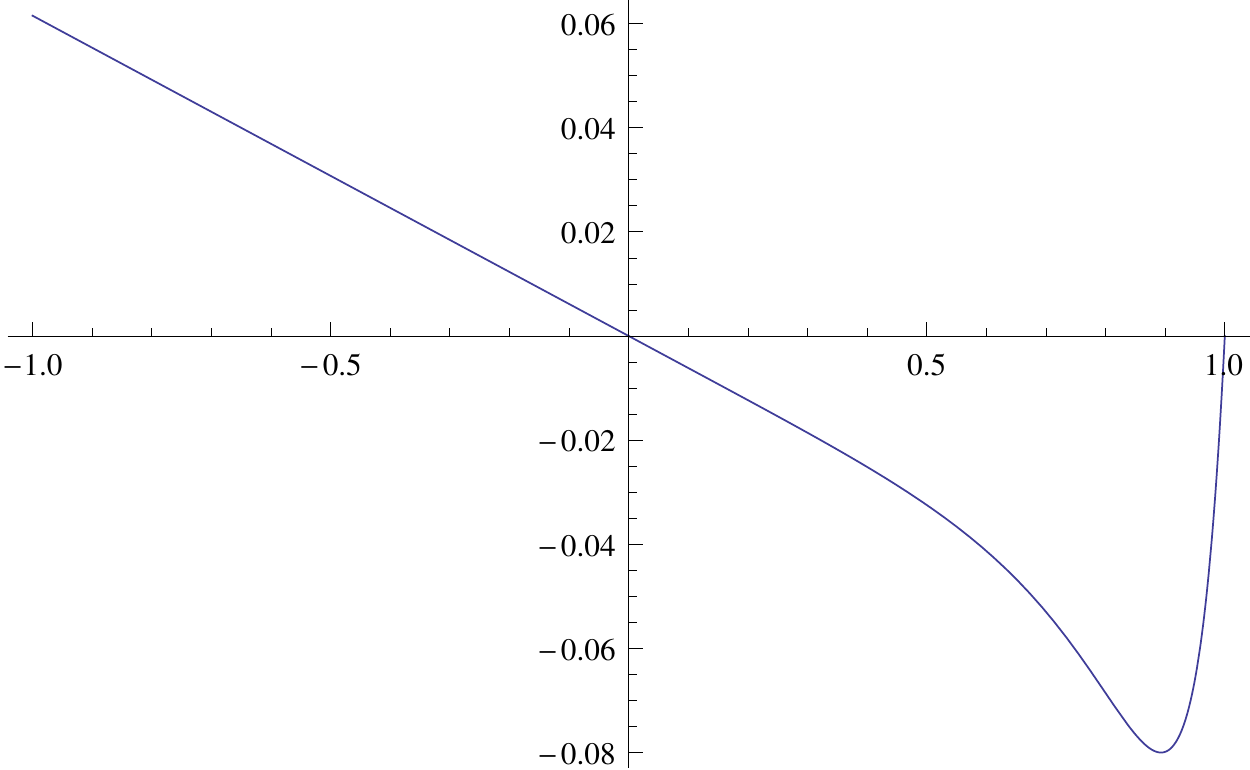} 
\vfill
\includegraphics[scale=0.5]{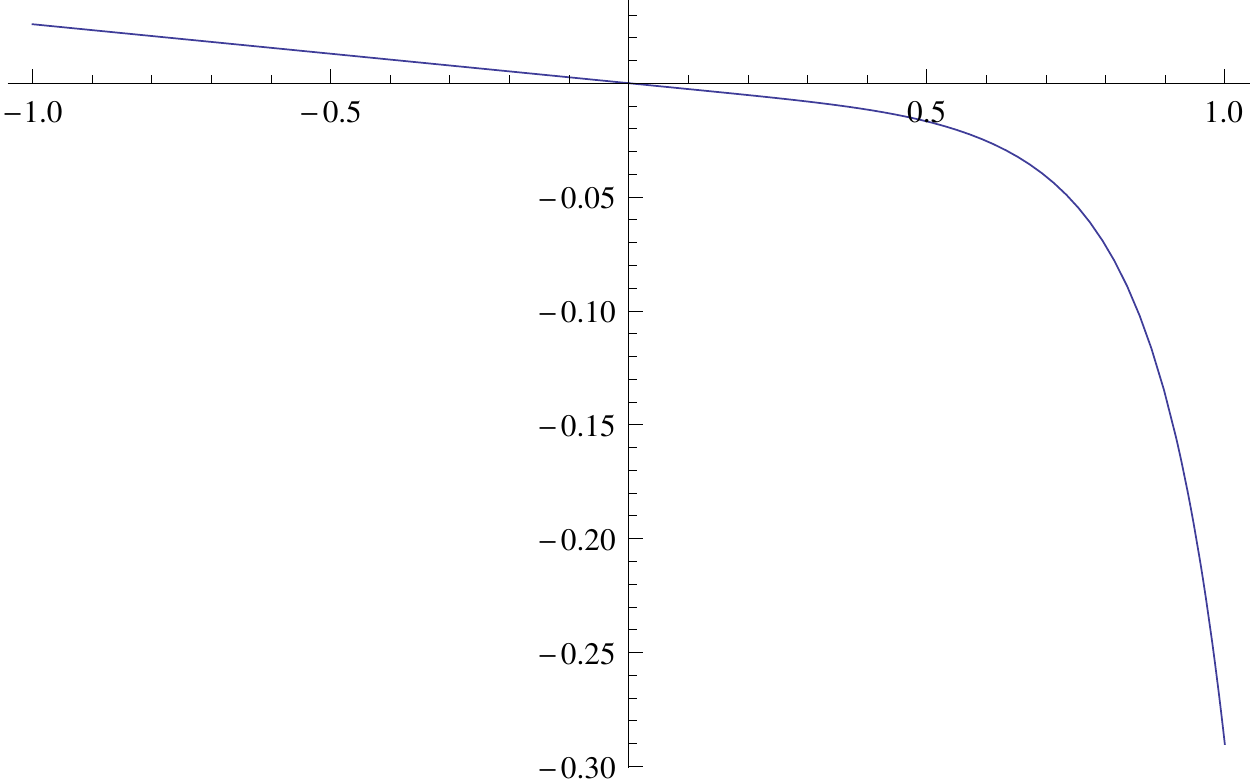} 
\vfill
\includegraphics[scale=0.5]{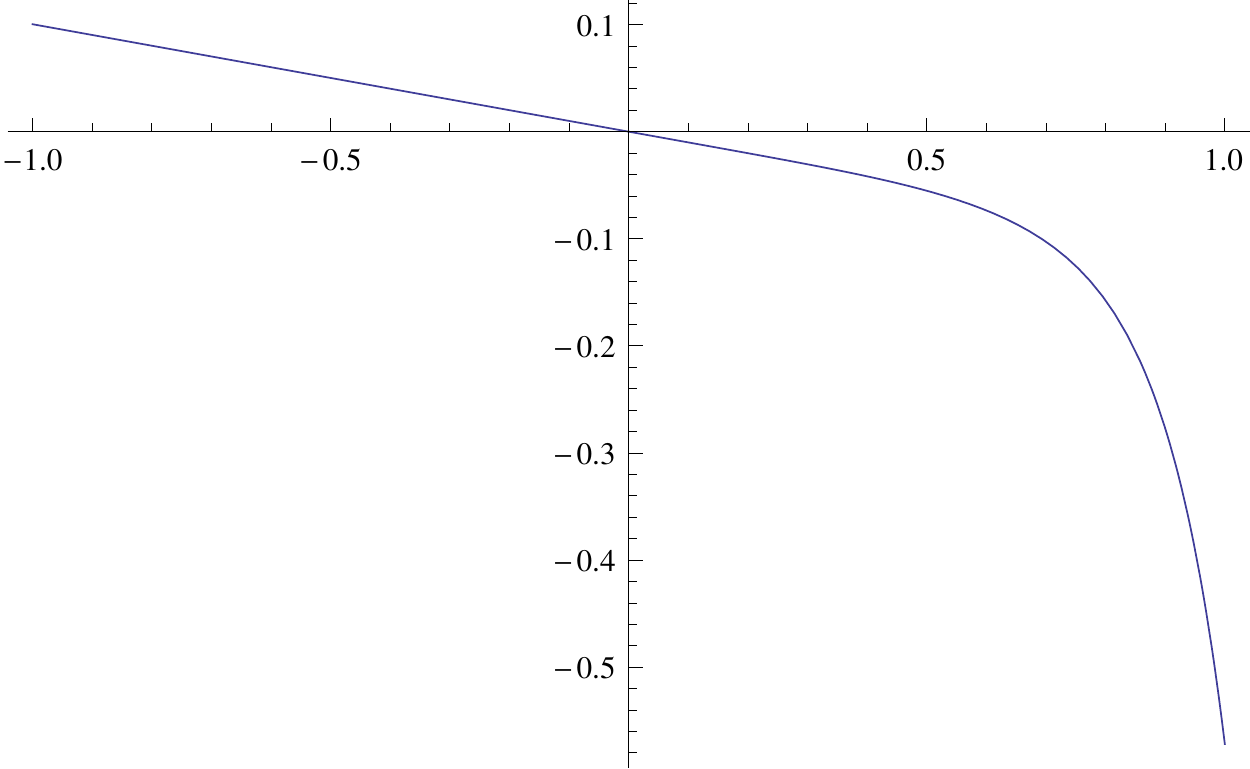} 
\vfill
\caption{$R_1(x)$, $f_1(x)$ and $\cA_1(x)$ (from top to bottom) against
  $x$ for $-1<x<1$. Note that the regular centre is at $x=1$ and the
  lightcone at $x=0$. In the outer region $x<0$, these are just the
  linear functions (\ref{R1xouter}) to (\ref{A1xouter}).}
\label{figure:Garfinklecorrections}
\end{figure}


\subsection{$\Lambda$ corrections of the null continuation}
\label{section:Lambdanull}


The $\Lambda$ corrections of the null continuation are given by
\begin{eqnarray}
\label{R1xouter}
R_1(x)&=& \frac{4^{1\over n} n^2}{16 (1-6n)} x,\\
f_1(x)&=&- \frac{4^{1\over n} n^3}{8(1-5n)(1-6n))} x,\\
\label{A1xouter}
\cA_1 (x)&=& \frac{4^{1\over n} n^2 (1-8n)}{16(1-5n)(1-6n)}x,
\end{eqnarray}
This is matched to the highest possible order to the particular
$\Lambda$ correction of the Garfinkle solution specified above. The
 $\Lambda$ corrections of the Garfinkle solution with the null
  continuation, in the gauge where they all vanish at $x=0$, are
shown in Fig.~\ref{figure:Garfinklecorrections}.


\subsection{Derivation of $\gamma$}
\label{section:gammaderivation}


We conjecture that for a yet unknown reason the true critical solution
has only one growing mode with Lyapunov exponent $\lambda_0$. Then the
scaling of the maxima and minima of Ricci and their location in proper
time (relative to the accumulation point and relative to each other)
can be calculated by the standard argument based on dimensional analysis
\cite{KoikeHaraAdachi1995,GundlachLRR}. We summarise it here for
completeness.

Assume that the first maximum of Ricci is reached when the solution moves
away from the critical solution, and that this happens when the one
growing mode has reached some $O(1)$ reference amplitude at which the growing
perturbation becomes nonlinear and stops growing exponentially in
$T$. This gives
\begin{equation}
|p-p_*|e^{\lambda_0T_{\rm nonlin}}\propto c_{2,{\rm
    top}} e^{\lambda_0T_{\rm nonlin}} \sim 1.
\end{equation}
Then, because all scales are proportional to $\exp(-T)$ and because
the Ricci scalar has dimension of inverse length squared,
\begin{equation}
R_{\rm max}\propto e^{2T_{\rm nonlin}}\propto |p-p_*|^{-2\gamma}, 
\end{equation}
where
\begin{equation}
\gamma={1\over \lambda_0}={2n\over 2n-1}={8\over 7}\simeq 1.1429,
\end{equation}
and we have assumed $n=4$ in the last equality. 
After the growing mode has become nonlinear, the evolution is
no longer CSS, but as we are now on very small scales, the
cosmological constant can locally be neglected with respect to the
scalar field gradient (squared) in the stress-energy tensor. Hence the
subsequent local evolution is approximately scale-invariant, and its
actual overall length scale is set by the length scale of the interim
data at $T_{\rm nonlin}$, which is $\ell \exp T_{\rm nonlin}$. We have
found numerically that this universal subsequent evolution has (at
least) two maxima and two minima before blowup. As the entire solution
scales, so will both the locations in proper time of these extrema (as
$|p-p_*|^\gamma$), and their values.


\subsection{Derivation of $\delta$}
\label{section:deltaderivation}


The ``standard'' argument \cite{KoikeHaraAdachi1995,GundlachLRR} would
now continue by noting that in near-supercritical evolutions the
above-mentioned universal evolution results in a black hole, and
that the linear size of this black hole scales with the overall
length scale, and hence as $|p-p_*|^\gamma$. In $d+1$ spacetime
dimensions, the black hole mass has units of (length)${}^{d-2}$, and
so scales as $|p-p_*|^\delta$ with $\delta=(d-2)\gamma$. However, in
$d=2$ this argument fails because $M$ is dimensionless. We also know
that no black holes can form from regular data for $\Lambda=0$, and so
it is clear that $\Lambda$ must play a role, leading to anomalous
scaling exponent. The following theoretical model appears to be
consistent with our numerical experiments.

Assume that the true critical solution is well approximated by the
$n=4$ Garfinkle solution inside the lightcone and its null
continuation outside the lightcone, plus their first
$\Lambda$ corrections. (For clarity of presentation only, we retain
the generic $n$.) Assume further that the true critical solution
has only one growing mode that is well approximated by the $m=7$ (top)
perturbation mode, while the $m=2,3$ modes disappear in the true
critical solution.

By definition, a MOTS is given by $\rb_{,\hat v}=0$, meaning that the
area radius does not grow on an outgoing null ray. (We use $\rb_{,\hat
  v}$ because $\hat v$ is a regular coordinate
at the lightcone while $\tilde v$ is not.) In terms of $R(x,T)$, this
is given by
\begin{equation}
\rb_{,\hat v}(\hat u,\hat v)=-e^{\left({1\over 2n}-1\right)T}R_{,x}(x,T).
\end{equation}

In the null-continued Garfinkle solution without any
$\Lambda$ corrections, every point in the $(r,t)$ plane outside the
lightcone is a MOTS, and so the AH is not well defined. Therefore we
add the first-order $\Lambda$ corrections. From
Fig.~\ref{figure:Garfinklecorrections} we see that everywhere except
close to the centre $R_1'(x)<0$ and so this will make $R_{,x}$ more
negative. But close to the centre where $R_1'(x)>0$, $|R_0'(x)|\gg
\exp(-2T)|R_1'(x)|$, and so $R_{,x}$ remains negative there,
too. Outside the lightcone, $R_0'(x)=0$ and $R_1'(x)<0$, so $R_{,x}$ is
also negative. Hence the leading $\Lambda$ correction removes all
MOTS, as already pointed out for the Garfinkle solution (inside the
lightcone) in \cite{CCF}.

Assuming that we have fine-tuned the initial data and have reached
large $T$, we now add the growing perturbation mode, but neglect all
decaying modes. Given that the $\Lambda$ corrections have removed MOTS
already in perturbation theory, we shall see that the growing
perturbation mode will bring them back for $p>p_*$. We begin with the
region outside the lightcone, where
\begin{eqnarray}
\label{Rxexpr}
R_{,x}(x,T)&=&e^{-2T}R_1'(x) 
+ c_{2,{\rm top}} e^{\lambda_0T} b_{\rm top}'(x) \\
\label{Rxouter}
&=& e^{-2T}\frac{4^{1\over n} n^2}{16 (1-6n)} 
 -c_{2,{\rm top}} e^{\lambda_0T} {C_{c,{\rm top}}\over 2}.
\end{eqnarray}
Here ``top'' denotes the most rapidly growing mode, with $m=2n-1$, and
hence $\lambda_0=1-{1\over 2n}$.  
We have assumed that the other growing modes, with $1<m<n$, of the
null-continued Garfinkle solution are not present in the true critical
solution. [However, from (\ref{bxnullcont}) we see that $b'(x)=0$ for
these modes, so they would not contribute to $R_{,x}(x,T)$
anyway.] Recall also that $R_0'(x)=0$ in the outer region, and so does
not contribute to $\rb_{,\hat v}$.

If there is only one growing mode, then its amplitude must be zero at
$p=p_*$, and so must be proportional to $p-p_*$ to leading
order. Hence, outside the lightcone, an AH is present for $c_2>0$,
corresponding by assumption to $p-p_*>0$. It is the outgoing null
surface $T=T_{{\rm AH},0}$, given by
\begin{equation}
\rb_{,\hat v}=0 \quad \Leftrightarrow \quad
p-p_*\propto c_{2,{\rm top}} \propto e^{-(2+\lambda_0)T_{{\rm AH},0}},
\end{equation}
where the proportionality signs hide irrelevant constant factors. But
on this null segment of the AH, 
\begin{equation}
M_{{\rm AH},0}={\rb^2\over \ell^2}=R^2e^{-2T}= {1\over
  4}e^{-2T_{{\rm AH},0}} \propto (p-p_*)^\delta,
\end{equation}
where
\begin{equation}
\label{mydelta}
\delta={2\over 2+\lambda_0}={4n\over 6n-1}={16\over 23}\simeq 0.6957,
\end{equation}
and we have assumed $n=4$ in the last equality. Note this is the mass
on the null part of the AH outside the lightcone, not yet the mass of
the EMOTS.  However, inside the lightcone the contribution of
$R_0'(x)$ to $R_{,x}(x,T)$ dominates the contributions of the
$\Lambda$ correction and the perturbation modes except just inside the
lightcone, where $R_0'(x)\to 0$ as $x\to 0_+$. We conclude without
explicit calculation that the AH must continue as a (probably
spacelike) surface running just inside the lightcone. Hence the EMOTS,
or minimum of $t_{AH}(r)$, must occur just inside the lightcone of the
critical solution.  Hence
\begin{equation}
M_{\rm EMOTS}\simeq M_{{\rm AH},0}.
\end{equation}
Moreover, while the EMOTS is slicing-dependent, the theoretical AH
curve $\rb_{,\hat v}=0$ plotted in the regular coordinates $(\hat
u,\hat v)$ makes a sharp bend from ``almost'' ingoing null to outgoing
null as it crosses the lightcone. This shape is confirmed by
  plots of the AH in coordinates $(t,r)$ in near-critical numerical
  evolutions. Hence almost any time slice will first intersect the AH
at this sharp bend, and so the location of the EMOTS and its mass
depend only weakly on the slicing.

Numerical data confirm that the EMOTS occurs just inside the lightcone, or
\begin{equation}
\tilde u_{\rm EMOTS}\ll \tilde v_{\rm EMOTS}< 0,
\end{equation}
For sub8 to sub20 Gaussian data, $\tilde v_{\rm EMOTS}/\tilde u_{\rm
  EMOTS}\propto (p-p_*)^{0.7}$. For sub20 to sub26 data, $\tilde
v_{\rm EMOTS}$ is approximately constant. The numerical data also
confirm that
\begin{equation}
M_{\rm EMOTS}\simeq {1\over 4}\left(-{\tilde u_{\rm EMOTS}\over \ell}\right)^2.
\end{equation}

Consider now the very special initial data consisting of the critical
solution plus a small amplitude $p$ times the growing mode. For
sufficiently large (but still very small) $p$, such that $T=T_{{\rm
    AH},0}$ intersects the initial data surface $t=0$, we expect a
MOTS to be present in the initial data. More precisely, if we evolved
these initial data backwards in time we would find that the EMOTS had
formed at some $t<0$, roughly where $T=T_{{\rm AH},0}$ intersects the
lightcone of the critical solution. We have argued above that the AH
extends from the EMOTS towards larger $r$ as the curve $T=T_{{\rm
    AH},0}$ with constant mass. Hence the MOTS at the intersection of
$t=0$ with the AH still has the same mass as the EMOTS did.

The value of $\delta=16/23$ derived here compares well with our
numerical value of $\delta \simeq 0.68(4)$, but we cannot be sure that
$\lambda_0=7/8$ exactly in the true critical solution. Hence we note
that, with $\gamma=1/\lambda_0$, (\ref{mydelta}) can be restated
 more robustly as
\begin{equation}
\label{deltagammalaw}
\delta={2\gamma\over 2\gamma+1}, 
\end{equation}
independently of the value of $\gamma$. What has gone into this
relation is the assumption that both the $\Lambda$ correction and the
growing mode make a constant contribution to $R_{,x}$ at leading order
-- something we would expect to hold to leading order in $x$ for
smooth functions.


\subsection{An exact continuation of the Garfinkle solution
  beyond the lightcone with $\Lambda<0$}
\label{section:outeransatz}


Three clear observations in near-critical evolutions of generic initial
data were that the critical solution outside the lightcone appears to
have $M\simeq 0$, $R\simeq 1/2$ and $\phi\simeq cT$. As we have seen,
these are exact properties of the null continuation, but the null
continuation is a solution only for $\Lambda=0$. This can be rectified
by incorporating the effects of $\Lambda$ perturbatively, as we have
done in Sec.~\ref{section:Lambdanull}. However, an
{\em exact} solution can also be found in which these qualitative
features hold, and which perturbatively reduces to the null
continuation with $\Lambda$ corrections.

Consider again the metric (\ref{xTmetric}). For general $\bcA(x,T)$
and $R(x,T)$, with the scalar field $\phi=c[T+f(x,T)]$, this ansatz is
generic, with $T={\rm const}$ and $x=0$ null by ansatz. As we saw, for
$\bcA=\bcA(x)$ and $R=R(x)$, compatible with $f=f(x)$, the metric is
CSS. Consider now instead
\begin{eqnarray}
\label{ccAdef}
\bcA&=&\ccA(\cx), \\
R&=&\cR(\cx), \\
\label{cfdef}
f&=&\cf(\cx), 
\end{eqnarray}
where we have defined the ``slow $x$'' 
\begin{equation}
\cx:=e^{-2T}x.
\end{equation}
Note the lightcone is at $x=\cx=0$. 

To understand the geometric significance of this ansatz, we express
the metric (\ref{xTmetric}) with (\ref{ccAdef}-\ref{cfdef}) in the
coordinates $(\cx,T)$ instead of $(x,T)$. The result is
\begin{equation}
\label{checkxTmetric}
ds^2=\ell^2\left[
e^{2\ccA}\left(d\cx-{\cx\over 2\cn}\,dT\right)dT+e^{-2T}\cR^2\,d\theta^2\right],
\end{equation}
where we have defined the constant
\begin{equation}
\label{ncheckdef}
\cn:={n\over 1-4n}.
\end{equation}
The functional form of this metric differs from (\ref{xTmetric}) only
by the absence of the overall factor $\exp(-2T)$ in the metric in the
$(x,T)$ plane. Hence, $K:=\partial/\partial T$ now acts as a homothetic
vector field only on the metric along the orbits of the circular
symmetry, but as a Killing vector field on its orthogonal
complement. We can also see immediately that the mass function
(\ref{BTZmass}) now takes the form
\begin{equation}
M=\mu(\cx)e^{-2T},
\end{equation}
and so is exponentially small.

It is straightforward to verify that the ansatz
(\ref{ccAdef}-\ref{cfdef}) transforms the five field equations {\em
  exactly} into ODEs in $\cx$. Moreover, by introducing the auxiliary
variables
\begin{equation}
\cF(x):=\cf'(\cx), \qquad \crho(\cx):={\cR'(\cx)\over \cR(\cx)},
\end{equation}
we reduce this system to the pair of first-order ODEs
\begin{eqnarray}
\cF'&=&{(\cn-1-\cx\crho)\cF-\cn\crho\over \cx}, \\
\crho'&=&{-\tilde c^2\cx(\cF^2+2\crho \cF)
+\crho[\cx\crho-2\cn(1+\tilde c^2)-1]\over \cx},
\end{eqnarray}
and the constraint
\begin{equation}
e^{2\ccA}={\tilde c^2\cx(\cF^2+2\crho \cF)-\crho[2\cx\crho-2\cn(2+\tilde
    c^2)]\over \cn}.
\end{equation}
Here the constant $\tilde c$ is related to the constant $c$ by
(\ref{ctildef}), and hence to the constant $n$ by (\ref{ct2nrelation}).

Consider for a moment the parameter $\cn$ as unrelated to $n$. Locally
in $\cx$, the reduced system then has a 2-parameter family of
solutions. The solution $(\ccA,\cR,\cf)$ follows by integration, where
an additive constant in $\cf$ and a constant factor in $\cR$ can be
fixed arbitrarily. Moreover, the system has a scale invariance
  that reflects an arbitrary overall factor in the definition of
  $\cx$, for a third integration constant.

If we want the solution to include a regular lightcone $\cx=0$, we we
are forced to choose
\begin{equation}
\cn=-{1\over 2(1+\tilde c^2)}.
\end{equation}
Modulo (\ref{ct2nrelation}), this is equivalent to
(\ref{ncheckdef}). The solution with a regular lightcone is then
unique, up to the above-mentioned three integration
constants. $\crho(x)>0$ throughout this solution, meaning that there
are no outer-trapped surfaces.

Choosing furthermore $n=4$, the equations become
\begin{eqnarray}
\cF'&=&{4\crho-\cF(19+15\cx\crho)\over 15\cx}, \\
\crho'&=&-{7\cF^2\over 8}-{7\cF\crho\over 4}+\crho^2, \\
e^{2\ccA}&=&{-105\cx\cF^2-210\cx\cF\crho+8\crho(23+30\cx\crho)\over
  32}.
\end{eqnarray}

We can use the three integration constants to match the regular light
cone solution to the Garfinkle solution at the lightcone continuously
(but of course not analytically). In particular, $\crho(0)$
  fixes the overall scale of $\cx$.

To compare our numerical solutions
  against this exact solution, we note that, from (\ref{dsdx}), we have
\begin{equation}
s_{,\cx}(\cx,T)=-{1\over 2}\ell e^{T+2\ccA(\cx)},
\end{equation}
and we can use this to define $\cx$ from the affine parameter $s$
along outgoing null curves, with $s=0$ on the lightcone of the
critical solution, so that $s(\cx=0)=0$. Hence to check for this
symmetry, we should plot $\bcA$, $R$ and $f$ against the similarity
coordinate
\begin{equation}
\label{csclambda}
\check\lambda:={s-s_{\rm lightcone}\over \ell e^T} =e^{-2T}(\bar\lambda-1),
\end{equation}
We compute $\check\lambda$ in the exact solution as
\begin{equation}
\check\lambda =-{1\over 2}\int_0^\cx e^{2\ccA(\cx)}\,d\cx.
\end{equation}

The regular lightcone solution blows up at $\cx\simeq 40$ (inside the
lightcone) with $R\to0$, meaning that in this ansatz we cannot have
both a regular lightcone and a regular centre. This is not surprising,
and we do not want to use this ansatz inside the lightcone anyway. It
also blows up at $\cx\simeq -10$ (outside the lightcone) with
$R\to\infty$ and $s\to\infty$, so this blowup is likely to be caused only
by infinity being mapped to a finite coordinate value.

Taylor-expanding about $\cx=0$, we obtain
\begin{eqnarray}
\check R(\cx) &=& \check R(0)+{d\check R\over d\cx}(0)\,\cx
+O(\cx^2) \nonumber \\ 
&=& \check R(0)+{d\check R\over d\cx}(0)\,e^{-2T}x+O\left(e^{-4T}x^2\right)
\nonumber \\ 
&=& R_0(0)+e^{-2T}{dR_1\over dx}(0)\,x+\dots \nonumber \\
\nonumber \\
&=&R_0(x)+e^{-2T}R_1(x)+\dots
\end{eqnarray}
[The third equality follows from matching, and the last equality
  because $R_0(x)$ is constant and $R_1(x)$ proportional to $x$.]
Hence to this order we recover the null continuation and its first
$\Lambda$ correction. We expect that higher orders in $\cx$ correspond
to higher $\Lambda$ corrections (fixing a suitable gauge at each order
in $\exp-2T$).

In our numerical simulations, we only access very small (negative)
values of $\cx$ and hence $\check\lambda$, because $\exp(-2T)$ is
small. This means that we cannot distinguish the exact solution from
its approximation to first order in $\Lambda$. In fact, other
deviations from the null continuation (the zeroth order) from other
sources are already larger than the first $\Lambda$ correction.

It is possible that there exists an analytic function $\xi(x,T)$ such
that $\xi\simeq x$ for $1>x>\epsilon$ and $\xi\simeq \cx$ for
$x<-\epsilon$, with a transition in a boundary layer of
width $\epsilon$ around the lightcone, and such that making $\bcA$,
$R$ and $f$ functions of $\xi(x,T)$ only transforms the field
equations into ODEs in $\xi$. However, if there is no $\xi$ that gives
rise to a global ODE system then it may still be possible to make a
simple ad-hoc ansatz for $\xi$, and expand the corresponding PDE
system in powers of $\exp(-2T)$ on both sides of the lightcone at
once, rather than separately, as we have done, thus maintaining
analyticity at the lightcone order by order in $\Lambda$.


\subsection{Construction of initial data for the amended Garfinkle solution}
\label{section:Garfinkledata}


To extract the free initial data $B(r)$, $B_{,t}(r)$, $\phi(r)$,
$\phi_{,t}(r)$ for the Cauchy code based on (\ref{PCmetric}), we need
to define a time slice $t=0$ through that solution, and a radial
coordinate $r$ on it, such that both are regular, and $t\pm r$ are
null coordinates. We choose $t=(\hat v+\hat u)/2+{\rm const.}$ and
$r=(\hat v-\hat u)/2$, where $\hat u$ and $\hat v$ in turn are related
to $x$ and $T$ by (\ref{xThatuv}).

The Garfinkle solution assumes $\Lambda=0$ whereas the form
(\ref{PCmetric}) of the metric has an inbuilt compactification at
timelike infinity that makes sense only if $\Lambda<0$. We therefore
make a smooth switchover from the initial data discussed above in an
interior region, to vacuum data $\phi_{,r}=\phi_{,t}=0$ in the gauge
$B_{,r}=B_{,t}=0$ in an exterior region. We then solve the constraint
equations (8) and (9) of \cite{PretoriusChoptuik} as algebraic
equations for $A_{,t}(r,0)$ and $A_{,r}(r,0)$, and then solve a
first-order quasilinear ODE for $A(r,0)$. This pair of algebraic
equations becomes singular on MOTS, and so we must include the first
$\Lambda$ corrections in the data to avoid MOTS. For the same reason,
we do not try to impose the gauge $B(r,0)=B_{,t}(r,0)=0$ that we have
used for evolving generic initial data.

The resulting initial data are parameterised by $T_{\rm initial}$, the
value of $T$ at $(r=0,t=0)$, which governs the magnitude of the
$\Lambda$ corrections, the value $r_{\rm lightcone}$ of $r$ where the
lightcone of the Garfinkle solution intersects $t=0$, and the location
$r_0$ and width $\Delta r$ of the switchover from Garfinkle to
trivial data. The switched initial data are obtained by multiplying
the initial data for $B_{,r}$, $B_{,t}$, $\phi_{,r}$ and $\phi_{,t}$
by a switch-off function $\chi_-(r)$ that goes smoothly to zero, then
integrating to get initial data for $B$ and $\phi$, and finally
solving the constraints to get initial data for $A$ and $A_{,t}$. 

We use the switch-off function $\chi_-$ defined by
\begin{eqnarray}
\chi_-(r)&:=&{1- \chi\left({r-r_0\over \Delta r}\right)\over 2}, \\
\chi(x)&:=&\tanh\left[18\left({1\over 4}x+{3\over
    4}x^3\right)\right].
\end{eqnarray}
The coefficients of $x$ and $x^3$ in $\chi(x)$ have been chosen so
that when we work in double precision the switching happens
effectively between $r_0-\Delta r$ and $r_0+\Delta_r$, while at the
same time minimising the first four derivatives of $\chi(x)$.


\section{Conclusions}


The presence of a negative cosmological constant in massless scalar
field collapse adds reflecting boundary conditions and breaks scale
invariance. However, simulations in 3+1 and higher dimensions have
shown that there is a regime in phase space where the outer boundary
conditions are irrelevant and the local dynamical effect of the
cosmological constant is neglible compared to the scalar field
stress-energy tensor. Critical collapse then proceeds as without a
cosmological constant, showing local discrete self-similarity and
scaling of the maximal curvature (for subcritical data) or initial
black hole mass (for supercritical data). (The word ``initial'' had to
be inserted here as the reflecting boundary conditions will lead to
continuing growth of the black hole mass.)

In 2+1 dimensions significant differences are clear a priori: Energy
and mass are dimensionless, and hence the dimensional analysis formula
(\ref{deltafromgamma}) linking Ricci and black hole mass scaling
cannot hold. There is a mass gap, with empty adS space having a mass
of $M=-1$, while regular initial data can form a black hole only if
they have $M>0$. Finally, black holes cannot form from regular initial
data at all in 2+1-dimensional gravity without a negative cosmological
constant. Hence the effect of the cosmological constant in critical
collapse cannot be just perturbative. 

Pretorius and Choptuik \cite{PretoriusChoptuik} fine-tuned four
1-parameter families of circularly symmetric initial data to the threshold of
prompt collapse (before reflection from the outer boundary) and found
universal continuous self-similarity (CSS), Ricci scaling and mass
scaling. This was confirmed in \cite{HusainOlivier}.

Setting $\Lambda=0$ as a heuristic starting point, Garfinkle
\cite{Garfinkle} found a countable family of analytic CSS solutions
and gave some numerical evidence that the $n=4$ member of the family
agrees with the near-critical evolutions of \cite{PretoriusChoptuik},
at least inside the lightcone of the accumulation point. Garfinkle and
Gundlach \cite{GarfinkleGundlach} showed that the Garfinkle solutions
have $n-1$ growing perturbation modes, thus apparently ruling out
Garfinkle's $n=4$ solution as the critical one.  Cavagli\`a, Cl\'ement
and Fabbri \cite{CCF} showed that the Garfinkle solution(s) can be
perturbatively corrected for $\Lambda<0$, and that this does not
change its perturbation spectrum. They also noted that when $\Lambda$
is taken into account the lightcone of the $n=4$ Garfinkle solution
is no longer a marginally outer-trapped surface (MOTS) -- something
that would otherwise independently rule it out as a critical solution.

In the numerical part of our paper, we have repeated the time
evolutions of \cite{PretoriusChoptuik}, using essentially the same
algorithm (coded independently), and re-analysed the data. In
  fine-tuned Gaussian data, we find a Ricci-scaling exponent of
$\gamma\simeq 1.23(4)$, compatible with the $\gamma\simeq 1.2\pm0.05$ of
\cite{PretoriusChoptuik}, and a mass-scaling exponent of $\delta\simeq
0.68(4)$, roughly compatible with the value of $\delta\simeq 0.81$
given by \cite{HusainOlivier}. (The value $\delta=2\gamma$ of
\cite{PretoriusChoptuik} is incorrect.)

We have also found excellent agreement between the $n=4$ Choptuik
solution and near-critical time evolutions inside the lightcone of the
accumulation point. In particular, we can definitely rule out any
other value of $n$. (Based on the fact that these have $n-1$ growing
modes, one might otherwise have suspected the $n=2$ solution to be the
critical one.) However, we can also definitely rule out the analytic
continuation of the $n=4$ Garfinkle solution beyond the lightcone as
the critical solution there.

Rather, the observed critical solution outside the lightcone seems to
be characterised by $M\simeq 0$, $\rb \simeq (-\tilde u)/2$, and
$\phi\simeq c\ln(-\tilde u)+{\rm const}$, where $\rb$ is the area
radius, $\tilde u$ retarded time normalised to proper time at the
centre and with its origin suitably adjusted, and $\phi$ the scalar
field.

In the theoretical part of our paper, we have shown that there is in
fact a simple exact solution of this kind for $\Lambda=0$, which can
be matched to the Garfinkle solution at the lightcone. We call this
the null-continued Garfinkle solution. We have also calculated the
leading-order $\Lambda$ correction of this solution and its
perturbations. As the adjustment to $\Lambda<0$ is done perturbatively
in powers of $\Lambda$, the perturbation spectrum remains unchanged
from the original Garfinkle solution. The $\Lambda$ corrections remove
the MOTS and thus one obstacle for this being the critical solution,
but the obstacle of three growing modes remains, and the matching
procedure has introduced a new obstacle, namely a lack of analyticity
at the lightcone -- our solution and its perturbations are only $C^3$
there. (All other known critical solutions are analytic at the
lightcone, and are in fact defined by this property.) 

Switching again to numerical time evolutions, we have taken this
``amended Garfinkle solution'' and taken five 1-parameter families of
initial data through it, including the addition of its putative three
growing modes. We do not find any evidence for three growing
modes. Rather, our results are compatible with our theoretical
predictions of $\gamma=8/7$ and $\delta=16/23$, based on a single
growing mode with $k=7/8$.

Based on the simple null continuation suggested by the numerics, we
have also constructed an exact solution for $\Lambda<0$ that gives the
null-continuation and its first $\Lambda$ correction when expanded to
first order in $\Lambda$, and we expect that this holds to all
orders. This is satisfactory from a theoretical point of view, but in
our numerical evolutions we do not have enough numerical accuracy to
see the difference between our exact outer solution and its approximation to
first order in $\Lambda$, so we have used only the approximation in the
main numerical part of this paper.

As discussed in Sec.~\ref{section:outeransatz} above, the missing
ingredient for constructing the exact critical solution is a way of
analytically gluing together the $\Lambda$-corrected Garfinkle
solution inside the lightcone to with new exact solution outside the
lightcone. We conjecture that this analytic gluing procedure will
somehow select $n=4$ and remove two of the three growing modes.

\acknowledgments

We would like to thank Piotr Bizo\'n for extensive helpful
discussions. This work was supported by the NCN grant
DEC-2012/06/A/ST2/00397, and in part by computing resoruces of ACC
Cyfronet AGH. CG would like to thank Jagellonian University, MFI
Oberwolfach, KITP and Chalmers Technical University for hospitality
during parts of this work. JJ acknowledges the hospitality of
University of Southampton where part of this work was done.


\end{document}